\documentclass[iop, dvipdfmx, revtex4]{emulateapj}
\usepackage{bm}
\usepackage{ulem}

\usepackage{mathrsfs}

\usepackage{graphicx}
\usepackage{epstopdf}

\begin{document}

\title{The Properties of Planetesimal Collisions under Jupiter's Perturbation and the Application to Chondrule Formation via Impact Jetting}

\author{Shoichi Oshino\altaffilmark{1,2}, Yasuhiro Hasegawa\altaffilmark{3}, Shigeru Wakita\altaffilmark{1,4}, and Yuji Matsumoto\altaffilmark{1,5,6} }
\altaffiltext{1}{Center for Computational Astrophysics, National Astronomical Observatory of Japan, Mitaka, Tokyo 181-8588, Japan}
\altaffiltext{2}{Institute for Cosmic Ray Research, University of Tokyo, Hida, Gifu 506-1205, Japan}
\altaffiltext{3}{Jet Propulsion Laboratory, California Institute of Technology, Pasadena, CA 91109, USA}
\altaffiltext{4}{Earth-Life Science Institute, Tokyo Institute of Technology, Meguro-ku, Tokyo 152-8550, Japan}
\altaffiltext{5}{Planetary Exploration Research Center, Chiba Institute of Technology, Narashino, Chiba 275-0016, Japan}
\altaffiltext{6}{Institute of Astronomy and Astrophysics, Academia Sinica, Taipei 10617, Taiwan}

\email{oshino@icrr.u-tokyo.ac.jp}

\begin{abstract}
Understanding chondrule formation provides invaluable clues about the origin of the solar system.
Recent studies suggest that planetesimal collisions and the resulting impact melts are promising for forming chondrules.
Given that the dynamics of planetesimals is a key in impact-based chondrule formation scenarios,
we here perform direct $N$-body simulations to examine how the presence of Jupiter affects the properties of chondrule-forming collisions.
Our results show that the absence/presence of Jupiter considerably changes the properties of high velocity collisions whose impact velocities are higher than 2.5 km s$^{-1}$;
high velocity collisions occur due to impacts between protoplanets and planetesimals for the case without Jupiter;
for the case with Jupiter, eccentricities of planetesimals are pumped up by the secular and resonant perturbations from Jupiter.
We also categorize the resulting planetesimal collisions and find that most of high velocity collisions are classified as grazing ones for both cases.
To examine the effect of Jupiter on chondrule formation directly,
we adopt the impact jetting scenario and compute the resulting abundance of chondrules.
Our results show that for the case without Jupiter, chondrule formation proceeds in the inside-out manner, following the growth of protoplanets.
If Jupiter is present, the location and timing of chondrule formation are determined by Jupiter's eccentricity, 
which is treated as a free parameter in our simulations.
Thus, the existence of Jupiter is the key parameter 
for specifying when and where chondrule formation occurs for impact-based scenarios.
\end{abstract}


\section{Introduction} \label{sec:intro}

The origin of the solar system is one of the important questions in astrophysics and planetary science.
A number of significant improvements have recently been done on the modeling of how the solar system formed and evolved to the current configuration 
\citep[e.g.,][]{2005Natur.435..459T,2011Natur.475..206W,2015Natur.524..322L}.
Such progress has been accelerated thanks to the discovery of a large number of extrasolar planetary systems \citep[e.g.,][]{2015ARAA..53..409W}, 
which proposes the importance of planetary migration driven either by planet-disk interaction \citep[e.g.,][]{1996Natur.380..606L,2004ApJ...604..388I,2014ApJ...794...25H,2016ApJ...832...41M}
or by $N$-body dynamics \citep[e.g.,][]{1996Sci...274..954R,2008ApJ...686..580C,2008ApJ...678..498N,2018ARAA..56..175D}.
Therefore, the formation of the solar system is currently viewed as a highly dynamical processe and 
the presence of Jupiter would play a critical role in both forming other solar system planets and shaping up their present orbital architecture.

The solar system stands out because one can access to its fossil record that is capsulated in primitive materials.
Chondrules are one of the famous examples that shed light on the origin of the solar system.
They are sub-mm to mm sized spherical materials found in chondritic meteorites \citep[e.g.,][]{sk05,s07}.
One conservative age estimate using Al and Mg isotopes suggests that chondrule ages cluster at $\sim 2\,\mathrm{Myr}$
after the formation of calcium-aluminum-rich inclusions \citep[CAIs, e.g.,][]{Villeneuve:2009aa,Villeneuve:2015aa}.\footnote{
Note that the age estimates using the Al and Mg isotopes provide relative values measured from the age of CAIs.
Absolute ages of chondrules can be estimated, using the Pb-Pb dating, 
which suggests that chondrule formation may have taken place and continued for 3-5 Myr after CAI formation 
\citep[e.g.,][]{2012Sci...338..651C,2017Bollard}.}
Given that the age of CAIs is about 4567 Myr \citep[e.g.,][]{2010EPSL.300..343A,2012Sci...338..651C}, 
it is expected that chondrule formation would intimately link with the early solar system formation in the solar nebula.

A number of mechanisms have been proposed for forming chondrules in the solar nebula 
\citep[e.g.,][]{1996Sci...271.1545S,2001ApJ...548.1029S,1993prpl.conf..939M,2000Icar..143...87D,1963Icar....2..152W,2001Icar..153..430I,2010ApJ...722.1474M}.
Recently, planetesimal collisions and the resulting impact melts have received considerable attention as a chondrule-forming process.
Two kinds of impact-based scenarios are proposed in the literature:
one is to consider low velocity collisions between planetesimals that are (partially) melted by the decay heat of short-lived radionuclides ($^{26}\mathrm{Al}$) 
and the subsequent splashes of melted ejecta \citep[e.g.,][]{2011E&PSL.308..369A,Sanders:2012aa,2018Icar..302...27L}.
The other is to consider high velocity collisions between planetesimals and/or between planetesimals and protoplanets, and the resulting jetting of ejecta 
\citep[]{2015Natur.517..339J,2016aHasegawa,2017ApJ...834..125W}.
Both scenarios can broadly reproduce the cooling rate ($10-10^3$ K hr$^{-1}$) of chondrules estimated from petrological experiments and chondrules' textures \citep[e.g.,][]{2012M&PS...47.1139D,Soulie:2017aa},
and may be applicable to the recent finding of a faster cooling rate of chondrules \citep[$>10^3$\,K\,hr$^{-1}$,][]{Villeneuve:2015aa}. 
For example, \citet{2014ApJ...794...91D,Dullemond:2016aa} estimated the cooling rate by considering the expanding plume from impact splash/jetting, and 
\citet{2015Natur.517..339J} developed a radiative transfer model for ejecta produced by impact jetting.
Impact-based scenarios would therefore be promising for better understanding of the formation mechanisms of chondrules.
However, more detailed studies are required for examining the validity of these scenarios.
Importantly, no mechanisms (including impact-based scenarios) can fully explain 
the rich properties of chondrules such as their thermal histories, abundances, and compositions \citep{2010GeCoA..74.4807R,2012M&PS...47.1139D,Soulie:2017aa}.

The dynamics of planetesimals is a key in impact-based scenarios, 
because the chondrule formation efficiency depends on the impact velocity ($v_{\rm imp}$) of planetesimal collisions, 
which is given as
\begin{equation}
v_{\rm imp} \equiv \sqrt{v_{\rm esc}^2 + v_{\rm ran}^2}, 
\label{eq:vimp}
\end{equation}
where $v_{\rm esc}$ is the surface escape velocity of two colliding bodies and $v_{\rm ran}$ is the random velocity of planetesimals.
The impact splash scenario predicts high sensitivity of produced chondrule mass to the impact velocity \citep[e.g.,][]{2011E&PSL.308..369A}.
On the other hand, a consistent conclusion is drawn from the impact jetting scenario that when the impact velocity exceeds $2.5\, \mathrm{km\,s}^{-1}$, 
about 1 \% of the impactor's mass becomes the progenitor of chondrules 
for both protoplanet-planetesimal and planetesimal-planetesimal collisions \citep{2015Natur.517..339J,2017ApJ...834..125W}.
This consistency in the impact jetting scenario has opened up a new window to explore chondrule formation in the framework of planet formation 
\citep[see Section \ref{sec:app} for both successes and problems in the impact jetting scenario]{2015Natur.517..339J,2016bHasegawa}.
These previous calculations, however, are conducted under the assumption that chondrule-forming regions are isolated from surrounding planet-forming regions. 
It is important to take into account the effect of nearby, growing planets such as Jupiter.

Here we perform direct $N$-body simulations to investigate how the presence of Jupiter affects 
the dynamics of planetesimals and the resulting collisions that may serve as chondrule-forming events.
Inclusion of Jupiter would be important because previous studies show that the perturbation from giant planets pumps up eccentricities and inclinations of planetesimals 
\citep[e.g.,][]{2000Nagasawa,2006Icar..184...39O}.
To examine the effect of Jupiter on chondrule formation explicitly, we will consider the impact jetting scenario and compute the resulting abundance of chondrules,
while our results are applicable to other impact-based, chondrule formation mechanisms.
In our $N$-body simulations, gravitational interactions between planetesimals, gas drag, and the perturbation from Jupiter are all included.

The plan of this paper is as follows. 
In Section \ref{sec:method}, we describe the numerical method and initial conditions of simulations.
In Section \ref{sec:Jup}, we consider both the resonant and secular perturbations from Jupiter and estimate their timescales.
In Section \ref{sec:result}, we show the results of our $N$-body simulations, 
focusing on the properties of planetesimal collisions such as impact velocities, eccentricities, and collision types.
In Section \ref{sec:app}, we apply the results of $N$-body simulations to the impact jetting scenario and estimate the produced abundance of chondrules.
In Section \ref{sec:dis}, we discuss the assumptions used in our simulations.
The brief conclusion of this work is provided in Section \ref{sec:conc}.

\section{Methods} \label{sec:method}

In this section, we describe the numerical method, setup, and assumptions of our simulations.

\subsection{Disk Model}

We adopt the minimum-mass solar nebula (MMSN) disk model \citep{1981PThPS..70...35H}.
In this model, the gas and solid surface densities are written as,
\begin{eqnarray}
\Sigma_{\rm gas} = 2400
  \left(  \frac{r}{1\ \rm au}\right)^{-3/2} \mathrm{g \, cm^{-2}} \label{eq:sgas},\\
\Sigma_{\rm solid} = 10 f_{\rm d} \left(  \frac{r}{1\ \rm au}\right)^{-3/2} \mathrm{g \, cm^{-2}} \label{eq:sdust},
\end{eqnarray}
where $f_{\rm d}$ is an increment factor.
Following \cite{2016aHasegawa}, we consider two values of $f_{\rm d}$: $f_{\rm d}=1$ and $f_{\rm d}=3$.
The former case is referred to as the standard disk and the latter is as the massive disk in this paper.
Note that even for the case that $f_{\rm d}=1$, the surface densities of both gas and solid are 1.4 times higher than those in the original MMSN model \citep[see][]{2000Icar..143...15K}. 
The presence of the nebular gas is important because the gas damps the eccentricity of planetesimals very efficiently (see below).
While a Jupiter-mass planet is included in a subset of our simulations, the interaction between the planet and the disk gas is not considered in this paper for simplicity.
The potential effect of the planet-disk interaction will be discussed in Section \ref{sec:dis}.

\subsection{$N$-body Simulations}

We compute the dynamics of planetesimals, growing protoplanets, and Jupiter (if included) around the sun under the presence of the nebular gas.

The equation of motion of the $i$-th particle is given as
\begin{equation}
\frac{\mathrm{d} {\bm r}_i^2}{\mathrm{d}t^2} = -\mathrm{G}M_* \frac{{\bm r}_i}{|{\bm r}_i|^3} - \sum_{i \neq j}^N \mathrm{G}m_j \frac{{\bm r}_i - {\bm r}_j}{|{ \bm r}_i - {\bm r}_j|^3} + {\bm F}_{{\rm gas},i},
\label{eq:motion}
\end{equation}
where ${\bm r}_i$ and $m_i$ are the position and mass of the particle, respectively, G is the gravitational constant, $M_*$ is the mass of the sun,
and ${\bm F}_{{\rm gas},i}$ is the gas drag force acting on the particle.
The first term of Equation (\ref{eq:motion}) is gravitational force arising from the sun, and
the second term is the sum of mutual gravitational interactions between particles.
The gas drag force is given as \citep{1976PThPh..56.1756A}
\begin{equation}
  {\bm F}_{{\rm gas},i} = - \frac{1}{2m_i}C_{\rm D} \pi R_i^2 \rho_{\rm g} u_i {\bm u}_i \label{eq:gas},
\end{equation}
where $C_{\rm D}=1$ is the gas drag coefficient, $R_i$ is the physical radius of the particle, and ${\bm u}_i$ is the relative velocity between the body and the gas.
The absolute value of the relative velocity ($u_i$) is given by $u_i=( (5/8)e_i^2 + (1/2) i_i^2 + \eta^2 )^{1/2} v_{{\rm K},i}$, 
where $e_i$ is the eccentricity, $i_i$ is the inclination, and $\eta$ is the fraction between the Kepler velocity ($v_{{\rm K},i}$) and gas velocity.
With our MMSN model, the spatial density of disk gas ($\rho_{\rm g}$) is written as
\begin{equation}
  \rho_{\rm g} = 2 \times 10^{-9} \left( \frac{r}{1\ \rm au}\right)^{-11/4} \mathrm{g \, cm^{-3}}.
\end{equation}

The orbital integration of particles is performed using the 4th-order Hermite scheme \citep{1992PASJ...44..141M} with the hierarchical timestep \citep{1991ApJ...369..200M}.
The mutual gravitational force is computed by directly summing up interactions of all pairs using GRAPE-DR that is the special-purpose computer for $N$-body simulations \citep{2007Makino}.
To compute the impact velocity precisely, we do not use the enhancement factor for planetesimal radius \citep{1996KI, 2008Morishima}.\footnote{Note that in most direct $N$-body simulations,
the radius enhancement factor of planetesimals is used for accelerating planetesimal collisions and shortening computational times.}
We assume the perfect merger in our simulations for simplicity; 
the merger arises when the physical radii of two particles overlap with each other.
As demonstrated below, however, various kinds of collisions will occur during protoplanet formation.
Hereafter, particles mean planetesimals and/or growing protoplanets.

\subsection{Classification of collisions} \label{sec:method_crit}

We introduce a classification of collisions that will be used in this paper.

To characterize collisions, we consider two quantities: the impact velocity ($v_{\rm imp}$, see Equation (\ref{eq:vimp})) and the impact parameter ($b$).
We calculate $v_{\rm imp} = \sqrt{(\bm{v}_{\rm t} - \bm{v}_{\rm i})^2}$ at the every moment of collisions, 
where $\bm{v}_{\rm i}$ and $\bm{v}_{\rm t}$ are the velocities of an impactor and a target, respectively.
The impact parameter is defined by $b = \sin \theta$, where $\theta$ is the impact angle \citep[]{2012ApJ...751...32S}.
Then, head-on collisions are denoted by $b \sim 0$ while oblique collisions are by $b > 0$.
Oblique collisions can be further divided into two types: non-grazing collisions and grazing ones.
Grazing collisions are realized when the impact parameter is larger than the size ratio of colliding bodies.
The corresponding critical impact parameter is given by \citep{2010ChEG...70..199A}
\begin{equation}
 b_{\rm crit} = \left( \frac{R_{\rm t}}{R_{\rm t}+R_{\rm i}} \right),
 \label{b_crit}
\end{equation}
where $R_{\rm t}$ is the radius of a target body and $R_{\rm i}$ is that of an impactor.
When the equal mass bodies collide, $b_{\rm crit}=0.5$, which is the minimum value of $b_{\rm crit}$.
As protoplanets grow, the value of $b_{\rm crit}$ for protoplanet-planetesimal collisions increases. 
It is expected that more non-grazing collisions should occur, following the growth of protoplanets.

Considering $v_{\rm imp}$ and $b_{\rm crit}$, we categorize collisions into three groups: perfect merger, partial accretion, and hit-and-run. 
In practice, we adopt the following simple criteria 
\citep[more detailed categorization for collision outcome is discussed in][]{Leinhardt&Stewart_ST2012, 2012ApJ...751...32S,Genda+2012}:
\begin{description}
  \item[Perfect merger] when $v_{\rm imp}\leq v_{\rm esc}$,
  \item[Partial accretion] when $v_{\rm imp}> v_{\rm esc}$ and $b<b_{\rm crit}$,
  \item[Hit-and-run] when $v_{\rm imp}> v_{\rm esc}$ and $b>b_{\rm crit}$.
\end{description}
We will use this classification and discuss what kinds of collisions are realized through protoplanet formation below.

In addition to the above classification, the impact velocity ($v_{\rm imp}$) itself is the key quantity for impact-based chondrule formation scenarios.
That is why the actual physical radius of planetesimals is used in our $N$-body simulations (see above). 
To examine the effect of Jupiter on chondrule formation via impact-based scenarios directly,
we will consider the impact jetting scenario \citep{2015Natur.517..339J,2017ApJ...834..125W}.
More specifically, we assume that when $v_{\rm imp}$ exceeds the critical velocity of 2.5\,km\,$^{-1}$,
the corresponding collisions are energetic enough to generate the progenitor of chondrules. 
These collisions are referred to as high velocity collisions in this paper (see Sections \ref{sec:result} and \ref{sec:app}).

\subsection{Numerical Setup and Assumptions}

The main purpose of this paper is to assess the effect of Jupiter's perturbation on the dynamics of planetesimals and the resulting collisions.
We, therefore, perform simulations with and without Jupiter.
Table \ref{table:parameter} summarizes our numerical setup.

For the case without Jupiter, we perform two simulations where the values of $f_{\rm d}$ are different.
These are labeled as MMSN1NJ and MMSN3NJ and correspond to the cases that $f_{\rm d}=1$ and $f_{\rm d}=3$, respectively.
For the case with Jupiter, we set the semimajor axis of Jupiter as $5.2 \, \mathrm{au}$ and treat the eccentricity of Jupiter ($e_{\rm J}$) as a free parameter.
This is because while the current $e_{\rm J}$ is measured at 0.0485 \citep[e.g.,][]{standish,standish_ch8}, 
its past value is unknown.
To  examine the effect of Jupiter's perturbation comprehensively,
we take $e_{\rm J}$ as 0, 0.05, and 0.1 (MMSN1JE0, MMSN1JE05 and MMSN1JE1, respectively) for the standard disk.
For the massive disk, we consider only the case that $e_{\rm J}=0.05$ (MMSN3JE05).
Note that whereas the formation timing of Jupiter is also poorly constrained \citep[e.g.,][]{1996Icar..124...62P,2009Icar..199..338L,2017PNAS..114.6712K},
we assume that Jupiter is fully formed when simulations begin.
By considering two extreme cases (with and without Jupiter),
we can bracket the role of Jupiter on the properties of planetesimal collisions.

We initially distribute 5000 equal-mass planetesimals from 2 au to 3 au, following the profile of $\Sigma_{\rm solid}$.
The mass of a planetesimal ($m_{\rm pl}$) is $1.79 \times 10^{24} \mathrm{\ g}$ 
for the standard disk ($f_{\rm d}=1$) and $5.36 \times 10^{24} \mathrm{\ g}$ for the massive disk ($f_{\rm d}=3$).
The radius of particles (either planetesimals or protoplanets) is computed with the assumption that their internal density is $\rho_{\rm p} = 3\ \mathrm{g\,cm^{-3}}$. 

The initial eccentricities and inclinations of planetesimals are given by Rayleigh distribution with 
dispersions of $\langle e^2\rangle^{1/2}=2 \langle i^2 \rangle^{1/2}=0.05$ \citep{1992Icar...96..107I}.
Our simulations follow the dynamical evolution of particles for $\gtrsim1$ Myr.

We do not consider gas disk evolution, disk turbulence, planet-disk interaction, and the effect of snow lines in our current simulations for simplicity.
The motivation and justification of neglecting these physical processes are discussed in Section \ref{sec:dis}.

\begin{table}[htb]
\caption{Summary of the initial setup} \label{table:parameter}
\centering
  \begin{tabular}{l|c|c|c} \hline
    NAME            & Jupiter                 & $f_{\rm d}$  & $m_{\rm pl}$                                      \\ \hline \hline
    MMSN1NJ     &  w/o                     & 1                  & $1.79 \times 10^{24} \mathrm{\ g}$   \\
    MMSN3NJ     &  w/o                     & 3                  & $5.36 \times 10^{24} \mathrm{\ g}$  \\
    MMSN1JE0   & $e_{\rm J}=0$      & 1                 & $1.79 \times 10^{24} \mathrm{\ g}$    \\
    MMSN1JE05 & $e_{\rm J}=0.05$ & 1                 & $1.79 \times 10^{24} \mathrm{\ g}$    \\
    MMSN1JE1   & $e_{\rm J}=0.1$   & 1                 & $1.79 \times 10^{24} \mathrm{\ g}$    \\
    MMSN3JE05 & $e_{\rm J}=0.05$  & 3                & $5.36 \times 10^{24} \mathrm{\ g}$  \\ \hline
  \end{tabular}
  \tablecomments{
  The column of Jupiter shows its eccentricities, and the "w/o" indicates that Jupiter is not included in the simulations.
  The column of $f_{\rm d}$ summarizes an adopted value of the increment factor (see Equation (\ref{eq:sdust})), 
  and the corresponding planetesimal mass ($m_{\rm pl}$) is listed in the fourth column.
  }
\end{table}

\section{Perturbation from Jupiter} \label{sec:Jup}

In this section, we consider how the dynamics of planetesimals is affected by the presence of Jupiter. 
There are two kinds of perturbations: the resonant perturbation and secular one.
To quantify these effects on our simulation results (see Section \ref{sec:result}),
we here estimate the timescales of these effects.

\subsection{Resonant part}\label{sec:res_part}

The eccentricity of a planetesimal located at a mean motion resonance with Jupiter's orbit gradually increases due to the gravitational force of Jupiter.
There are the 3:1 mean motion resonance at $2.50 \, \mathrm{au}$, 4:1 at $2.06 \, \mathrm{au}$, 5:2 at $2.82 \, \mathrm{au}$, 
7:2 at $2.26 \, \mathrm{au}$, and 7:3 at $2.96 \, \mathrm{au}$ from $2 \, \mathrm{au}$ to $3 \, \mathrm{au}$.
Here we adopt the circular restricted three-body problem to calculate the libration timescale of planetesimals.
As an example, we consider the 3:1 mean motion resonance.

Suppose that a perturbation source (Jupiter for this case) with a small value of eccentricities is located beyond the orbit of a test particle (i.e., a planetesimal)
and that the orbital planes of these two bodies are common.
Then the resonant term of the disturbing function is given as \citep[the second term of Equation (8.26) in][]{SSD} 
\begin{eqnarray}
\langle \mathcal{R}_{\rm res} \rangle &=& \frac{G M_{\rm J}}{a_{\rm J}} \left[ \mathscr{F}_{\rm d}(\alpha) e^{|j_4|} \cos \varphi \right] ,
\end{eqnarray}
where $M_{\rm J}$ and $a_{\rm J}$ are the mass and semimajor axis of Jupiter, respectively,
$\alpha$ is the ratio of semimajor axes between Jupiter and the planetesimal,
$\mathscr{F}_{\rm d}$ is the $\alpha$ dependent part of the direct term in the disturbing function, 
$e$ is the eccentricity of the perturbed particle, and $j_4$ is the order of the resonance. 
The resonant angle ($\varphi$) is expressed by \citep[Equation (8.27) in][]{SSD} 
\begin{eqnarray}
\varphi &=& j_1 \lambda_{\rm J} + j_2 \lambda + j_4 \varpi ,
\end{eqnarray}
where $\lambda$ is the mean longitude and $\varpi$ is the longitude of pericenter in $j_1$:$j_2$ resonance.

We now estimate the libration timescale of $\varphi$. 
Since $\lambda=n(t-\tau)+\varpi \equiv nt+\epsilon$, where $n$ is the mean motion, $t$ is a time, $\tau$ is the time of a pericenter passage, and $\epsilon = - n \tau+\varpi$, 
the second time derivative of $\varphi$ can be written as
\begin{eqnarray}
{\ddot \varphi} &=& j_2 {\dot n} + j_2 {\ddot \epsilon} + j_4 {\ddot \varpi }.
\end{eqnarray}
The contributions of ${\ddot \epsilon}$ and ${\ddot \varpi }$ are negligible because they are $\mathcal{O}((M_{\rm J}/M_*)^2)$, while ${\dot n}\sim \mathcal{O}(M_{\rm J}/M_*)$.
Using Lagrange equation, the time derivative of mean motion is written as 
\begin{eqnarray}
\dot{n} &=& -\frac{3}{a^2}\frac{\partial \langle \mathcal{R}_{\rm res} \rangle}{\partial \lambda} \nonumber , \\
&=& 3j_2 C_r ne^{|j_4|} \sin \varphi ,
\end{eqnarray}
where $C_r=(M_{\rm J}/M_*) n \alpha \mathscr{F}_{\rm d} $.
Under $\varphi \ll 1$, we obtain the libration equation of the resonant angle as \citep[Equation (8.46) in][]{SSD} 
\begin{equation}
\ddot{\varphi} = j_2  \dot{n} = -\omega_0^2 \varphi ,
\end{equation}
where $\omega_0^2 = -3j_2^2 C_r n e^{|j_4|}$ is always a positive quantity.

Finally, the libration timescale ($T_{\rm lib}$) of the 3:1 mean motion resonance (i.e., at $a=2.5$ au) is written as 
\begin{equation}
T_{\rm lib} \equiv \frac{2 \pi}{\omega_0} \simeq 1.3 \times 10^4 \left( \frac{e}{10^{-2}} \right)^{-1} \, \mathrm{yr}.
\label{eq:T_lib}
\end{equation}

\subsection{Secular part}\label{sec:sec_part}

The secular perturbation excites oscillation in eccentricities of planetesimals.
We derive the timescale of the secular perturbation between planetesimals and Jupiter.

Assuming the circular restricted three-body problem, the secular term of the disturbing function can be given as \citep[Equation (7.20) in][]{SSD} 
\begin{equation}
\mathcal{R}_{j,{\rm sec}} = n_j a_j^2 \left[ \frac{1}{2}A_{jj}(h_j^2 + k_j^2) + A_{jk}(h_jh_k + k_jk_k) \right] ,
\end{equation}
where $h_j=e_j \sin \varpi_j$, $k_j=e_j \cos \varpi_j$, and the indices 1 and 2 are a planetesimal and Jupiter, respectively \citep[also see][]{2000Nagasawa}. 
The components of the matrix $A$ are given by \citep[Equation (7.9) and (7.10) in][]{SSD} 
\begin{eqnarray}
A_{jj} &=& \frac{n_j}{4} \frac{m_k}{M_*+m_j} \alpha^2 b^{(1)}_{3/2},\\
A_{jk} &=& -\frac{n_j}{4} \frac{m_k}{M_*+m_j} \alpha^2 b^{(2)}_{3/2},
\end{eqnarray}
where $b^{(1)}_{3/2}$ and $b^{(2)}_{3/2}$ are the Laplace coefficients.
Since the perturbed equations are written as \citep[Equation (7.25) in][]{SSD} 
\begin{equation}
\dot{h_j}=\frac{1}{n_j a_j^2} \frac{\partial \mathcal{R}_{j,{\rm sec}}}{\partial k_j}, \,\dot{k_j}=-\frac{1}{n_j a_j^2} \frac{\partial \mathcal{R}_{j,{\rm sec}}}{\partial h_j},
\end{equation}
the solutions are given by \citep[Equation (7.28) in][]{SSD}
\begin{equation}
h_j=\sum_{i=1}^2 e_{ji} \sin(g_i t+\beta_i), \, k_j=\sum_{i=1}^2 e_{ji} \cos(g_i t+\beta_i) ,
\end{equation}
where the frequencies $g_i\ (i = 1, 2)$ are the eigenvalues of the matrix $A$, and $\beta_i$ is a phase determined by the initial condition.

We take planetesimals as mass less test particles, i.e., $m_1=0$, since $M_{\rm J}\gg m_1$.
Then the eigen frequencies can be derived from $g^2-A_{11}g=0$.
Given that 
\begin{eqnarray}
e_j & = & ( h_j^2 + k_j^2 )^{1/2}  \\ \nonumber
       & = & ( e_{j1}^2 + e_{j2}^2 + 2e_{j1}e_{j2}\cos{(\beta_1-\beta_2 + (g_1-g_2)t)} )^{1/2}, 
\end{eqnarray}
the timescale of the secular perturbation from Jupiter is 
\begin{eqnarray}
T_{J, \rm{sec}} &=& 2\pi/(g_1 - g_2)=2\pi/A_{11} .
\end{eqnarray}
Finally, the timescale of the secular perturbation from Jupiter becomes
\begin{equation}
\label{eq:secular}
T_{\rm J, sec} \sim 6.6 \times 10^4 \left( \frac{a}{2\,  \mathrm{au}} \right)^{-3/2} \left( 1+0.28\left( \frac{a}{2\,  \mathrm{au}} \right)^{2}+\mathcal{O}(\alpha^4) \right)^{-1} .
\end{equation}

We will compare these timescales with our numerical results in Sections \ref{sec:perturb} and \ref{sec:w_J}.

\section{Results} \label{sec:result}

We present the results of $N$-body simulations and discuss the properties of planetesimal collisions that occur in these simulations.
We first show the results for the case without Jupiter in Section \ref{sec:wo_J}. 
The effect of the perturbation from Jupiter is examined in Section \ref{sec:perturb}.
The results for the case with Jupiter are shown in Section \ref{sec:w_J}.
Our model parameters are listed in Table \ref{table:parameter}.
The data of our simulation results will be available upon request.

\subsection{Planetesimal collisions without Jupiter}\label{sec:wo_J}

In this section, we present the results for the case without Jupiter.
Given that runaway and oligarchic growth of protoplanets via planetesimal collisions are well explored in the literature 
\citep[e.g.,][]{1989Icar...77..330W,1996KI,1998Icar..131..171K,2003Icar..161..431T},
we do not discuss this and focus on the properties of planetesimal collisions.

We first discuss when and what value (high vs low) of the impact velocity of collisions are realized during the protoplanet's growth.
Figure \ref{fig:ITmass-ratio-woJ} depicts the results of the MMSN1NJ case (the standard disk) and those of the MMSN3NJ case (the massive disk) on the left and right panels, respectively.
On each panel, all the collisions are shown as a function of time.
Following the definition made in Section \ref{sec:method_crit},
high velocity collisions whose impact velocities exceed 2.5\,km\,s$^{-1}$ are denoted by the circles.
If the impact velocity is less than 2.5\,km\,s$^{-1}$,
then the corresponding collisions are labeled by the triangles.
Certain colors are allocated to these collisions to show the ratio between the impact velocity and escape one (see the top color bar).
The impact velocity is given from simulations and the escape velocity is calculated based on the total mass of the impactor and target.
To further specify what kinds of collisions (either planetesimal-planetesimal ones or protoplanet-planetesimal ones) take place,
the mass ratio between impactors and targets is shown on the left side of the vertical axis.
On the right side of the vertical axis, the mass of the largest particle at a certain time is traced by the black line with the Earth mass unit.
Our simulations show that high velocity collisions do not occur in the early stage, which is $\la 10^{6}$ yr for the MMSN1NJ case and $\la 3\times10^5$ yr for the MMSN3NJ case.
Instead, these collisions are realized as protoplanets grow.
We find that collisions with the impact velocity of $> 2.5\, \mathrm{km\, s}^{-1}$ become possible only after the largest particle mass exceeds about $0.01 \mathrm{M_{\Earth}}$.
This is because when protoplanets become more massive than $1.8 \times 10^{-2} \mathrm{M_{\Earth}}$, 
their escape velocities are higher than $2.5\, \mathrm{km\, s}^{-1}$.
This trend is confirmed in observing the mass ratio of the target to the impactor:
most of high velocity collisions occur when the mass ratio becomes larger than $\sim 10$.
Since the growth timescale of protoplanets is proportional to $f_{\rm d}^{-1}$, 
the high velocity collisions take place in the MMSN3NJ case three times earlier than the MMSN1NJ case.
These results are consistent with those of previous studies \citep{2015Natur.517..339J, 2016aHasegawa}.

When the high velocity collisions occur,
it is expected that small impactors tend to have higher eccentricities than large targets.
To examine this expectation,
Figure \ref{fig:ecc-ratio-woJ} shows the eccentricity of impactors and targets just before their collisions.
The left and right panels represent the results of the MMSN1NJ and MMSN3NJ cases, respectively.
In these plots, the eccentricities of impactors are labeled on the vertical axis, and those of targets are on the horizontal one.
The color bars denote the mass ratios of targets to impactors.
As done in Figure \ref{fig:ITmass-ratio-woJ}, 
the high velocity ($v_{\rm imp} > 2.5\, \mathrm{km\, s}^{-1}$) collisions are shown by the circles and other collisions are by the triangles.
For reference, the dotted line is drawn to specify collisions in which both impactors and targets have the same eccentricity.
These plots confirm that all high velocity collisions concentrate in the upper left region, that is, the eccentricities of impactors are larger than those of targets.
On the other hand, low velocity collisions distribute almost equally on both sides of the dotted line.
Our results, therefore, suggest that impactor planetesimals are attracted by the gravity of target protoplanets,
which increases the eccentricity of impactors and leads to high velocity collisions.
The detailed behavior of impactors' eccentricities is discussed in the next section.

We now turn our attention to the locations of collisions.
Figure \ref{fig:pos-col-woJ} shows the spatial distribution of collisions as a function of time.
For clear presentation, we adopt the range of $10^4$ - $2\times10^6\, \mathrm{yr}$ for the horizontal axis.
The color bar denotes the ratio of the impact velocity to the escape velocity.
Our results show that high velocity collisions tend to occur from inside to outside in the solar nebula.
This is because inner protoplanets grow up faster than outer ones.
One can confirm that the impact velocities of high velocity collisions are almost equal to the escape velocity and 
hence that protoplanet growth regulates the value of the impact velocity.
These results are consistent with those of \cite{2015Natur.517..339J}.
We also find that the radial distribution of collisions spreads with time.
This is the consequence that the spatial distribution of planetesimals diffuses radially with time due to their viscous stirring and the gas drag.
Given that particles are more massive in the MMSN3NJ case than the MMSN1NJ case (see Table \ref{table:parameter}),
this spread is more enhanced for the former case.

%\begin{turnpage}
\begin{table*}
%\begin{minipage}{10cm}
%\begin{center}
\centering
%\resizebox{10.0cm}{!}{
\caption{Summary of collisions in our simulations}
\label{table:collisions}
%\scalebox{0.75}{
{\tiny
\tabcolsep = 0cm
\begin{tabular}{|c|c|c|c|c|c|c|c|c|c|c|c|c|c|} 
\hline 
                              &                       & \multicolumn{2}{|c|}{MMSN1NJ} & \multicolumn{2}{|c|}{MMSN3NJ} & \multicolumn{2}{|c|}{MMSN1JE0} & \multicolumn{2}{|c|}{MMSN1JE05} & \multicolumn{2}{|c|}{MMSN1JE1} &  \multicolumn{2}{|c|}{MMSN3JE05}    \\   \hline                           
Type of Collisions & Classification &  \begin{tabular}{c} Number \\ ($n$) \end{tabular}  &  \begin{tabular}{c} $n_{b \leq b_{\rm crit}}$ \\ $/ n_{b>b_{\rm crit}}$ \end{tabular} & \begin{tabular}{c} Number \\ ($n$) \end{tabular}  &  \begin{tabular}{c} $n_{b \leq b_{\rm crit}}$ \\ $/ n_{b>b_{\rm crit}}$ \end{tabular} & \begin{tabular}{c} Number \\ ($n$) \end{tabular} &  \begin{tabular}{c} $n_{b \leq b_{\rm crit}}$ \\ $/ n_{b>b_{\rm crit}}$ \end{tabular} & \begin{tabular}{c} Number \\ ($n$) \end{tabular}  &  \begin{tabular}{c} $n_{b \leq b_{\rm crit}}$ \\ $/ n_{b>b_{\rm crit}}$ \end{tabular} & \begin{tabular}{c} Number \\ ($n$) \end{tabular} &  \begin{tabular}{c} $n_{b \leq b_{\rm crit}}$ \\ $/ n_{b>b_{\rm crit}}$ \end{tabular} & \begin{tabular}{c} Number \\ ($n$) \end{tabular} &  \begin{tabular}{c} $n_{b \leq b_{\rm crit}}$ \\ $/ n_{b>b_{\rm crit}}$ \end{tabular} \\ \hline \hline
All                               &                                                                                         & 1809 &           & 2607  &         & 1271 &           & 1222 &          & 1313 &          & 2090 &          \\
                                   &                                                       $b \leq b_{\rm crit}$   & 531   &          & 864    &          & 351   &           & 331   &          & 400   &          & 617   &          \\
                                   &                                                       $b > b_{\rm crit}$       & 1278 & 0.42  & 1743  & 0.50  & 920   & 0.38   & 891   & 0.37  & 913   & 0.44  & 1473 & 0.42  \\ \hline
Perfect merger           & $v_{\rm imp} \leq v_{\rm esc} $ \& $b \leq b_{\rm crit}$ & 462   &          & 773   &           & 317   &           & 275   &          & 279   &          & 510   &           \\
                                   & $v_{\rm imp} \leq v_{\rm esc} $ \& $b > b_{\rm crit}$      & 1116  & 0.41  & 1503 & 0.51   & 778   & 0.41   & 698   & 0.39  & 635   & 0.44  & 1183 & 0.43    \\ \hline
Partial accretion         & $v_{\rm imp} > v_{\rm esc} $ \&     $b \leq b_{\rm crit}$  & 69      &          & 91     &           & 34     &           & 56     &          & 121   &          & 107   &            \\ 
Hit-and-run                 & $v_{\rm imp} > v_{\rm esc} $ \&     $b > b_{\rm crit}$      & 162    & 0.43  & 240   & 0.38  & 142   & 0.24    & 193   & 0.29  & 278   & 0.44  & 290  & 0.37    \\ \hline \hline
Collisions with            &                                                                                         & 7       &           & 195    &         & 3        &           & 14      &          & 55     &          & 66    &          \\
$v_{\rm imp} >2.5$ km s$^{-1}$     &                                  $b \leq b_{\rm crit}$   & 5      &           & 115    &          & 0        &           & 7       &          & 13     &          & 32    &          \\
                                   &                                                       $b > b_{\rm crit}$       & 2      & 2.5     & 80      & 1.44  & 3        & 0        & 7       & 1       & 42     & 0.31  & 34     & 0.94  \\ \hline
Perfect merger$^a$    & $v_{\rm imp} \leq v_{\rm esc} $ \& $b \leq b_{\rm crit}$  & 5       &           & 115    &           & 0       &           & 0       &          & 0       &          & 24    &           \\
                                   & $v_{\rm imp} \leq v_{\rm esc} $ \& $b > b_{\rm crit}$      & 2       & 2.5     & 80      & 1.44   & 0       &           & 0       &          & 0       &          & 17    & 1.41    \\ \hline
Partial accretion$^a$ & $v_{\rm imp} > v_{\rm esc} $ \&     $b \leq b_{\rm crit}$   & 0      &            & 0        &           & 0       &           & 7       &          & 13    &          & 8      &            \\ 
Hit-and-run$^a$         & $v_{\rm imp} > v_{\rm esc} $ \&     $b > b_{\rm crit}$      & 0       &           & 0         &          & 3       & 0         & 7       & 1       & 42    & 0.31  & 17    & 0.47     \\ \hline \hline
Collisions with            &                                                                                         & 1019 &           & 1170  &         & 801   &           & 813   &          & 832   &          & 1116 &          \\
Mass Ratio = 1           &                                                       $b \leq b_{\rm crit}$   & 261   &          & 305    &          & 197   &           & 206   &          & 224   &          & 273   &          \\
                                   &                                                       $b > b_{\rm crit}$       & 758   & 0.34  & 865    & 0.35  & 604   & 0.33   & 607   & 0.34  & 608   & 0.37  & 843 & 0.32  \\ \hline
Perfect merger$^b$    & $v_{\rm imp} \leq v_{\rm esc} $ \& $b \leq b_{\rm crit}$   & 217   &          & 246   &           & 175   &           & 164   &          & 140    &          & 200   &           \\
                                   & $v_{\rm imp} \leq v_{\rm esc} $ \& $b > b_{\rm crit}$      & 639    & 0.34   & 723  & 0.34   & 498   & 0.35   & 454   & 0.36  & 393   & 0.36  & 649   & 0.31    \\ \hline
Partial accretion$^b$ & $v_{\rm imp} > v_{\rm esc} $ \&     $b \leq b_{\rm crit}$  & 44       &          & 59     &           & 22     &           & 42     &          & 84    &          & 73    &            \\
Hit-and-run$^b$         & $v_{\rm imp} > v_{\rm esc} $ \&     $b > b_{\rm crit}$      & 119    & 0.37  & 142    & 0.42  & 106   & 0.21    & 153   & 0.27  & 215   & 0.39  & 194  & 0.38    \\ \hline \hline
Collisions with            &                                                                                         & 700   &           & 1437  &         & 470   &           & 409   &          & 481   &          & 974   &          \\
Mass Ratio $> 1 $      &                                                       $b \leq b_{\rm crit}$   & 270   &          & 559    &          & 154   &           & 125   &          & 176   &          & 344   &          \\
                                   &                                                       $b > b_{\rm crit}$       & 520   & 0.52  & 878    & 0.64  & 316   & 0.49   & 284   & 0.44  & 305   & 0.58  & 630   & 0.55  \\ \hline
Perfect merger$^c$    & $v_{\rm imp} \leq v_{\rm esc} $ \& $b \leq b_{\rm crit}$   & 245   &          & 527   &           & 142   &           & 111   &          & 139     &          & 310   &           \\
                                   & $v_{\rm imp} \leq v_{\rm esc} $ \& $b > b_{\rm crit}$      & 477    & 0.51   & 780  & 0.68   & 280   & 0.51   & 244   & 0.45  & 242   & 0.57  & 534   & 0.58    \\ \hline
Partial accretion$^c$ & $v_{\rm imp} > v_{\rm esc} $ \&     $b \leq b_{\rm crit}$  & 25       &          & 32     &           & 12     &           & 14     &          & 37     &          & 34     &            \\ 
Hit-and-run$^c$         & $v_{\rm imp} > v_{\rm esc} $ \&     $b > b_{\rm crit}$      & 43      & 0.58  & 98      & 0.33  & 36      & 0.33    & 40   & 0.35  & 63     & 0.59  & 96     & 0.35    \\ \hline 
\end{tabular}
%}
}

$^a$ The collisions whose impact velocities are larger than 2.5 km s$^{-1}$  are only considered.

$^b$ The collisions which satisfy the condition that the mass ratio of a target to an impactor is unity, are only considered.

$^c$ The collisions which satisfy the condition that the mass ratio of a target to an impactor is larger than unity, are only considered.

%\end{center}
%\end{minipage}
\end{table*}
%\end{turnpage}
%\clearpage

We then discuss the type of collisions that occurred in our simulations, using the classification discussed in Section \ref{sec:method_crit}.
To proceed, we count the number of collisions as functions of $v_{\rm imp}/v_{\rm esc}$, $b/b_{\rm crit}$, and the mass ratio of a target to an impactor, 
which are summarized in Table \ref{table:collisions}.

Figure \ref{fig:histgram-woJ} shows the resulting collision types as a function of time.
The results of the MMSN1NJ and MMSN3NJ cases are plotted on the left and right panels, respectively.
Our simulations show that perfect merger collisions (the green hatched bar) are dominant for high and low velocity collisions in both cases.
This arises because the impact velocity is governed by the escape velocity for most of the collisions in these numerical setups (also see Figure \ref{fig:ITmass-ratio-woJ}).
Both partial accretion (the red hatched bar) and hit-and-run (the blue hatched bar) collisions are realized with time.
This is the outcome of viscous stirring acting on planetesimals.
While these collisions are minor,
hit-and-run collisions dominate over partial accretion ones in both cases.
This can be understood as follows:
the number of collisions should depend on the cross-section of colliding bodies.
Using $b_{\rm crit}$ (see Equation (\ref{b_crit})), 
the collisional cross-section for partial accretion is proportional to $b_{\rm crit}^2$, and that for hit-and-run is proportional to $1-b_{\rm crit}^2$.
Since $b_{\rm crit}$ takes the minimum value ($b_{\rm crit}=1/2$) for collisions between equal-mass particles and 
the maximum value ($b_{\rm crit}=1$) for collisions between big protoplanets and small planetesimals,
the ratio of the collision number of partial accretion to hit-and-run varies from 1:3 ($b_{\rm crit}=1/2$) to 1:0 ($b_{\rm crit}=1$).
Both partial accretion and hit-and-run collisions occur at low impact velocities (Figure \ref{fig:histgram-woJ}),
suggesting that these are collisions between planetesimals with equal or similar masses.
Consequently, the number of hit-and-run collisions exceeds that of partial accretion ones by a factor of about 3.
Table \ref{table:collisions} supports this consideration.

We finally consider the distribution of collisions in the $v_{\rm imp}/v_{\rm esc}-b/b_{\rm crit}$ plane (see Figure \ref{fig:heatmap-woJ}).
Our plots confirm the above discussion: grazing collisions ($b/b_{\rm crit} > 1$, perfect merger or hit-and-run) dominate 
over non-grazing collisions ($b/b_{\rm crit} \leq 1$, perfect merger or partial accretion) at low impact velocities.
On the other hand, non-grazing collisions become more important for high velocity collisions.
This is simply because these are collisions between big protoplanets and small planetesimals.

\subsection{Eccentricity evolution of particles with and without Jupiter} \label{sec:perturb}

In this section, we discuss the eccentricity evolution of particles under the absence/presence of Jupiter's perturbation.
Both the resonant and secular perturbations are examined.
We explore both the cases with and without Jupiter for completeness.

\begin{figure*}
\plottwo{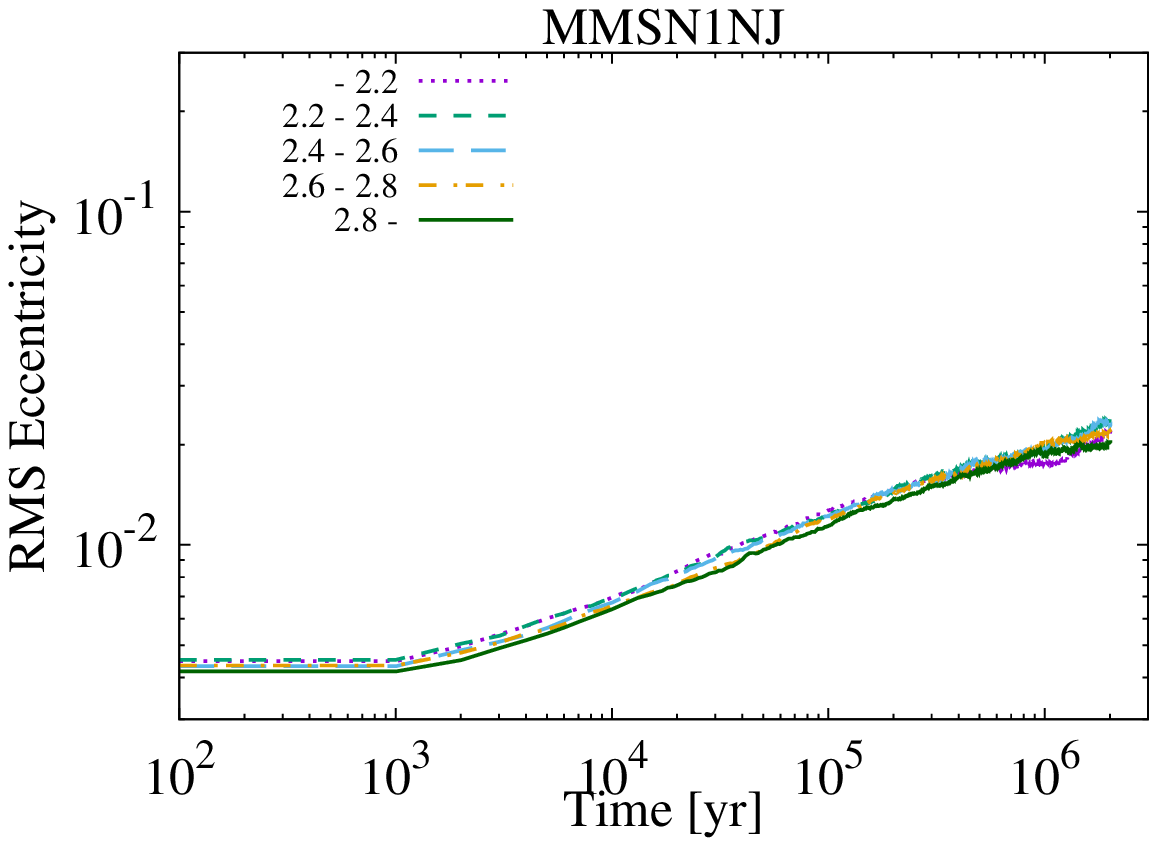}{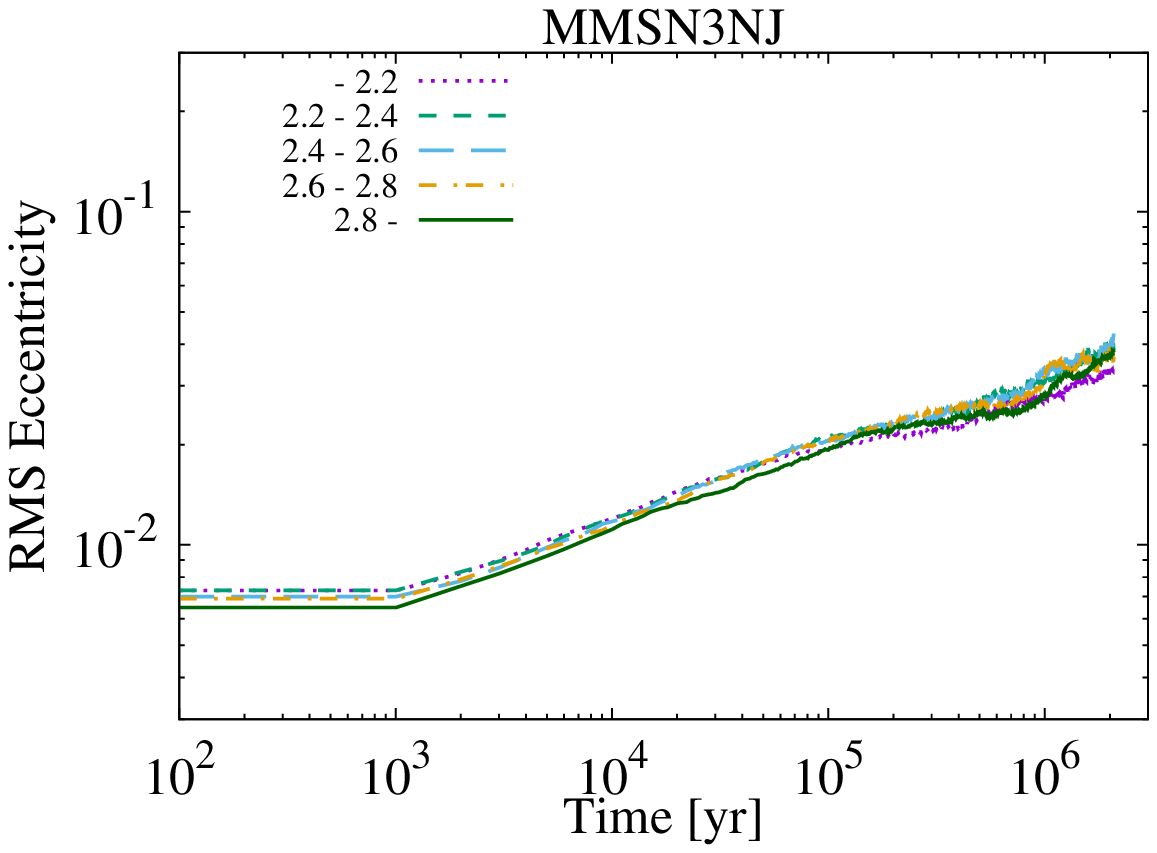}
\caption{The root mean square (RMS) eccentricities of particles as a function of time for the MMSN1NJ (left) and MMSN3NJ (right) cases.
The purple dotted line is for particles located at $< 2.2$ au, the green dashed is at 2.2-2.4 au, the blue long-dashed is at 2.4-2.6 au, 
the orange dot-dashed is at 2.6-2.8 au, and the green solid is at $> 2.8$ au.}
\label{fig:rms-box-woJ}
\end{figure*}

We first consider the cases without Jupiter (MMSN1NJ and MMSN3NJ).
Figure \ref{fig:rms-box-woJ} shows the time evolution of the root mean square (RMS) eccentricities of particles.
To examine which the perturbation of Jupiter (resonant vs secular) would be more important for the dynamics of particles,
we segment the radial extent into five zones: 
$a<2.2\mathrm{\, au},\ 2.2\mathrm{\, au} \le a <2.4\mathrm{\, au},2.4\mathrm{\, au} \le a <2.6\mathrm{\, au},\ 2.6\mathrm{\, au} \le a <2.8\mathrm{\, au},\ 2.8\mathrm{\, au} \le a$.
Our results show that the RMS eccentricities of particles gradually increase at all the locations.
Since Jupiter is not included in the simulations,
this increase is caused by viscous stirring.
There are slight fluctuations in RMS eccentricities, which are induced by protoplanets.

\begin{figure*}
\plottwo{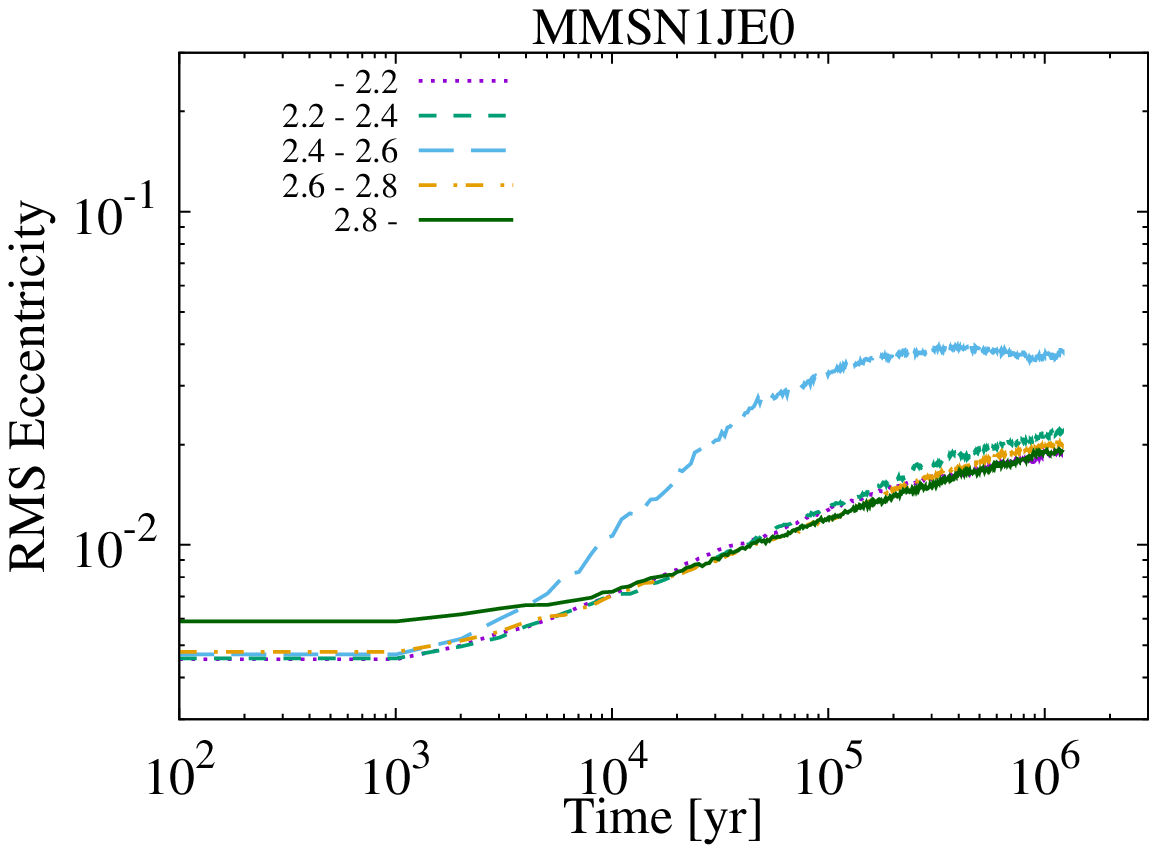}{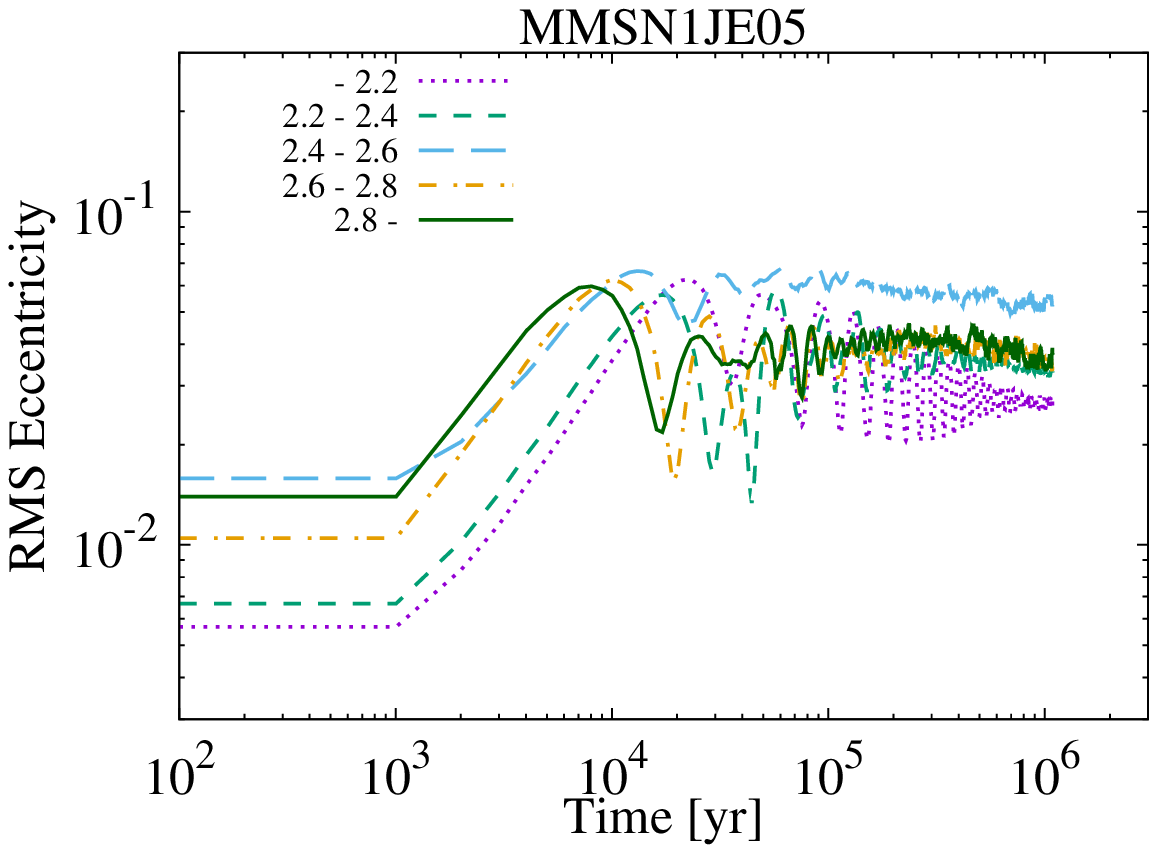}
\plottwo{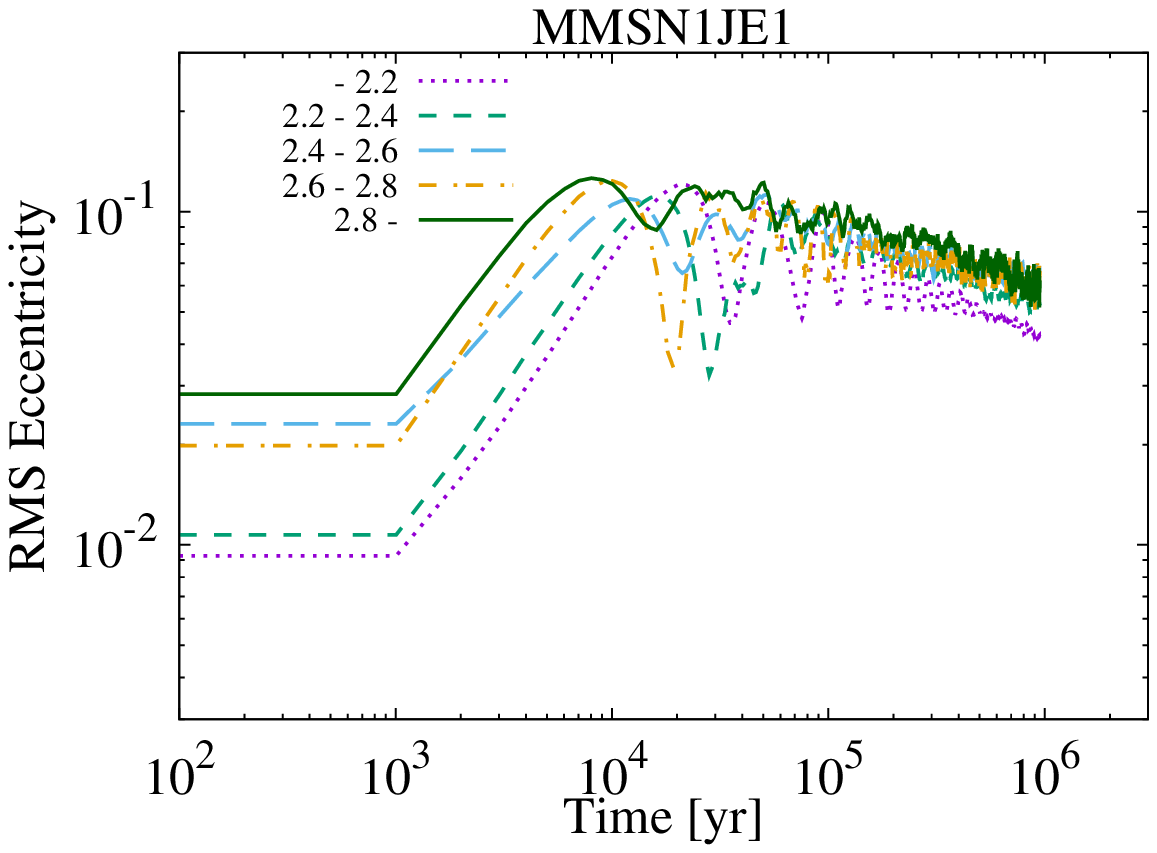}{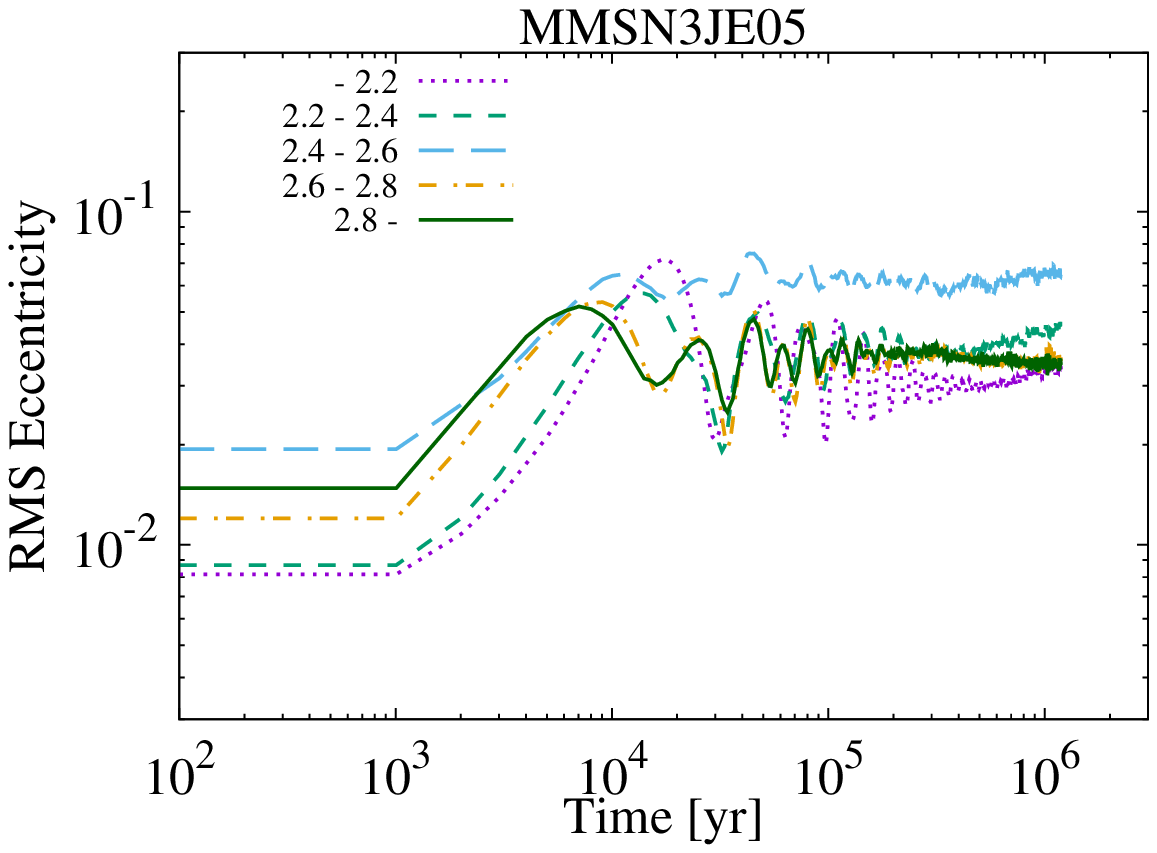}
\caption{RMS eccentricities of particles as a function of time for the MMSN1JE0 (top left), MMSN1JE05 (top right), MMSN1JE1 (bottom left), and MMSN3JE05 (bottom right) cases
as in Figure \ref{fig:rms-box-woJ}.}
\label{fig:rms-box-wJ}
\end{figure*}

We then discuss the cases with Jupiter (MMSN1JE0, MMSN1JE05, MMSN1JE1, and MMSN3JE05).
Figure \ref{fig:rms-box-wJ} shows the time evolution of the RMS eccentricities of particles.
We begin with the MMSN1JE0 case where the eccentricity of Jupiter is zero (see the top left panel).
Our results show that planetesimals located at $2.4\mathrm{\, au} \le a <2.6\mathrm{\, au}$ have the highest value of the RMS eccentricity.
This segment corresponds to the location of the 3:1 mean motion resonance with Jupiter (see Section \ref{sec:Jup}).
Except for the 3:1 resonant location, 
the evolution of RMS eccentricities is similar to that in the MMSN1NJ case (see Figure \ref{fig:rms-box-woJ}).
Our simulations, therefore, indicate that when the eccentricity of Jupiter is zero,
only the lowest order resonance plays an important role in pumping up eccentricities of particles.
The perturbations arising from higher order resonances than 3:1 is not effective, 
since the resonant perturbation depends on $e^{|j_4|}$, where $j_4$ is the order of the resonance (Section \ref{sec:res_part}). 
The secular perturbation also does not affect the eccentricities of particles for this case.

We now discuss the MMSN1JE1 case to examine the role of the secular perturbation (see the bottom left panel).
In this case, Jupiter has the highest value ($\sim 0.1$) of eccentricity (see Table \ref{table:parameter}).
We find that the RMS eccentricity of particles evolves similarly at all the locations, is pumped up faster, and has the largest value among all the cases.
This occurs because the secular perturbation plays a dominant role in regulating particles' eccentricities (see below for more discussion).
Our results also show that the eccentricities are pumped up to 0.1 and damped gradually after $\sim 10^5$ yr.
This eccentricity damping is caused by the disk gas.
The timescale of gas damping is given as
\begin{eqnarray}
T_{\rm gas}&=& \frac{|{\bm u}|}{{\bm F}_{\rm gas}}\nonumber \\
&=& 3.8\times10^{5}\mbox{\, yr}
  \left( \frac{C_{\rm D}}{1.0} \right)^{-1} 
  \left( \frac{m}{8.93\times10^{23}\mbox{ g}} \right)^{1/3} \nonumber \\ &&\times 
  \left( \frac{\rho_{\rm p}}{3\mbox{ g cm}^{-3}} \right)^{2/3}
  \left( \frac{a}{2\, \rm au}\right)^{13/4}
  \left( \frac{e}{0.05} \right)^{-1},
\end{eqnarray}
when $e\sim i/2$ and $e\gg\eta$. 
Thus, the RMS eccentricities of particles become about 0.05 at $10^6$ yr.

We consider the right two panels of Figure \ref{fig:rms-box-wJ} (MMSN1JE05 and MMSN3JE05) where $e_{\rm J}=0.05$.
In these cases, the eccentricities of particles are affected by both the resonant and secular perturbations.
Our results show that the RMS eccentricities of particles located in $2.4\mathrm{\, au} \le a <2.6\mathrm{\, au}$ are higher than the others.
This is because the resonant perturbation at 3:1 resonance is stronger than secular perturbation.
Except for this location, the RMS eccentricities oscillate similarly to those in the MMSN1JE1 case.
We find that the maximum eccentricity values are about 0.05, reflecting the eccentricity of Jupiter.
The RMS eccentricities of the particles at $a\leq 2.2$ au start increasing after $5\times 10^5$ yr for the MMSN3JE05 case,
which is caused by growing protoplanets.

\begin{figure*}
\includegraphics[bb=0 0 404 249]{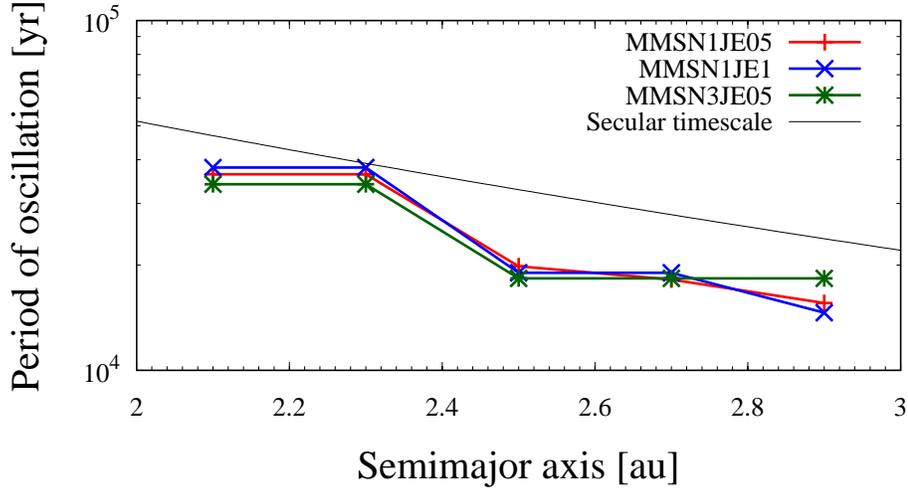}
\caption{Oscillation periods of the root mean square eccentricities of particles located at $< 2.2$ au, 2.2-2.4 au, 2.4-2.6 au, 2.6-2.8 au, and $>2.8$ au.
The red line with plus symbols (+) denotes the results of the MMSN1JE05 case, the blue one with crosses (x) is for the MMSN1JE1 case, 
and the green one with stars (*) is for the MMSN3JE05 case.
The power-law indices of the slopes are $-2.87\pm 0.74$ for the MMSN1JE05 case, $-3.09 \pm 0.85$ for the MMSN1JE1 case, and $-2.43 \pm0.75$ for the MMSN3JE05 case.
The black line represents the analytically estimated timescale of the secular perturbation from Jupiter (see Equation (\ref{eq:secular}) in Section \ref{sec:sec_part}).}
\label{fig:secular}
\end{figure*}

Here, we confirm that eccentricity oscillations are excited by the secular perturbation of Jupiter.
To proceed, we conduct Fourier transform to calculate oscillation periods and to compare these periods with the theoretical prediction (see Equation (\ref{eq:secular})).
Figure \ref{fig:secular} shows the oscillation periods of the RMS eccentricity for MMSN1JE05, MMSN1JE1, and MMSN3JE05 cases.
We choose these three cases since Jupiter has non-zero eccentricity and induces the secular perturbation.
The oscillation periods are derived from eccentricities of particles in each of five zones.
To simplify the analysis, only the highest peaks of periods are plotted.
Any features that have timescales shorter than 1000 years cannot be seen in our analysis, because we have obtained the data of simulations at 1000-year interval.
Our analysis shows that the periods of oscillations are almost the same values at each zone for these three cases.
This arises because the timescale of the secular perturbation from Jupiter does not depend on the value of $e_{\rm J}$.
We also find that the resulting slopes are between $-2.4$ to $-3.1$.
Thus, our results are broadly consistent with the prediction from the secular perturbation theory, which is shown by the black line.

We finally discuss how eccentricity pump-up caused by Jupiter's perturbation affects the population of high velocity collisions.
Given that the local RMS value of the relative velocity between planetesimals is given by $\simeq ev_{\rm K}$
\citep{1993LS}, 
the maximum impact velocity that can be achieved by planetesimal collisions is $\sim 2 e v_{\rm K}$.
Adopting the high impact velocity of $>2.5$\,km\,s$^{-1}$ (see Section \ref{sec:method_crit}) and that $v_{\rm K} \simeq 21$\,km\,s$^{-1}$ at 2 au,
the minimum required value of particles' eccentricities is 0.06.
As shown in Figures \ref{fig:rms-box-woJ} and \ref{fig:rms-box-wJ}, some particles attain this value of eccentricity in the MMSN1JE05, MMSN1JE1 and MMSN3JE05 cases,
while there are no such particles for the MMSN1NJ, MMSN3NJ, and MMSN1JE0 cases.
Thus, the presence of Jupiter would affect the properties of planetesimal collisions considerably in the former cases, which is discussed below.

\subsection{Planetesimal collisions with Jupiter}\label{sec:w_J}

In this section, we present the results for the case with Jupiter (see Table \ref{table:parameter}).
As done in Section \ref{sec:wo_J}, 
we discuss when and what value (high vs low) of the impact velocity of collisions occur (Figure \ref{fig:ITmass-ratio-wJ}), 
what value of eccentricities of impactors and targets are realized (Figure \ref{fig:ecc-ratio-wJ}),
the spatial distribution of collisions with time (Figure \ref{fig:pos-col-wJ}),
the collision types as a function of time (Figure \ref{fig:histgram-wJ}),
and the collisional outcomes on the impact velocity and impact parameter plane (Figure \ref{fig:heatmap-wJ}).

First, we examine how the presence of Jupiter affects the time evolution of planetesimal collisions.
Figure \ref{fig:ITmass-ratio-wJ} shows that collisions between particles with comparable masses can achieve high ($>2.5$ km s$^{-1}$) impact velocities for the case with Jupiter.
These collisions arise when eccentricities of planetesimals are pumped up by Jupiter (see below).
Our results also show that protoplanet growth is not altered by Jupiter very much (see the black line).
Since growing protoplanets contribute to the impact velocity by increasing the escape velocity,
the high velocity collisions take place not only by planetesimal-planetesimal collisions but also by protoplanet-planetesimal collisions at later time
(see the bottom right panel of Figure \ref{fig:ITmass-ratio-wJ}).

Second, we discuss the eccentricities of impactors and targets just before collisions.
Figure \ref{fig:ecc-ratio-wJ} summarizes the results.
As expected, these plots show that high velocity collisions with lower mass ratios exist and these collisions distribute almost equally on both sides of the dotted line.
This is the direct reflection that if the eccentricity of either an impactor or a target is excited by Jupiter's perturbation,
the resulting collision can reach the high impact velocity ($>2.5$\,km\,s$^{-1}$).
Our results also show that eccentricities of particles tend to increase with increasing Jupiter's eccentricity (see the MMSN1JE0, MMSN1JE05, and MMSN1JE1 cases).
We find that some of the high velocity collisions occur at particles' eccentricities of about 0.3 for the MMSN1JE05 case.
This value agrees with the previously reported value of the eccentricity of planetesimals located at 3:1 resonance with Jupiter \citep{1982Wisdom,1983Wisdom}.
As shown in Figure \ref{fig:ITmass-ratio-wJ}, 
the high velocity collisions with the mass ratio of $>10$ appear only in the MMSN3JE05 case (see the bottom right panel in Figure \ref{fig:ecc-ratio-wJ}).

Third, we consider the spatial distribution of collisions as a function of time to examine the effect of Jupiter's perturbation on the distribution.
Figure \ref{fig:pos-col-wJ} shows the results.
In the MMSN1JE0 case, 
we find that the high velocity collisions occur only around the 3:1 mean motion resonance with Jupiter 
and such collisions are realized earlier than the case without Jupiter (MMSN1NJ, see Figure \ref{fig:pos-col-woJ}).
This is simply because the resonant perturbation pumps up particles' eccentricities, as discussed in Section \ref{sec:perturb}.
In the MMSN1JE1 case, the secular perturbation from Jupiter excites the eccentricity of particles at all the locations (see Figure \ref{fig:rms-box-wJ}).
If particles are located around the 3:1 resonant point, then they can obtain higher eccentricities due to the combined effects of the secular and resonant perturbations from Jupiter.
This eccentricity pump-up occurs with the timescale of $T_{\rm lib}$, which is a few $10^4$ yr (see Equation (\ref{eq:T_lib})).
That is why high velocity collisions initiate around the 3:1 mean motion resonance at $ \sim 3\times10^4$ yr (see the bottom left panel).
As time goes on, the high velocity collisions spread towards various regions 
since the secular perturbation is strong enough to achieve high impact velocities in this case.
In the MMSN1JE05 and MMSN3JE05 cases, 
our results show that most of the high velocity collisions distribute near the 3:1 resonance.
This arises because the {effect of} secular perturbation in these cases is weaker than the MMSN1JE1 case.
After $5 \times 10^5 \, \mathrm{yr}$, 
protoplanet-planetesimal collisions also become energetic enough to achieve high ($>2.5$ km s$^{-1}$) impact velocities in the MMSN3JE05 case.

Fourth, we categorize the collisional outcome into three groups as a function of time (see Figure \ref{fig:histgram-wJ}, also see Section \ref{sec:method_crit}). 
As discussed above, the presence of Jupiter affects the properties of the high velocity collisions considerably.
For instance, all the high velocity collisions are classified as hit-and-run collisions in the case of MMSN1JE0, which reflects the 3:1 resonant perturbation.
As Jupiter's eccentricity increases, the number of partial accretion and hit-and-run collisions with high impact velocities also increases, 
since both the resonant and secular perturbations become stronger (see the MMSN1JE05 and MMSN1JE1 cases, also see Table \ref{table:collisions}).
When the massive disk is considered (see the MMSN3JE05 case), 
perfect merger collisions contribute to the population of high velocity impacts at later times because of protoplanet growth.
Our results also show that the ratio of the collision number of partial accretion to hit-and-run is about 1:3 (see Table \ref{table:collisions}).
This suggests that these collisions are the outcome of impacts between planetesimals with equal or similar masses.
Note that care is needed for interpreting Table \ref{table:collisions} because the number of collisions is very small for some cases.

Fifth, we examine the distribution of collisions in the $v_{\rm imp}/v_{\rm esc}-b/b_{\rm crit}$ plane. 
Figure \ref{fig:heatmap-wJ} confirms the above discussion: Jupiter's perturbation produces partial accretion and hit-and-run collisions with high impact velocities.

Thus, our results suggest that both the resonant and secular perturbations from Jupiter can accelerate the onset of high ($> 2.5$ km s$^{-1}$) velocity impacts 
via planetesimal-planetesimal collisions.

\section{Application to chondrule formation via the impact jetting scenario} \label{sec:app}

We here apply the results of our $N$-body simulations to the impact jetting scenario to examine the effect of Jupiter's perturbation on chondrule formation, explicitly.

\subsection{Background}

The importance of planetesimal collisions with high impact velocities and the resulting jetting to chondrule formation was demonstrated 
by the pioneering work of \citet{2015Natur.517..339J}.

In this scenario, chondrules are produced through the following three steps \citep{2015Natur.517..339J,2016aHasegawa}: 
in Step 1, planetesimals collide with a protoplanet at the impact velocity of $\ga 2.5\, \mathrm{km\,s}^{-1}$;
in Step 2, molten silicate materials are ejected from the collisional surface and become sub-mm to mm sized droplets;
in Step 3, these ejected droplets cool down at the rate of $10 - 1000$\,K\,hr$^{-1}$ in the solar nebula.
The key quantity in the impact jetting scenario is the impact velocity ($v_{\rm imp}$) of planetesimal collisions.
The critical impact velocity of $2.5\, \mathrm{km\,s}^{-1}$ is needed for chondrule formation.
Such high velocity collisions are readily achieved when planetesimals collide with protoplanets that are more massive than Moon-mass objects \citep{2016aHasegawa}.
Thus, the impact jetting scenario indicates that chondrules are the natural outcome of planet formation, rather than the leftover of planet-forming materials.

The intimate coupling of the formation of both chondrules and planets in the impact jetting scenario stimulated the follow-up studies.
For instance, \citet{2016bHasegawa} combined the impact jetting process with subsequent accretion of chondrules onto existing planetesimals 
and investigated under what condition of the solar nebula, the currently available meteoritic data are reproduced \citep[also see][]{moh17}.
These include the thermal history, abundance, the formation timing of chondrules, 
and the magnetic field strength of the nebula derived from chondrules in Semarkona meteorite \citep{2014Fu}.
\citet{2016bHasegawa} found that the produced chondrules can possess the above properties when the mass of planetesimals are $\la 10^{24} \mathrm{\ g}$ and the nebula gas is $\la 5$ times more massive than the MMSN model.
The previous studies, therefore, imply that impact jetting would be a promising process for understanding chondrule formation and hence the origin of the solar system.

\subsection{The resulting abundance of chondrules and their formation timing and locations}\label{sec:m_ch}

Given that a history of planetesimal collisions are obtained through our $N$-body simulations,
we here compute the abundance of chondrules that are produced by these collisions and the resulting jetting.
We also discuss their formation timing and locations.

Following the previous studies of \citet{2015Natur.517..339J,2017ApJ...834..125W},
we assume that 1 \% of the impactors' mass becomes the progenitor of chondrules if the corresponding collisions occur at the impact velocity of $>2.5$ km s$^{-1}$.
Equivalently, such high velocity impacts generate the progenitor of chondrules 
with the mass of $1.79 \times 10^{22} \mathrm{\ g}$ and of $5.36 \times 10^{22} \mathrm{\ g}$ per collision for the standard and massive disks, respectively.
Note that these criteria are obtained from numerical simulations of head-on collisions (see Section \ref{sec:issues}).

Figure \ref{fig:ccm} shows the cumulative chondrule mass as a function of time 
and its cumulative distribution as a function of the target-impactor mass ratio for all the cases on the left and right panels, respectively.
The horizontal black dotted line on the left panel denotes the current mass of the main asteroid belt.
The cumulative chondrule mass is computed by counting the total number of high velocity collisions ($v_{\rm imp} > 2.5$ km s$^{-1}$) with time.
As expected from the discussion in Sections \ref{sec:wo_J} and \ref{sec:w_J}, the onset of chondrule formation becomes earlier when Jupiter exists (see the left panel):
the formation starts at $\sim 2 \times 10^6 \, \mathrm{yr}$ for the MMSN1NJ case and about $5 \times 10^5 \,\mathrm{yr}$ for the MMSN3NJ case,
while it begins at $\sim 3$ - $5\times10^4$ yr for the MMSN1JE05, MMSN1JE1, and MMSN3JE05 cases and about $2\times 10^5$ yr for the MMSN1JE0 case.
In the former cases without Jupiter, the onset timing is determined by protoplanet growth,
and for the latter with Jupiter, the eccentricity of Jupiter regulates the timing since it controls the strength of the secular perturbation.

Also, Figure \ref{fig:ccm} shows the cumulative mass, which indicates 
what kinds of collisions (planetesimal-planetesimal ones vs protoplanet-planetesimal ones) are important for chondrule formation (see the left panel).
The sharp rise at $\sim 2 \times 10^6 \, \mathrm{yr}$ and $\sim 5\times 10^5 \, \mathrm{yr}$ for the MMSN1NJ and MMSN3NJ cases, respectively, 
traces the growth of the innermost protoplanet and is determined by only the initial $<100$ collisions.
The subsequent increase in the cumulative mass for the MMSN3NJ case suggests that protoplanet-planetesimal collisions provide the major contribution to chondrule formation.
The same features are seen in the MMSN3JE05 case.
For the cases with Jupiter (MMSN1JE0, MMSN1JE05, MMSN1JE1, and MMSN3JE05), 
the cumulative mass increases with similar slopes, 
which implies that planetesimal-planetesimal collisions occur at a nearly constant rate in these cases.
We can further confirm the importance of protoplanet-planetesimal and/or planetesimal-planetesimal collisions 
by plotting the cumulative distributions of the chondrule mass as a function of the mass ratio between impactors and targets (see the right panel).
For the case without Jupiter, collisions with large mass ratios (i.e., protoplanet-planetesimal collisions) are important
while for the case with Jupiter, collisions with small mass ratios (i.e., planetesimal-planetesimal collisions) are dominant.

\begin{figure*}
\plottwo{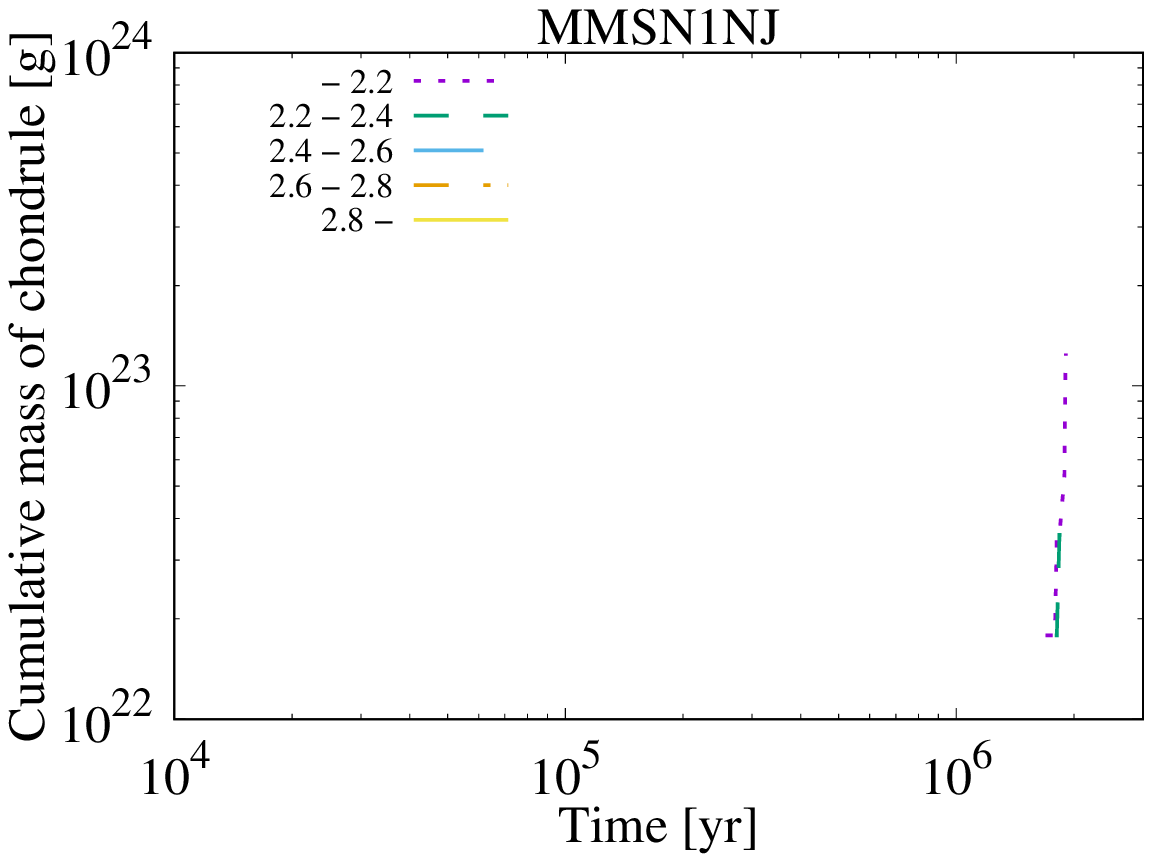}{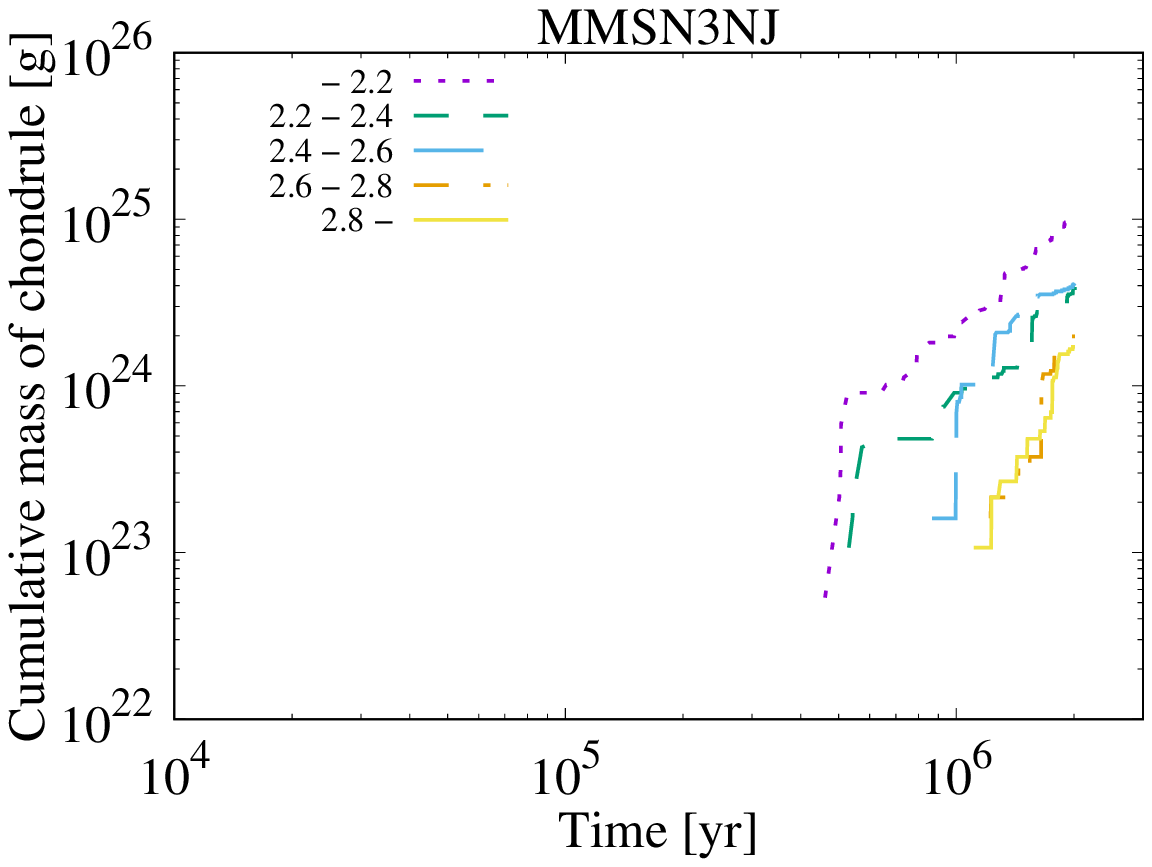}
\plottwo{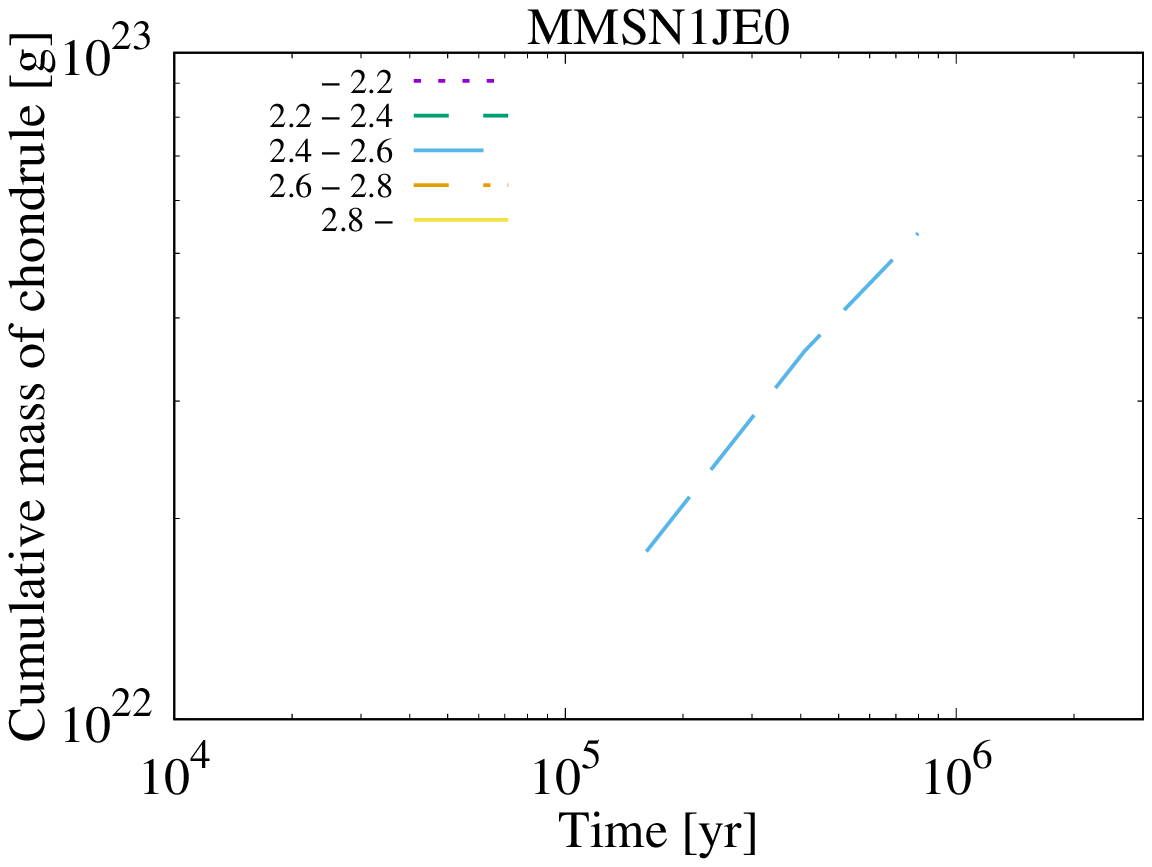}{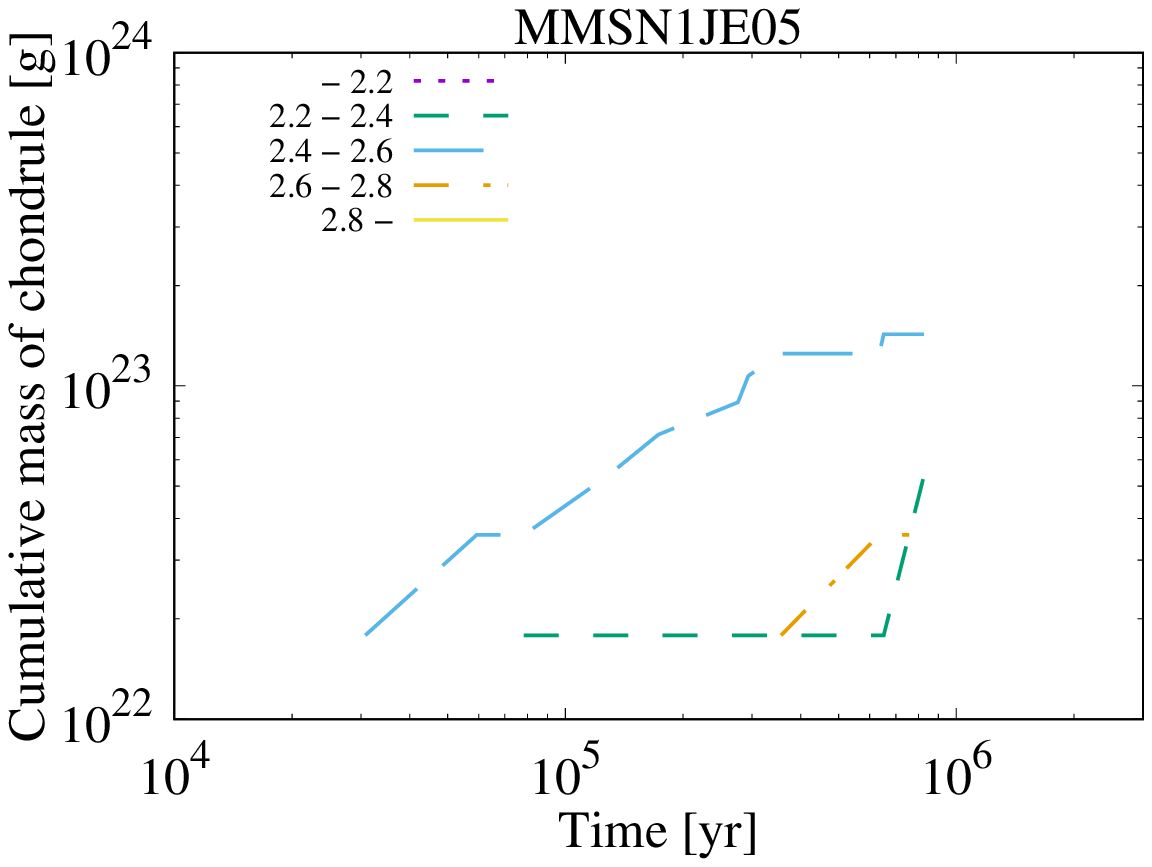}
\plottwo{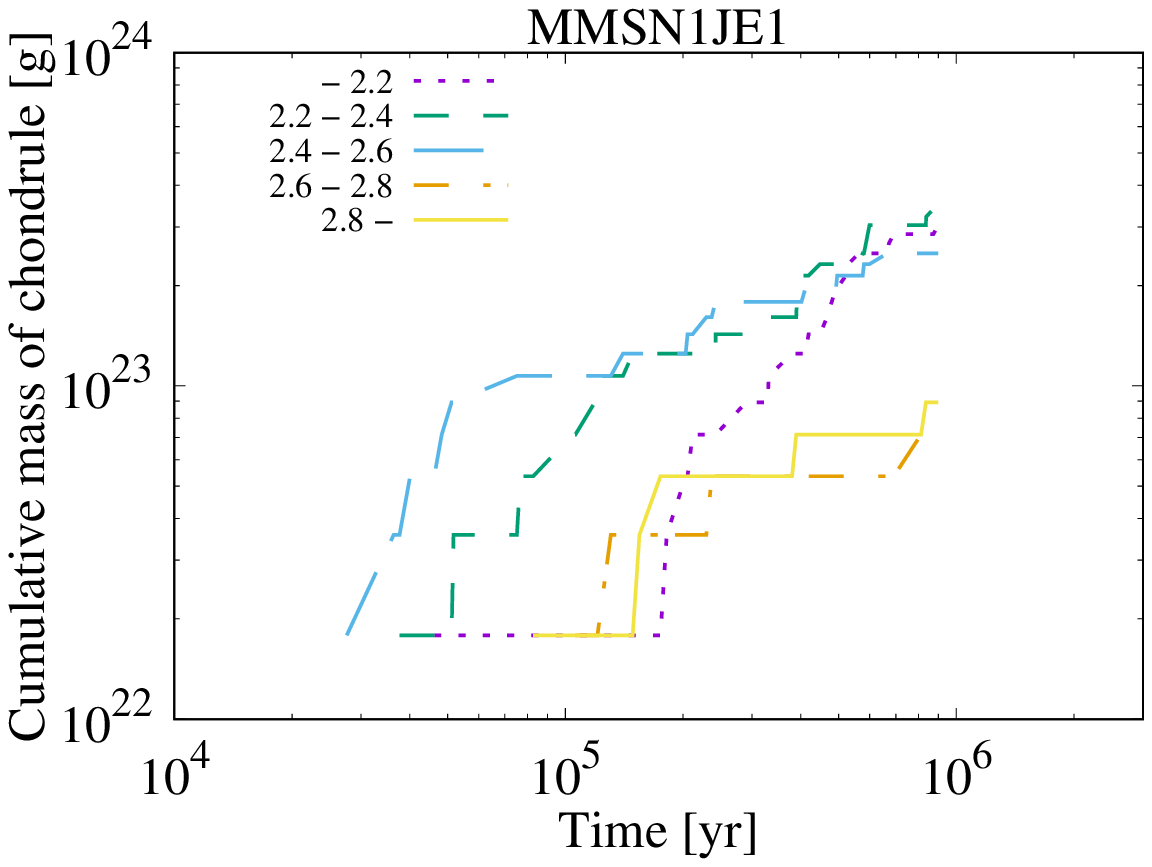}{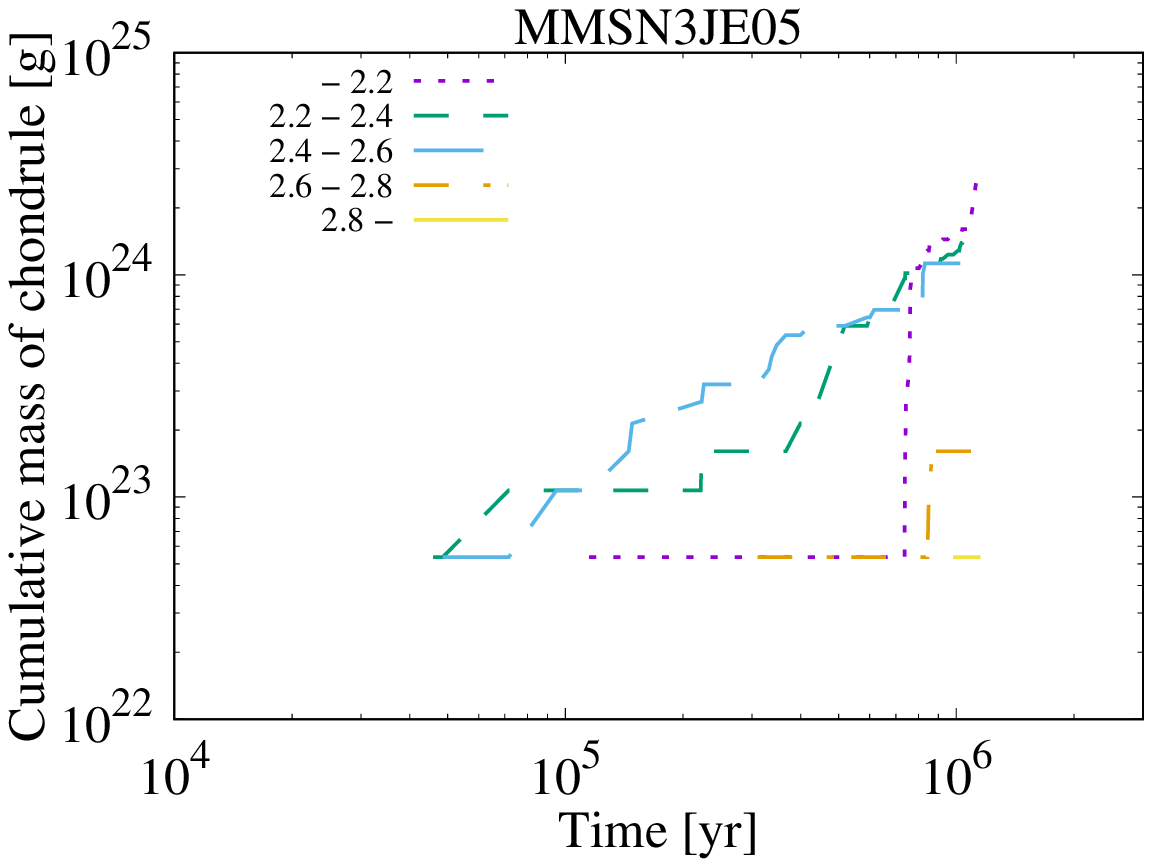}
\caption{Time evolution of the cumulative mass of chondrules formed at $<2.2$ au, 2.2-2.4 au, 2.4-2.6 au, 2.6-2.8 au, and $>2.8$ au.
From the top to bottom panels, the results for the MMSN1NJ and MMSN3NJ cases, those for the MMSN1JE0 and MMSN1JE05 cases, 
and those for the MMSN1JE1 and MMSN3JE05 cases are shown, respectively.}
\label{fig:ccm-box}
\end{figure*}

We now discuss how the time evolution of the cumulative chondrule mass differs at different locations.
As done in Section \ref{sec:perturb}, we consider five zones.
Figure \ref{fig:ccm-box} summarizes our results.
From the top to bottom panels, the results of the MMSN1NJ and MMSN3NJ cases, those of the MMSN1JE0 and MMSN1JE05 cases, 
and those of the MMSN1JE1 and MMSN3JE05 cases are shown, respectively.
These plots provide the behavior of chondrule formation that is expected from the properties of planetesimal collisions (Sections \ref{sec:wo_J} and \ref{sec:w_J}).
For the case without Jupiter, chondrules form in the inside-out manner and their amounts are larger in the inner region (see the top panels).
For the case with Jupiter, the active chondrule-forming sites are determined by Jupiter's eccentricity
since Jupiter's eccentricity regulates the strength of the secular perturbation (see the middle and bottom panels).
It is obvious that chondrule formation is the most efficient at 2.4 - 2.6 \,au, where both the resonant and secular perturbations can pump up eccentricities of planetesimals.

\subsection{Implications and issues of the impact jetting scenario} \label{sec:issues}

We here discuss the implications and issues that are contained in the impact jetting scenario.

\subsubsection{The production rate of chondrules}

As shown above, we have computed the abundance of chondrules 
based on the previous studies which show that head-on collisions with the impact velocity of $> 2.5$ km s$^{-1}$ generate the progenitor of chondrules 
whose abundance is about 1 \% of the impactors' mass \citep{2015Natur.517..339J, 2017ApJ...834..125W}.
While our results indicate that the impact jetting scenario can produce the enough amount of chondrules (see Figure \ref{fig:ccm}),
some issues are present in this approach.

The most crucial issue is that adopting these criteria does not reflect the fact that various kinds of collisions occur in the early solar system evolution.
As shown in Figures \ref{fig:heatmap-woJ} and \ref{fig:heatmap-wJ}, 
a wide range of impact velocities and angles are realized over the course of protoplanet formation \citep[also see][]{Movshovitz:2016aa}.
Importantly, our simulations demonstrate that oblique collisions provide the dominant contribution to the population of energetic collisions that can serve as chondrule-forming events.

It is currently unclear what kinds of relationships exist between the amount of produced chondrules and impact properties (e.g., velocity and angle) for oblique collisions.
Especially, the dependence of the impact angle on the chondrule abundance is controversial;
on one hand, it is suggested that oblique collisions decrease both the mass of ejecta and the internal temperature of colliding bodies \citep[e.g.,][]{2014MPS...49.2252D,2017Icar..294..234G}. 
On the other hand, it is also reported that ejecta's mass from oblique impact jetting becomes larger than that from head-on one \citep{Sugita:1999aa}.
Note that the dependence of the impact velocity on the chondrule abundance has recently been studied by \citet{2017ApJ...834..125W} 
under the assumption of head-on collisions.

Thus, it is important to derive a more realistic chondrule-forming criterion from numerical simulations of oblique collisions,
which should be done by three-dimensional impact simulations in the future. 

\subsubsection{The composition of produced chondrules}

The compositions of planetesimals and protoplanets are the important quantity for the impact jetting scenario,
because these objects serve as the parent bodies of chondrules.

Previous studies assume that the composition of these bodies are chondritic before/after collisions
based on the consideration that (potentially) unmelted lids may be present on the surface of protoplanets \citep{2015Natur.517..339J,2017ApJ...834..125W}.
Following their work \citep[also see][]{Weiss:2013aa}, 
we adopt the same assumption for calculations in Section \ref{sec:m_ch}.
However, it is currently not justified whether this assumption would hold under the realistic situation.
For example, the presence of internal heat source (e.g., $^{26}$Al) and surrounding atmospheres of massive planetesimals/protoplanets
would restrict the stability of the lid to a few cm at most \citep{Young:2019aa,Schlichting:2018aa},
which is too thin to form the enough amount of chondrules.
One may consider that rapid accretion of primitive pebbles may re-create the upper lid \citep{Visser:2016aa}.
Such efficient accretion would, in turn, increase the surface temperature of these bodies \citep{Brouwers:2018aa},
which would reduce the stability/thickness of the lid.
In addition, chemical differentiation in massive planetesimals and protoplanets occurs rapidly \citep{2018Icar..302...27L},
so that this process would take place before planetesimal collisions are realized.
Thus, both massive planetesimals and protoplanets have high chances of experiencing thermal evolution and 
the resulting ejecta originating from these bodies tend to have achondritic compositions \citep[e.g.,][]{Wilson:2017aa,Lichtenberg:2019aa}.

What about impact melts produced by collisions between small planetesimals?
As shown in our simulations, the presence of nearby, massive planets (such as Jupiter) increases the frequency of these collisions 
that satisfy the criteria of chondrule formation via impact jetting (see Figure \ref{fig:heatmap-wJ}).
In addition, \citet{2017ApJ...834..125W} showed that planetesimal-planetesimal collisions can also act as chondrule-forming events
by simulating these collisions.
The main issue in planetesimal-planetesimal collisions is 
that the formation mechanisms and timing of planetesimals are still unknown \citep[e.g.,][]{2014prpl.conf..547J},
and hence that their internal properties are poorly constrained.
In other words, if it would be known that planetesimals form via instability processes \citep[such as streaming instability, e.g.,][]{Johansen:2015aa,Klahr:2016aa,Simon:2017aa},
then their formation timescale would be very short.
For this case, one could expect that the early formed planetesimals would experience differentiation due to the decay heat of short-lived radionuclides
\citep[e.g.,][]{2012AREPS..40..113E,Ricard:2017aa}.
On the contrary, planetesimals formed at later times would be undifferentiated \citep[e.g.,][]{Gail:2014aa,Wakita:2018aa,Lichtenberg:2019aa}
and may be able to keep chondritic compositions.
Thus, a better understanding of planetesimal formation is required to conduct a more consistent simulation in which
$N$-body simulations and thermal evolution of growing bodies are coupled together.

In summary, further works are needed to verify whether the ejecta produced by the impact jetting scenario can satisfy the composition of chondrules.

\subsubsection{Formation timing of chondrules}

The formation timing of chondrules is one of the important constraints on their formation mechanisms.
We here discuss the implication of our calculations about chondrule formation.

We first point out that the formation timing of planetesimals and Jupiter is still unknown, as discussed above.
This information is however crucial to align the $t=0$ of our simulations with chronology studies of chondrules.
We second emphasize that the age estimate of chondrules itself has some diversity \citep{Connolly:2016aa}.
It is currently under active debate what is the main origin of this diversity (e.g., different formation mechanisms and/or locations).
Keeping these caveats and other issues discussed in the above sections in mind, 
we discuss how our results would be consistent with the age of chondrules.

To proceed, we consider the following four age estimates:
1) the age of chondrules clusters around $2\,\mathrm{Myr}$ after the formation of CAIs \citep{Villeneuve:2009aa};
2) parent bodies of iron meteorites likely formed at $<1\,\mathrm{Myr}$ after CAIs \citep{Goldstein:2009aa,Kruijer:2014aa};
3) parent bodies of chondrites likely formed around $2\,\mathrm{Myr}$ after CAIs \citep[e.g.,][]{Harrison:2010aa,Blackburn:2017aa,Wakita:2018aa}; and
4) CB chondrites likely formed at $\sim 3-5\,\mathrm{Myr}$ after CAIs \citep[e.g.,][]{Krot:2005aa,2017Bollard}.

Given that our simulations show that chondrule formation starts at $1\,\mathrm{Myr}$ and $0.1\,\mathrm{Myr}$ 
for the cases without and with Jupiter, respectively (see Figure \ref{fig:ccm}),
a potentially consistent picture can be drawn from the impact jetting scenario as below.
After the formation of CAIs, about $<1\,\mathrm{Myr}$ was needed for forming planetesimals (Estimate 2).
Once planetesimals were present, chondrule formation took place, following the formation of protoplanets.
Based on our results for the case without Jupiter, 
the impact jetting scenario predicts that the age of chondrules would be $\sim 2\,\mathrm{Myr} (= <1 + 1 \mathrm{Myr})$ after the CAI formation.
This estimate is roughly consistent with that of both chondrules (Estimate 1) and their parent bodies (Estimate 3).
As time went on, protoplanet formation completed at the asteroid belt region and chondrule formation via protoplanet-planetesimal collisions also halted there.
At the same time, however, Jupiter formation took place in the solar nebula.
If its formation timescale was about $3 \,\mathrm{Myr}$ after CAI formation, 
then chondrules found in CB chondrites could be explained by chondrule formation via planetesimal-planetesimal collisions (Estimate 4).
This is because our results show that chondrule formation starts at $0.1\,\mathrm{Myr}$ for the case with Jupiter.
In fact, previous studies suggested that 
the formation of the youngest chondrules in CB chondrites was triggered by planetesimal collisions excited by giant planets \citep{Krot:2005aa,Bollard:2015aa,2017Bollard}.
This picture is therefore consistent with previous work on CB chondrites \citep{Johnson:2016aa} 
if Jupiter underwent slow migration \citep[][see Section \ref{sec:dis}]{Kanagawa+2018a, Kanagawa2019}.

Thus, the impact jetting scenario can provide a potentially consistent chronology of chondrules,
while detailed work is needed to verify this simple argument.

\section{Discussions} \label{sec:dis}

Cautions are needed to interpret our results since several physical processes are not included in the simulations.
These are the gas disk evolution, the disk turbulence, the planet-disk interaction, the initial size distribution of planetesimals, the collisional fragmentation, and the effect of the snow lines.

Neglect of the gas disk evolution would be justified because our simulations end at $\gtrsim1 \,\mathrm{Myr}$,
while disk observations suggest that the typical gas disk lifetime is about $\gtrsim 3 \,\mathrm{Myr}$ \citep[e.g.,][]{2007ApJ...662.1067H,m09,2011ARA&A..49...67W}.
Thus, the effect of disk evolution on our simulation results would be minimal.

For the disk turbulence, we point out that Jupiter's perturbation would dominate over the eccentricity pump up by disk turbulence \citep[see their Section 2.4]{2016bHasegawa} 
if the level of turbulence estimated from the magnetic field strength recorded in chondrules \citep{2014Fu} is reasonable.
Consequently, we consider that disk turbulence would not change our results for the case with Jupiter very much.

We do not include the planet-disk interaction for simplicity.
It is, however, expected that a giant planet creates a gap in the gas disk due to the planet-disk interaction \citep{Lin&Papaloizou1986a,Lin&Papaloizou1986b,Crida+2006}.
The presence of the gas gap might reduce the gas surface density around the asteroid belt region \citep[e.g.,][]{ld06},
if the gap would be wide and deep enough \citep[e.g.,][]{Kanagawa+2015, Kanagawa+2016}.
The gap also acts as dust trapping at the outer edge of the gap \citep{Weber+2018,Kanagawa+2018, Kanagawa2019}.
If this is the case, planetesimals within the Jupiter's orbit do not grow due to pebble accretion \citep{Ormel&Klahr2010, Lambrechts&Johansen2012} but grow via their collisions.
It, therefore, may be reasonable to consider protoplanet formation via planetesimal collisions for the case with Jupiter.
The planet-disk interaction also leads to migration and eccentricity damping of massive bodies 
\citep[e.g.,][]{1980ApJ...241..425G,2000MNRAS.318...18N,2012ARAA..50..211K}.
For migration, its speed is highly sensitive to disk structures and is currently not constrained well 
\citep[e.g.,][]{2010MNRAS.401.1950P,2011MNRAS.413..286H,2014AA...570A..75B}.
Accordingly, it is not clear how our results would be altered by migration.
For the eccentricity damping, its most important effect would be on Jupiter's eccentricity.
In our setup, we adopt a certain value of $e_{\rm J}$ and only consider a gas drag force.
Under the presence of the nebular gas, it is expected that $e_{\rm J}$ decreases with time.
We, however, note that the eccentricity of massive bodies can be pumped up even in the gas disks
if the bodies open up a gap in the gas disks and the gas gap is heated by stellar irradiation \citep{2014ApJ...782..113T}.
More detailed simulations with these effects are needed.

We have assumed that planetesimals initially have equal-masses, and computed the time evolution of their size distribution due to planetesimal collisions.
The size distribution of the asteroid belt, however, can be reproduced better 
if planetesimals have an initial size distribution between 100\,km and $\sim$ 1000\,km, which should be similar to that of the current asteroid belt \citep{Morbidelli+2009}.
The size distribution of planetesimals does not change the picture of runaway and oligarchic growth \citep{Kobayashi+2016}.
If Jupiter is present, then the eccentricities of planetesimals are determined by its perturbation and less affected by the initial size of planetesimals (Section \ref{sec:perturb}).
The impact velocity distribution would not change due to the initial size distribution of planetesimals.
When the eccentricities of planetesimals are determined by Jupiter, 
the number of chondrule-forming collisions would increase if more planetesimals are present in $N$-body simulations.
In summary, it is necessary to adopt a more realistic initial size distribution of planetesimals 
for better characterizing planetesimal-planetesimal collisions and the resulting chondrule formation.

We have neglected the collisional fragmentation for simplicity.
As shown in our simulations, however, various kinds of collisions occur.
It is thus important to consider the outcome of collisions more realistically.
The presence of fragments would play a considerable role not only in computing protoplanet formation accurately, 
but also in reliably estimating the composition of parent bodies of chondrites.
For the former, recent studies suggest that accretion of pebble-sized bodies accelerates the formation of protoplanets \citep{Ormel&Klahr2010,Lambrechts&Johansen2012},
compared with the canonical model of runaway and oligarchic growth.
Accordingly, the formation timescale of protoplanets would be affected by 
the abundance of small-sized fragments and other solids (such as primitive dust and/or other sources of impact melts) if present.
For the latter, it would be reasonable to consider that the composition of parent bodies of chondrites is determined 
by the local abundances of chondrules, fragments generated by collisions, primitive dust, and/or other sources of impact melts (e.g., by splashing).
Thus, it is fundamental to carefully take into account the collision outcome and the resulting abundance of various forms of solids.

Finally, we have assumed in our simulations that all planetesimals are rocky and do not consider the effect of the snow line.
This is because we focus on the formation of chondrules, which are silicate particles.
It is known that the location of snow lines can move radially according to disk evolution and radial drift of dust particles \citep[e.g.,][]{2011Oka,2011MNRAS.417.1236H,2015ApJ...815..109P}.
The snow line location is not clear at the time of chondrule formation.
Given that the chondrule forming condition via impact jetting is derived from collisions between planetesimals composed of dunite \citep{2015Natur.517..339J, 2017ApJ...834..125W}
and that we have adopted such a condition in the current simulations,
consistency is maintained only when the effect of snow lines is neglected. 

Our future work will consider the above physical processes and effects to draw a better picture of chondrule formation via impact-based scenarios.

\section{Conclusions} \label{sec:conc}

\begin{figure*}
\plottwo{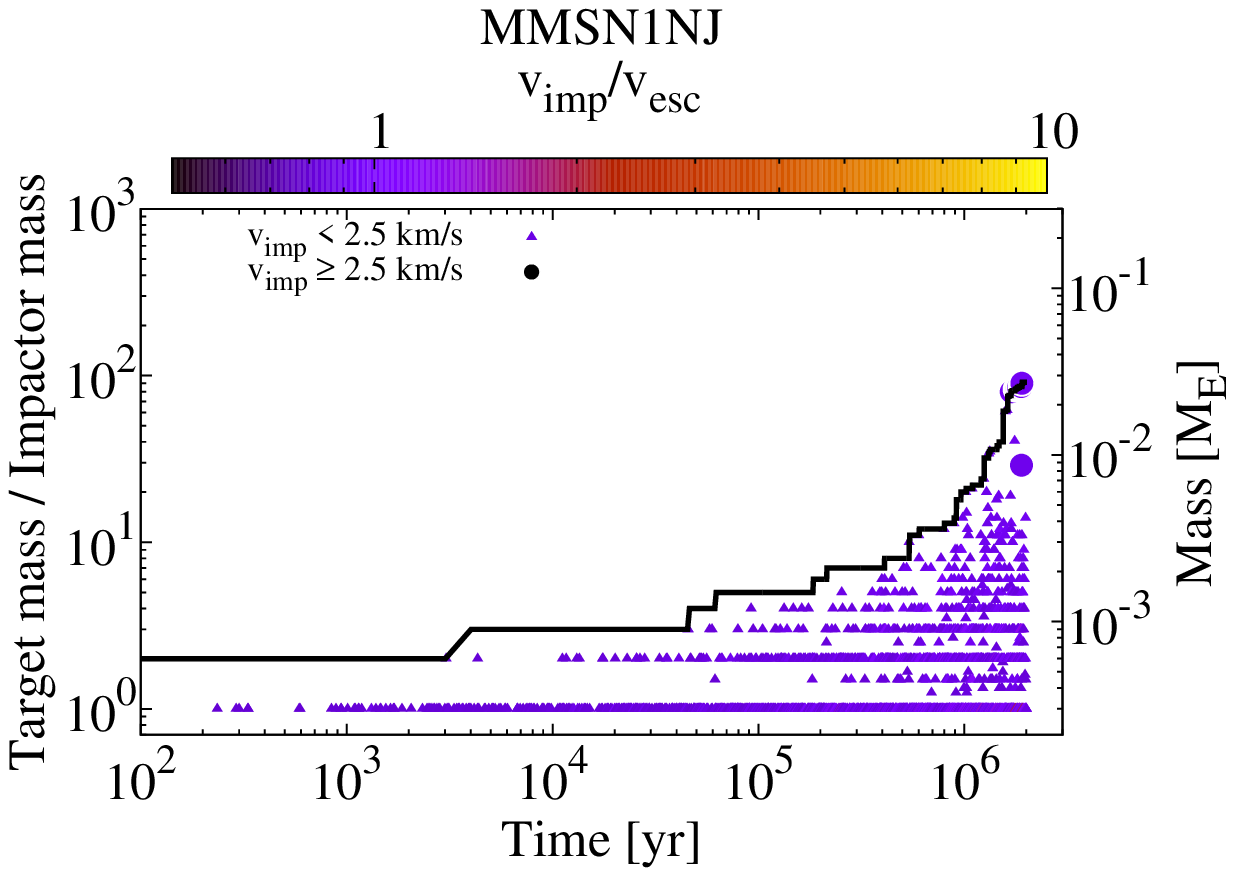}{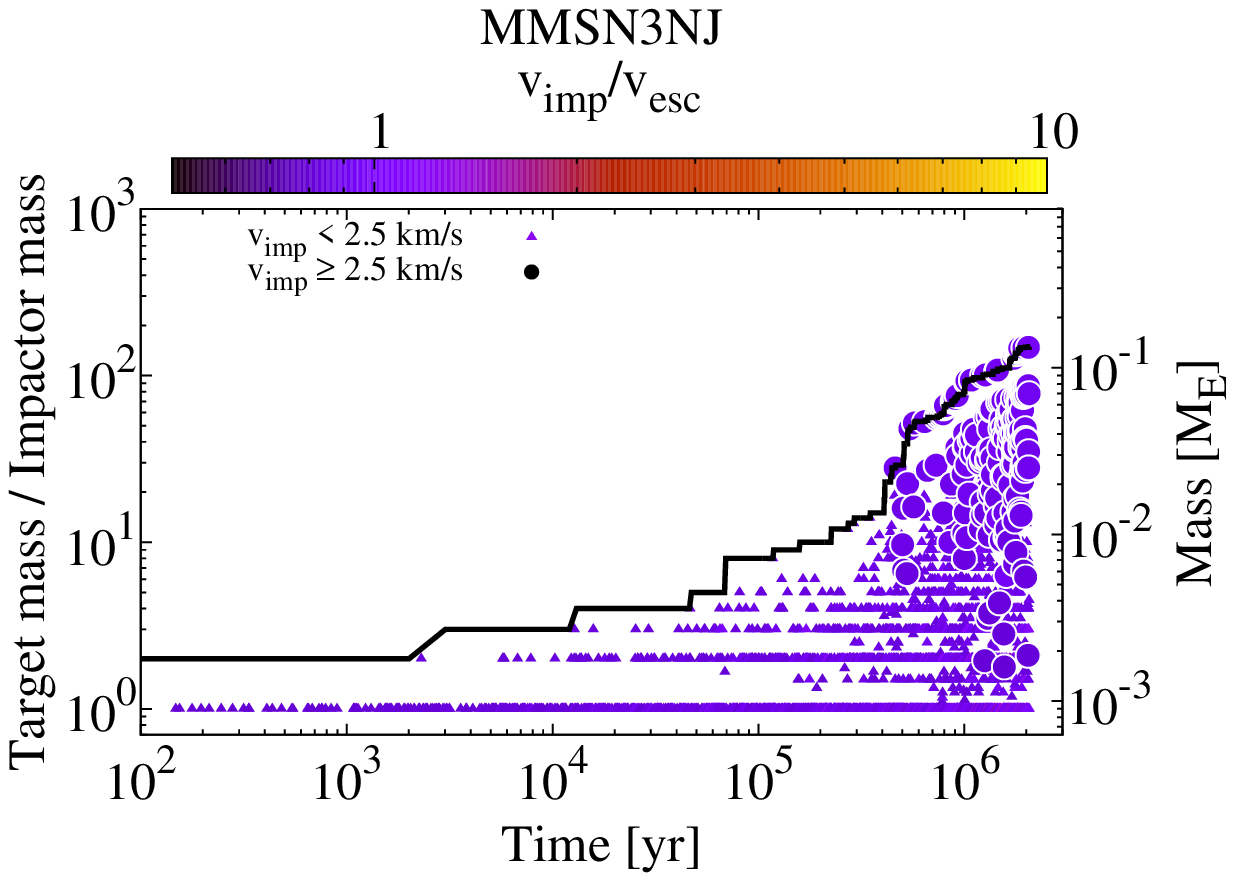}
\caption{The simulation results for the MMSN1NJ case are shown on the left panel and those for the MMSN3NJ case are on the right panel.
The mass ratios of targets to impactors in collisions are shown as a function of time (see the left side of the vertical axis).
The color bars denote the ratio of the impact velocity to the escape velocity.
The circles represent collisions whose impact velocities exceed $2.5\, \mathrm{km s}^{-1}$.
The triangles are for collisions whose impact velocities are less than $2.5\, \mathrm{km s}^{-1}$.
The black line is the mass of the largest particle (see the right side of the vertical axis).}
\label{fig:ITmass-ratio-woJ}
\end{figure*}

\begin{figure*}
\plottwo{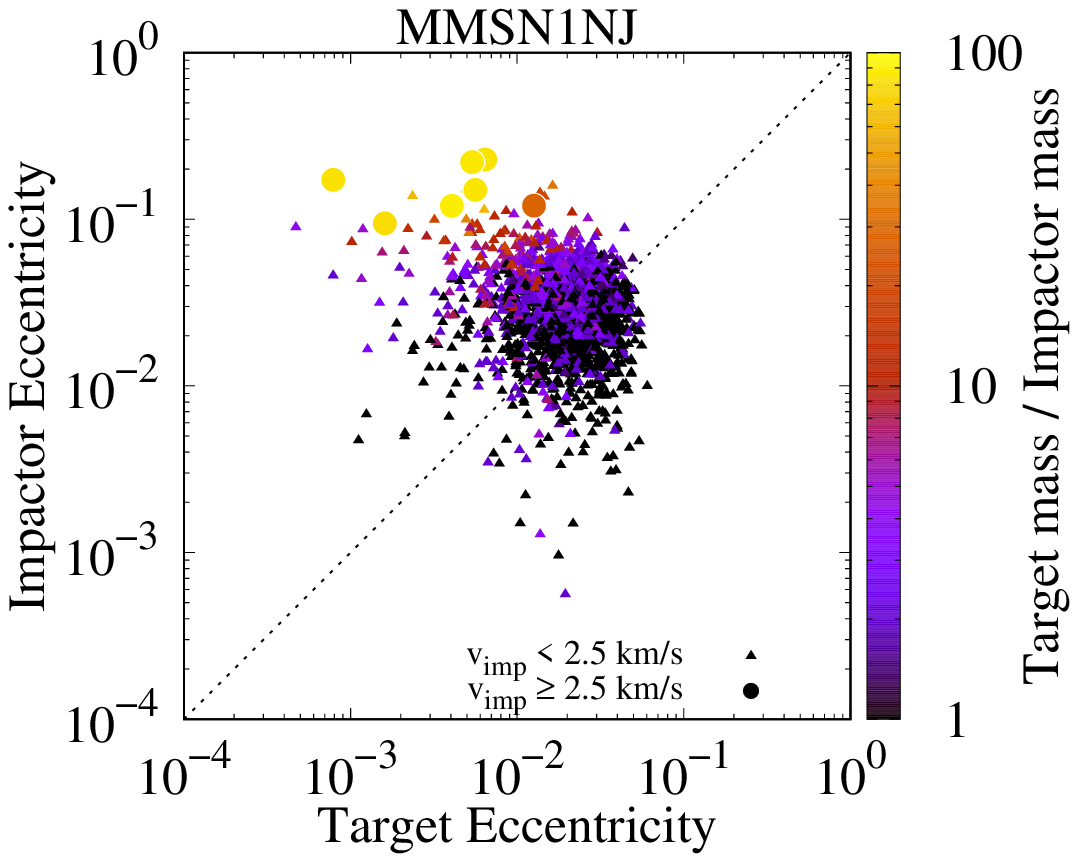}{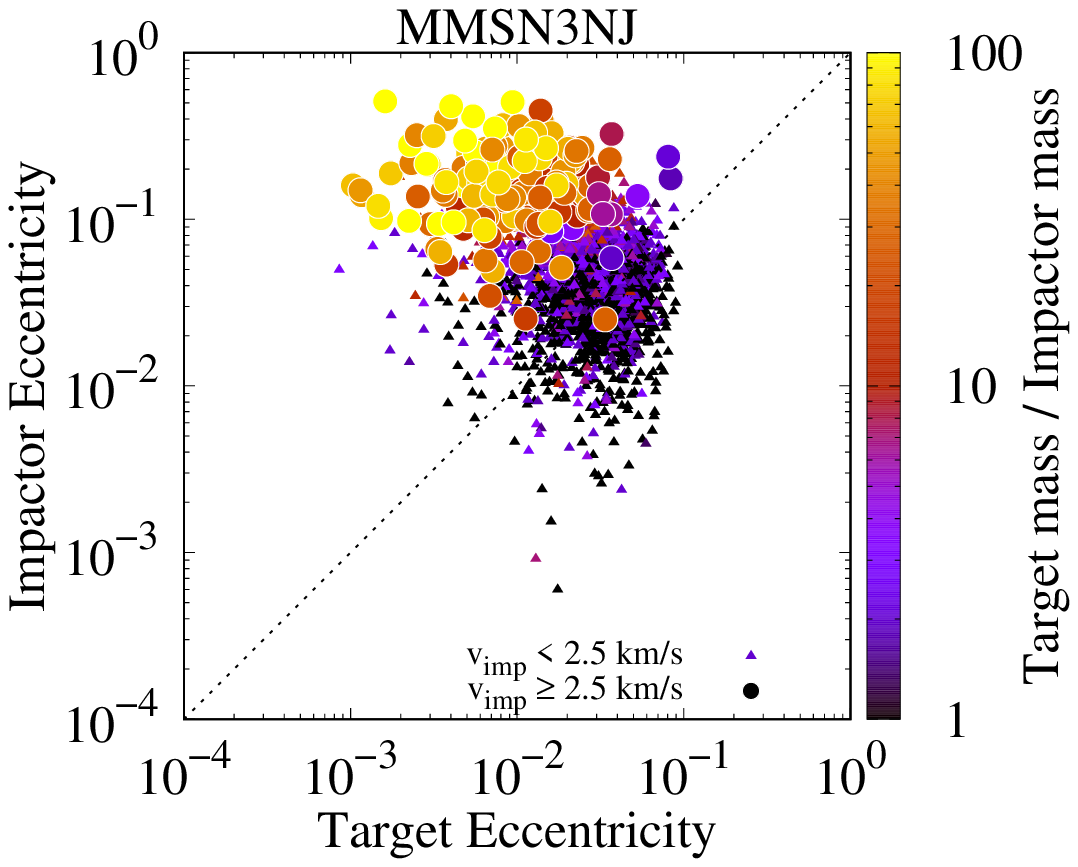}
\caption{The eccentricities of impactors vs those of targets just before collisions in the MMSN1NJ (the left panel) and MMSN3NJ (the right panel) cases.
The color bars denote the mass ratio of targets to impactors.
As in Figure \ref{fig:ITmass-ratio-woJ}, the circles are collisions with the impact velocity higher than $2.5\, \mathrm{km\, s}^{-1}$, and triangles are for collisions with the impact velocity less than $ 2.5\, \mathrm{km\, s}^{-1}$.
The dotted line indicates that the eccentricity of impactors is equal to that of targets.}
\label{fig:ecc-ratio-woJ}
\end{figure*}

\begin{figure*}
\plottwo{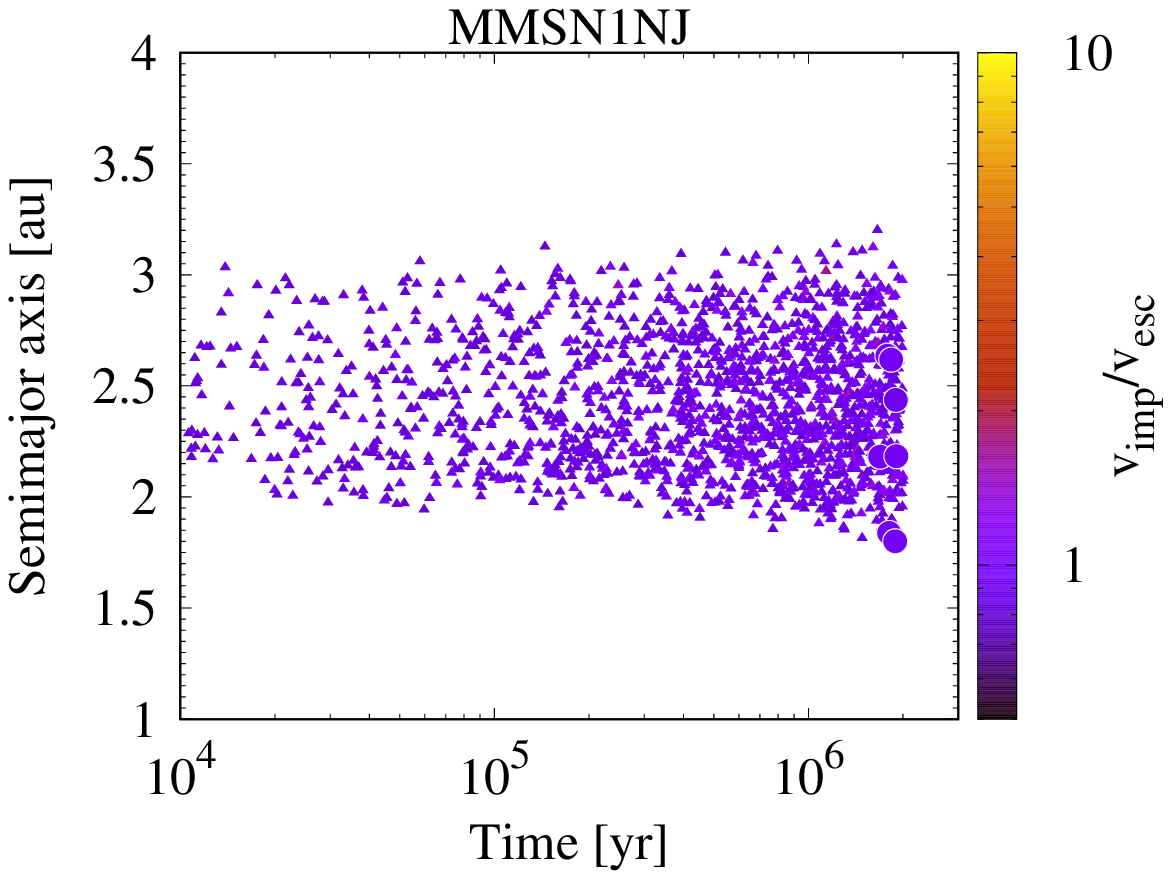}{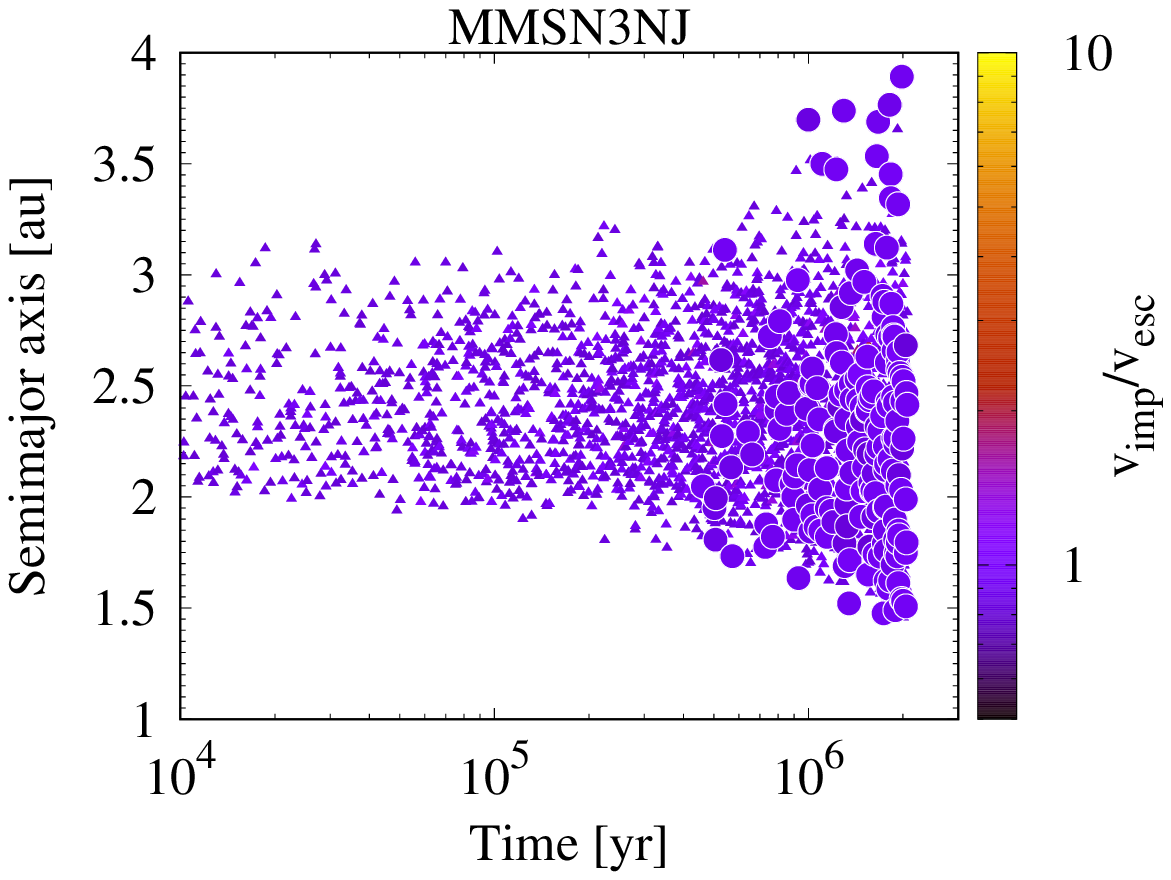}
\caption{The locations of collisions as a function of time for the MMSN1NJ (the left panel) and MMSN3NJ (the right panel) cases.
The color bars denote the ratio of the impact velocity to the escape velocity.
As in Figure \ref{fig:ITmass-ratio-woJ}, the circles represent collisions with the impact velocity higher than $ 2.5\, \mathrm{km\, s}^{-1}$,
and triangles are for collisions with the impact velocity less than $2.5\, \mathrm{km\, s}^{-1}$.}
\label{fig:pos-col-woJ}
\end{figure*}

\begin{figure*}
\plottwo{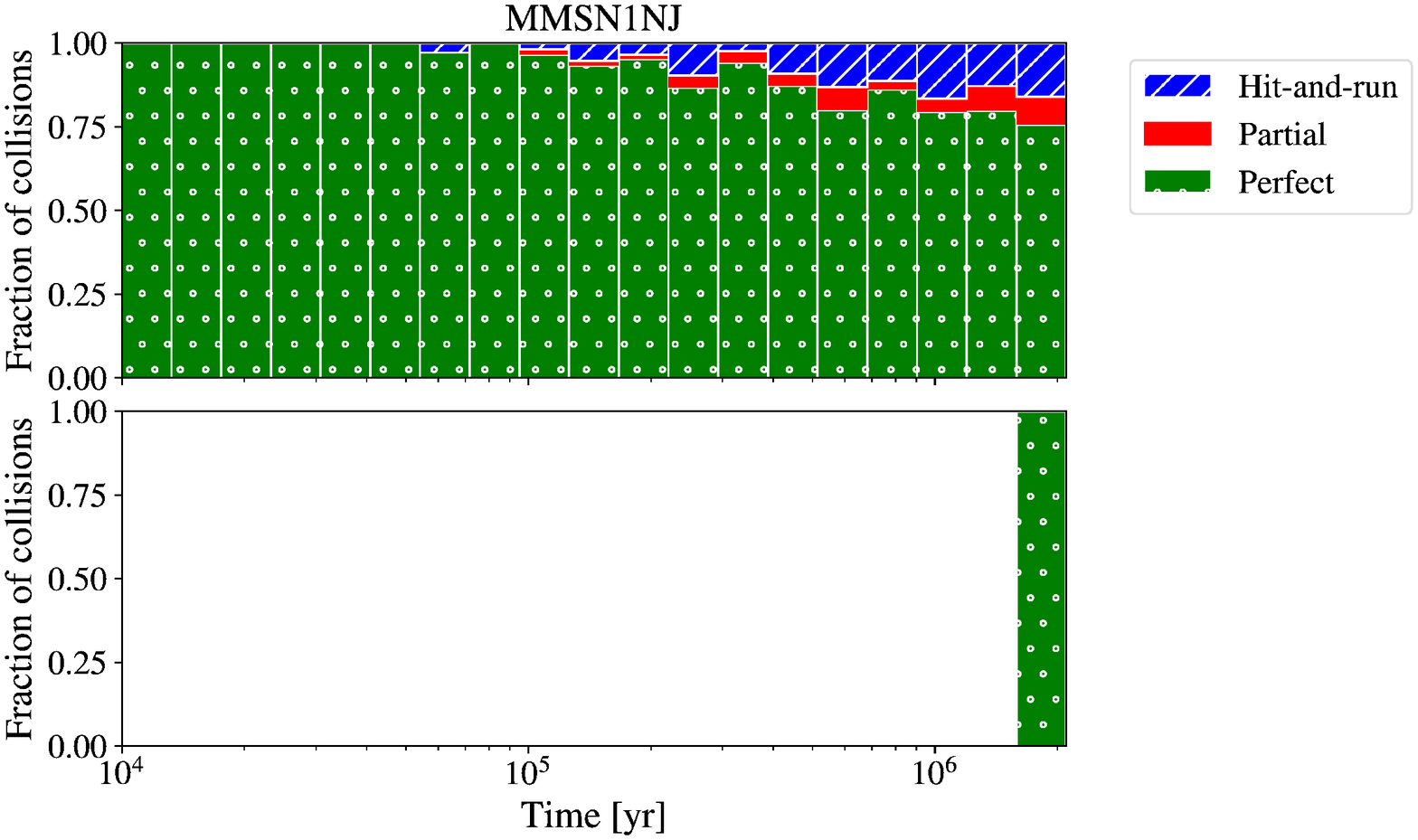}{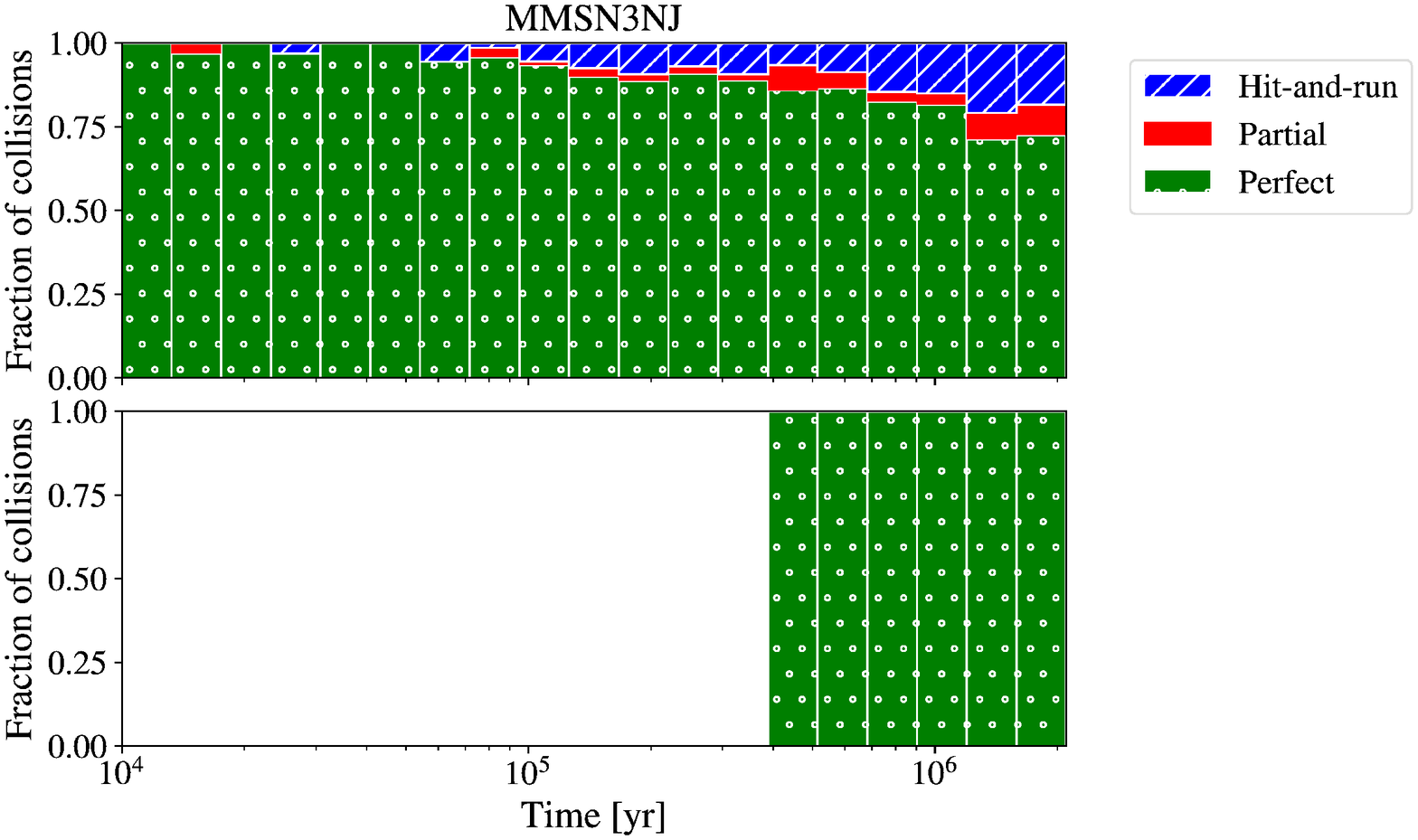}
\caption{
The fractional number of collisions as a function of time for the MMSN1NJ (the left panel) and MMSN3NJ (the right panel) cases.
On the top panels, collisions with the impact velocity of $\leq 2.5$ km s$^{-1}$ are shown, while those with the impact velocity of $> 2.5$ km s$^{-1}$ are on the bottom panels.
The green hatched bar denotes perfect merger, the red one is for partial accretion, and the blue one is for hit-and-run.
}
\label{fig:histgram-woJ}
\end{figure*}

\begin{figure*}
\plottwo{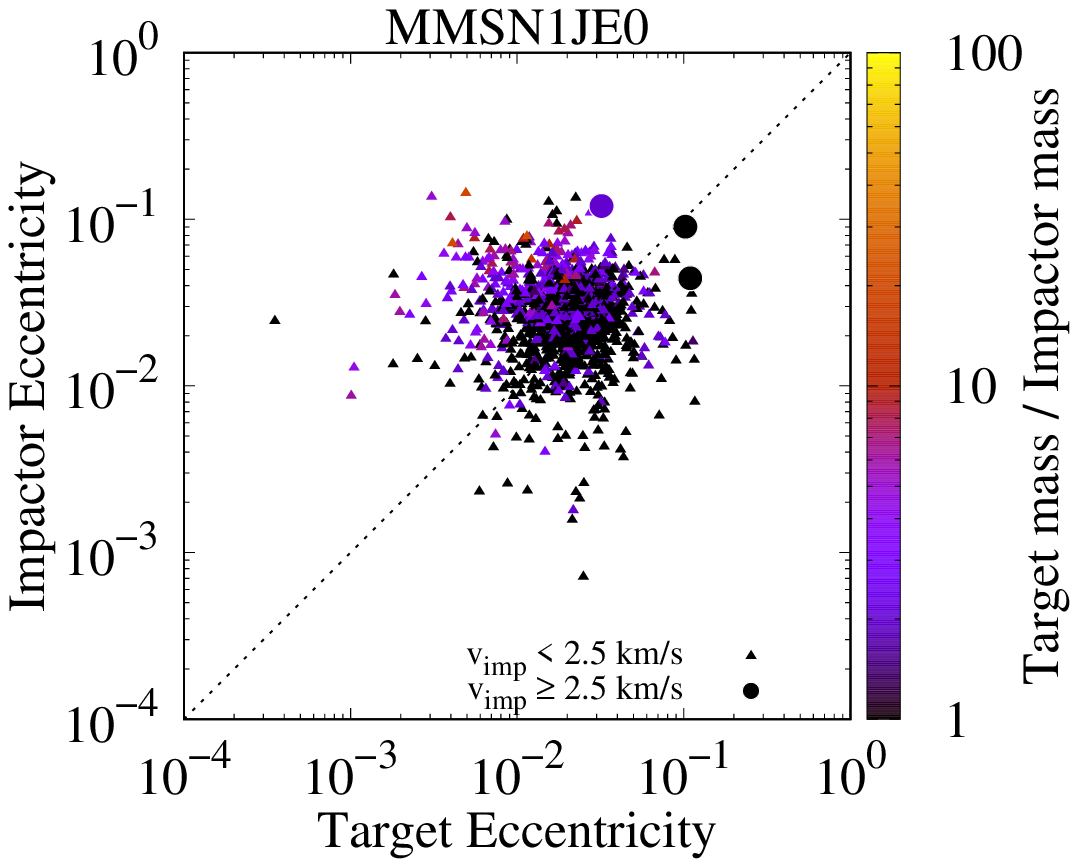}{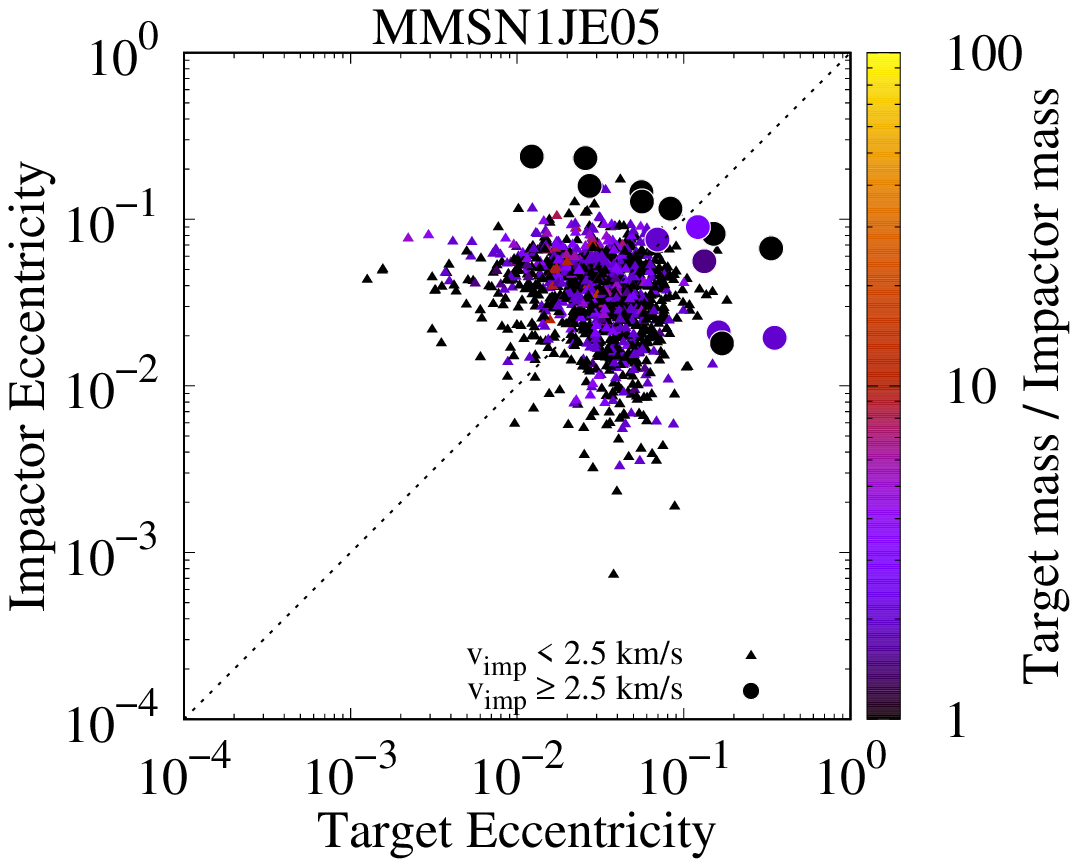}
\plottwo{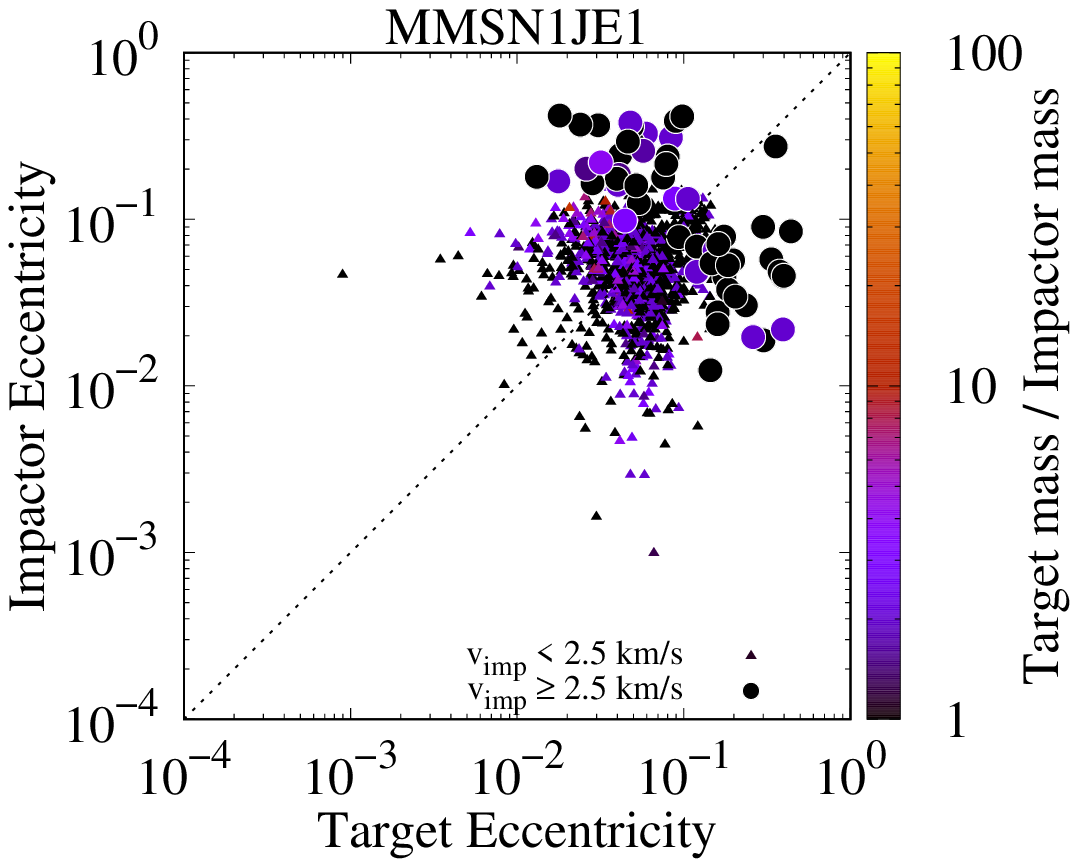}{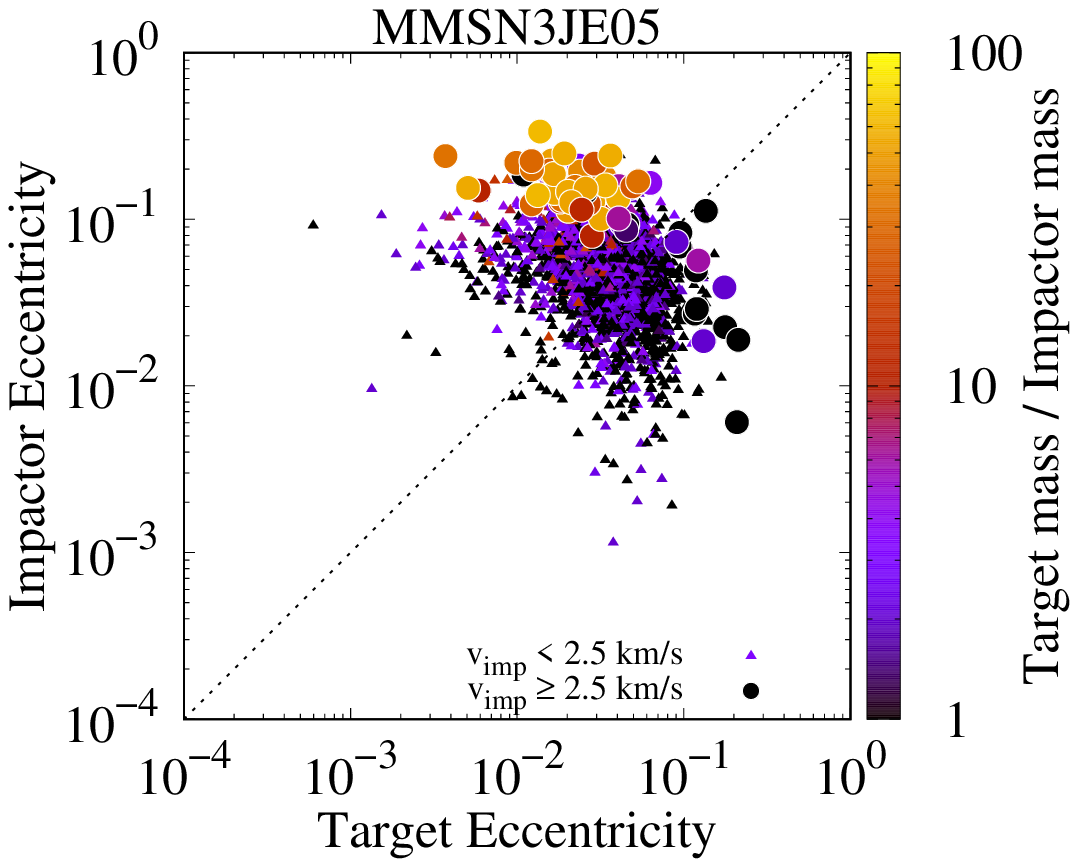}
\caption{Eccentricities of impactors vs those of targets just before collisions as in Figure \ref{fig:ecc-ratio-woJ}.
The results for the MMSN1JE0, MMSN1JE05, MMSN1JE1, and MMSN3JE05 cases are shown on the top left, top right, bottom left and bottom right panels, respectively.}
\label{fig:ecc-ratio-wJ}
\end{figure*}

\begin{figure*}
\plottwo{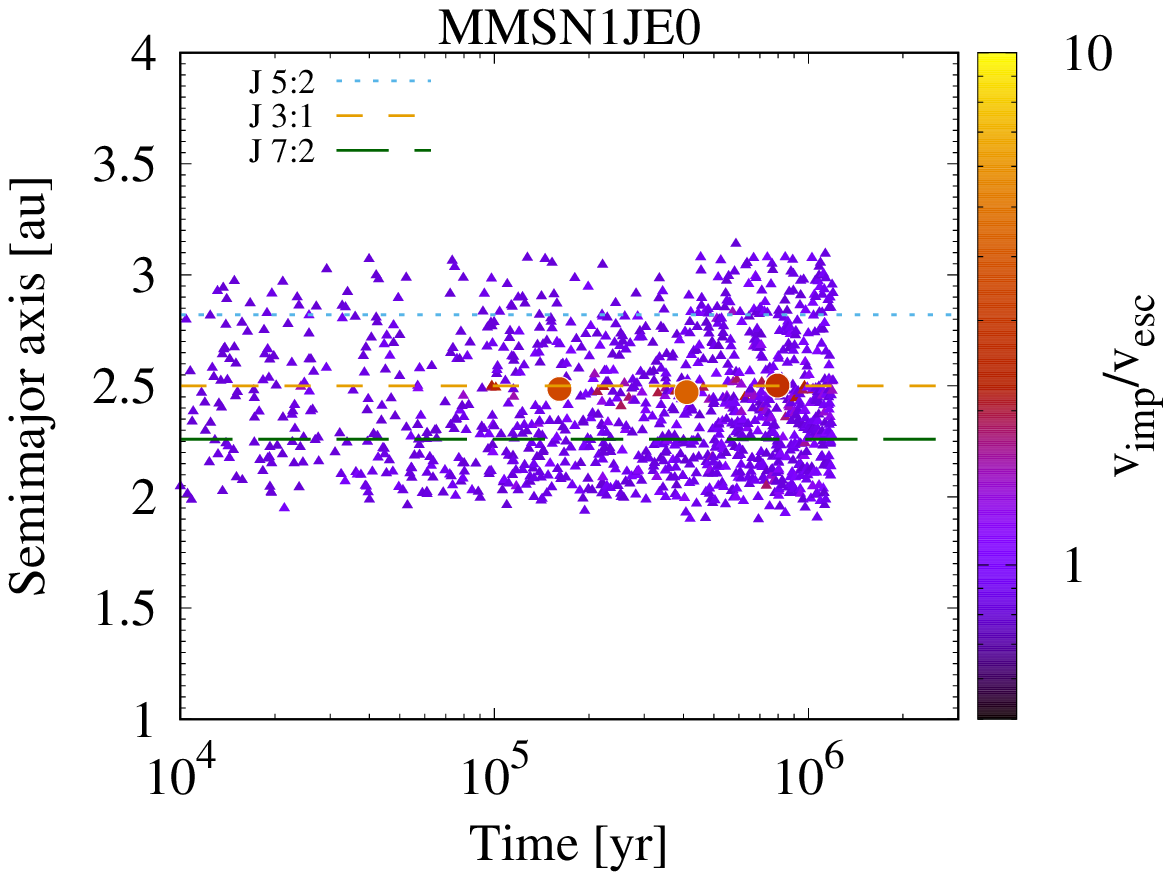}{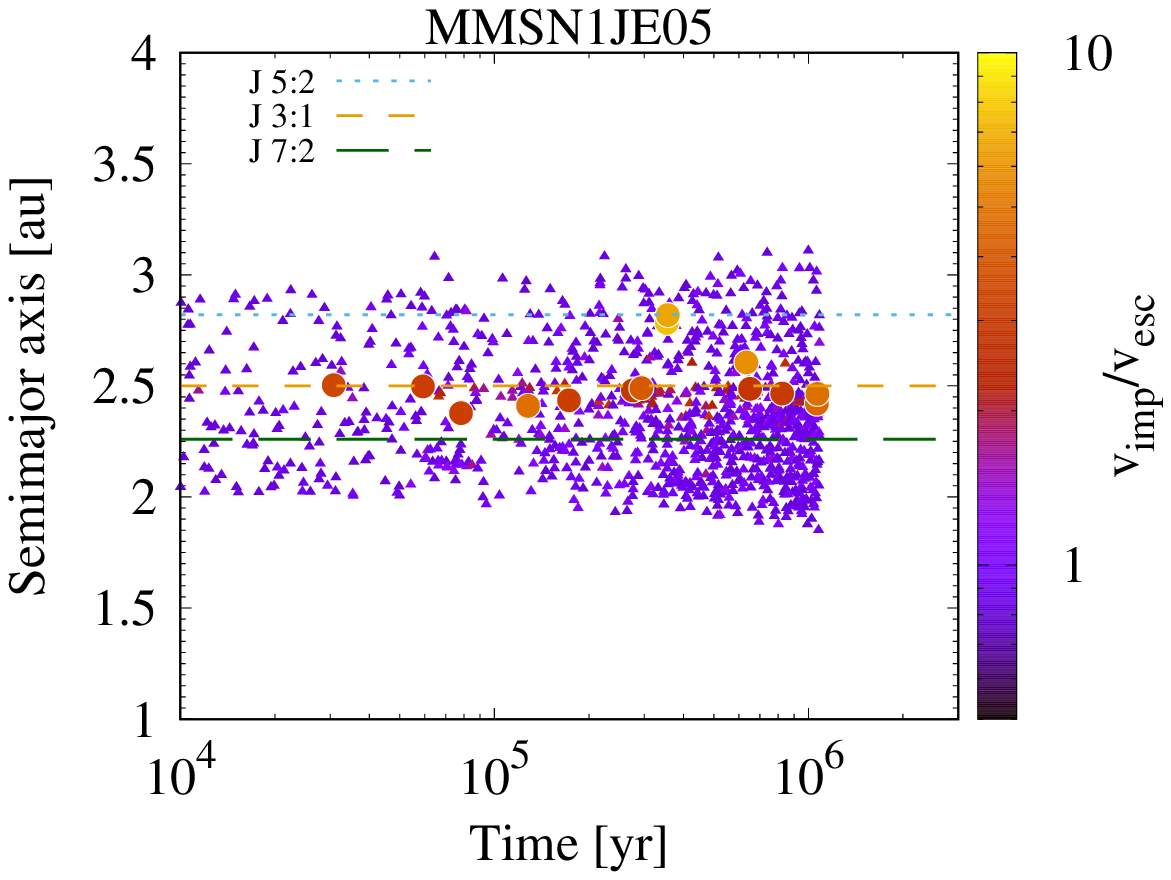}
\plottwo{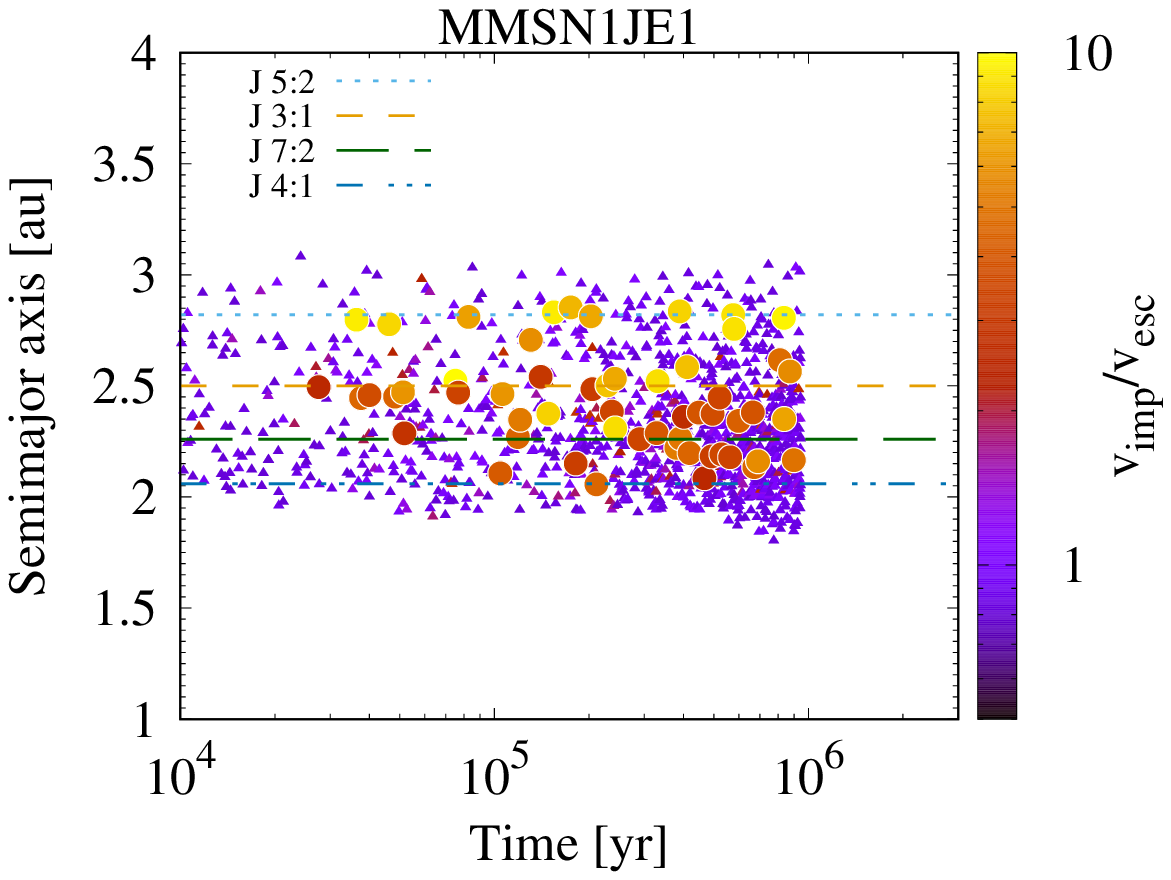}{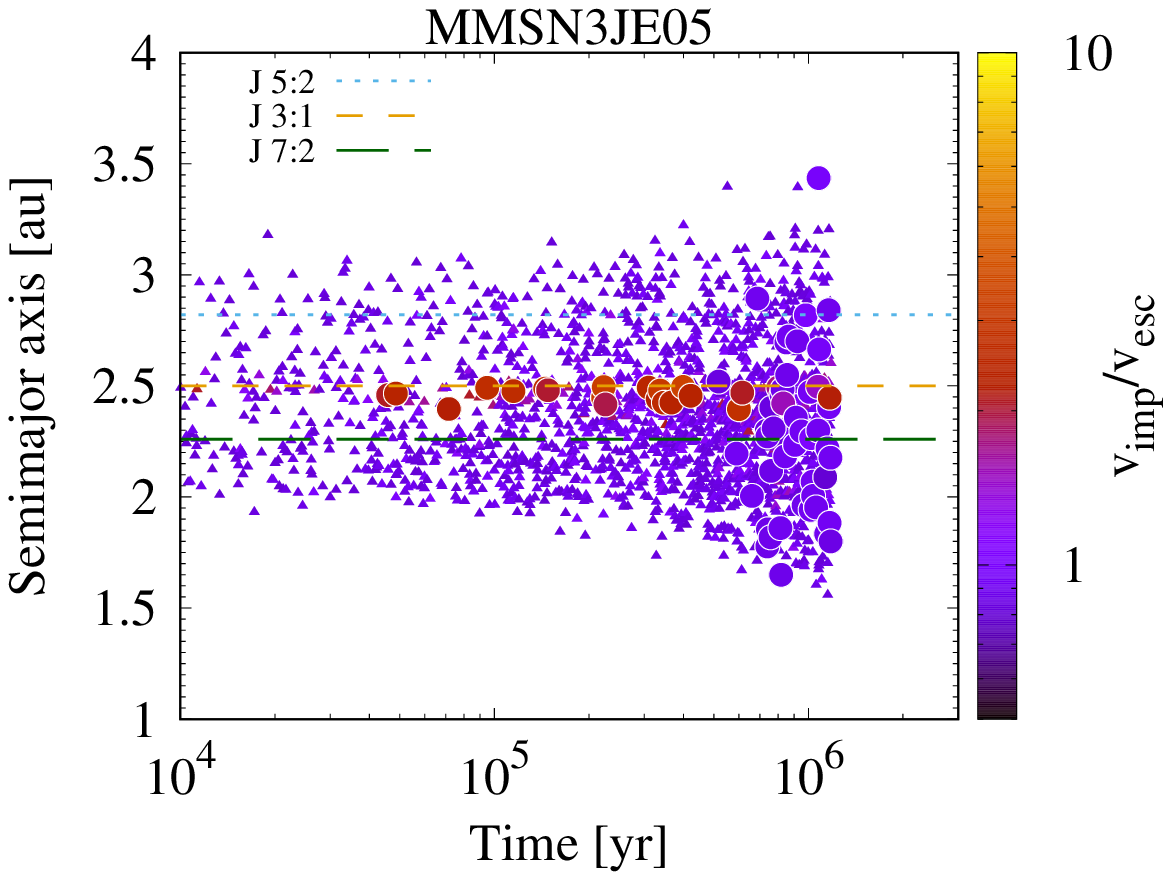}
\caption{Locations of collisions as a function of time as in Figure \ref{fig:pos-col-woJ}.
The results for the MMSN1JE0, MMSN1JE05, MMSN1JE1, and MMSN3JE05 cases are shown on the top left, top right, bottom left and bottom right panels, respectively.
The horizontal lines denote the locations of the mean motion resonances with Jupiter.}
\label{fig:pos-col-wJ}
\end{figure*}

We have performed direct $N$-body simulations to examine the effect of Jupiter's perturbation on planetesimal collisions.
We have confirmed that high velocity ($v_{\rm imp}>2.5$ km s$^{-1}$) impacts are realized 
when growing protoplanets become massive enough for the case without Jupiter (see Figure \ref{fig:ITmass-ratio-woJ}).
This is because the contribution of the escape velocity to the impact velocity becomes large enough to reach the impact velocity of $>2.5$ km s$^{-1}$.
Thus, most of the high velocity collisions originate from protoplanet-planetesimal collisions (see Figure \ref{fig:histgram-woJ}),
which is consistent with the previous studies \citep{2015Natur.517..339J, 2016aHasegawa}.

\begin{figure*}
\plottwo{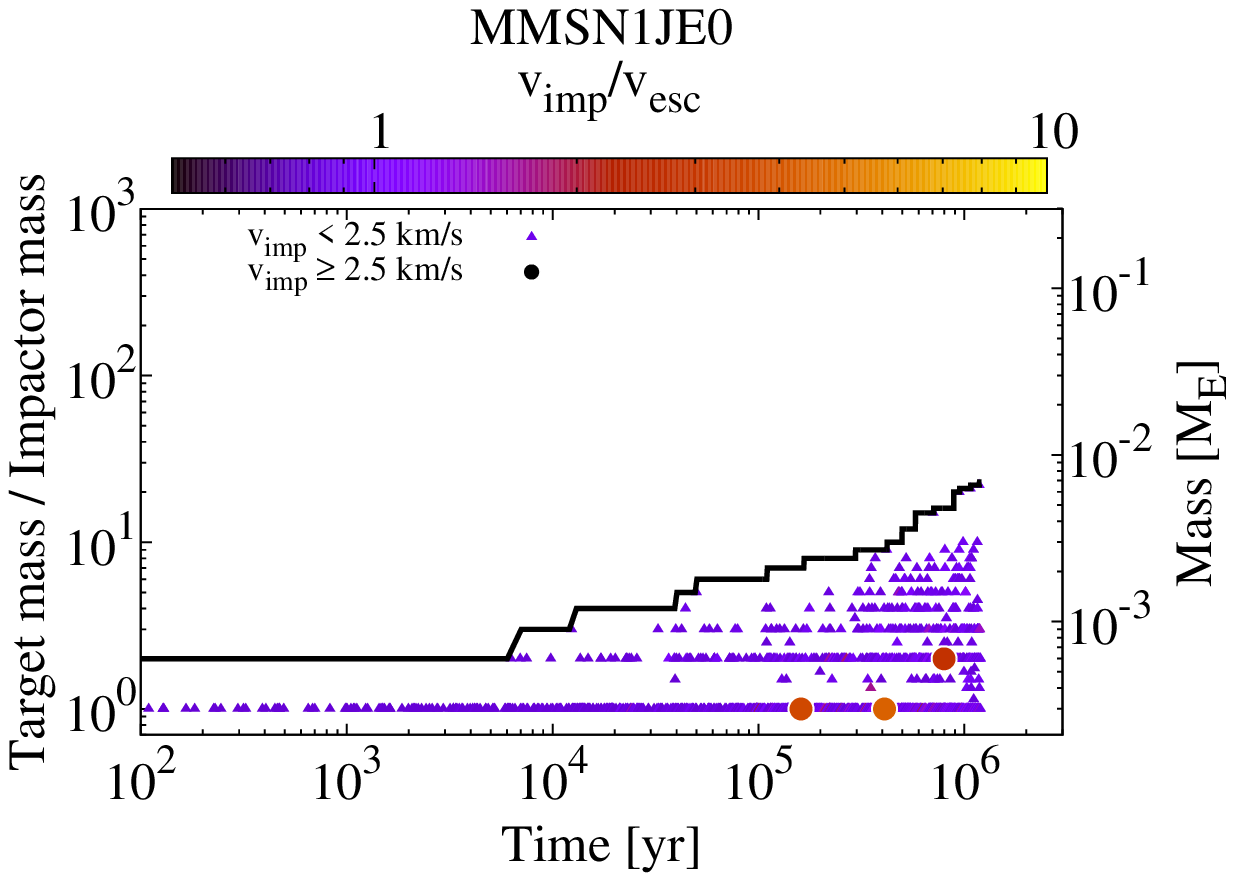}{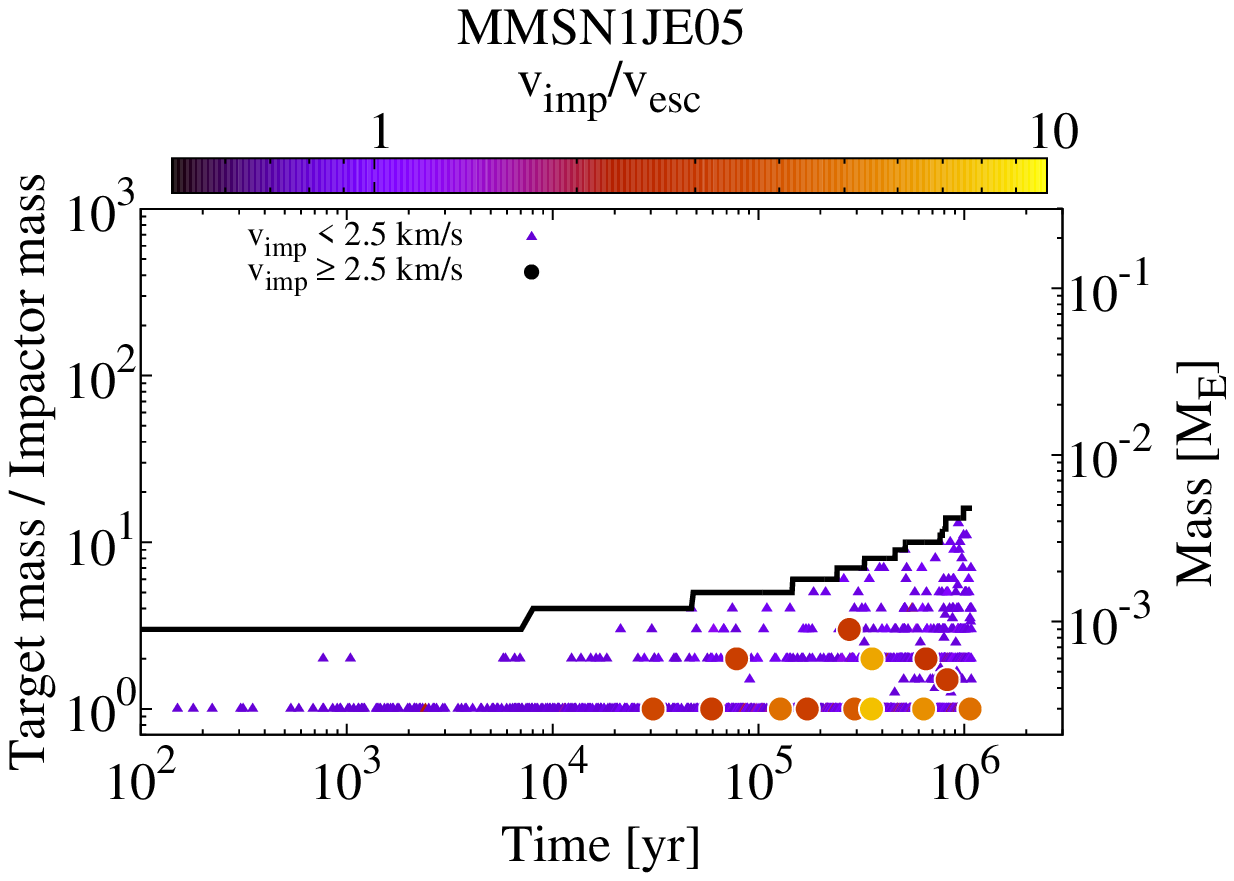}
\plottwo{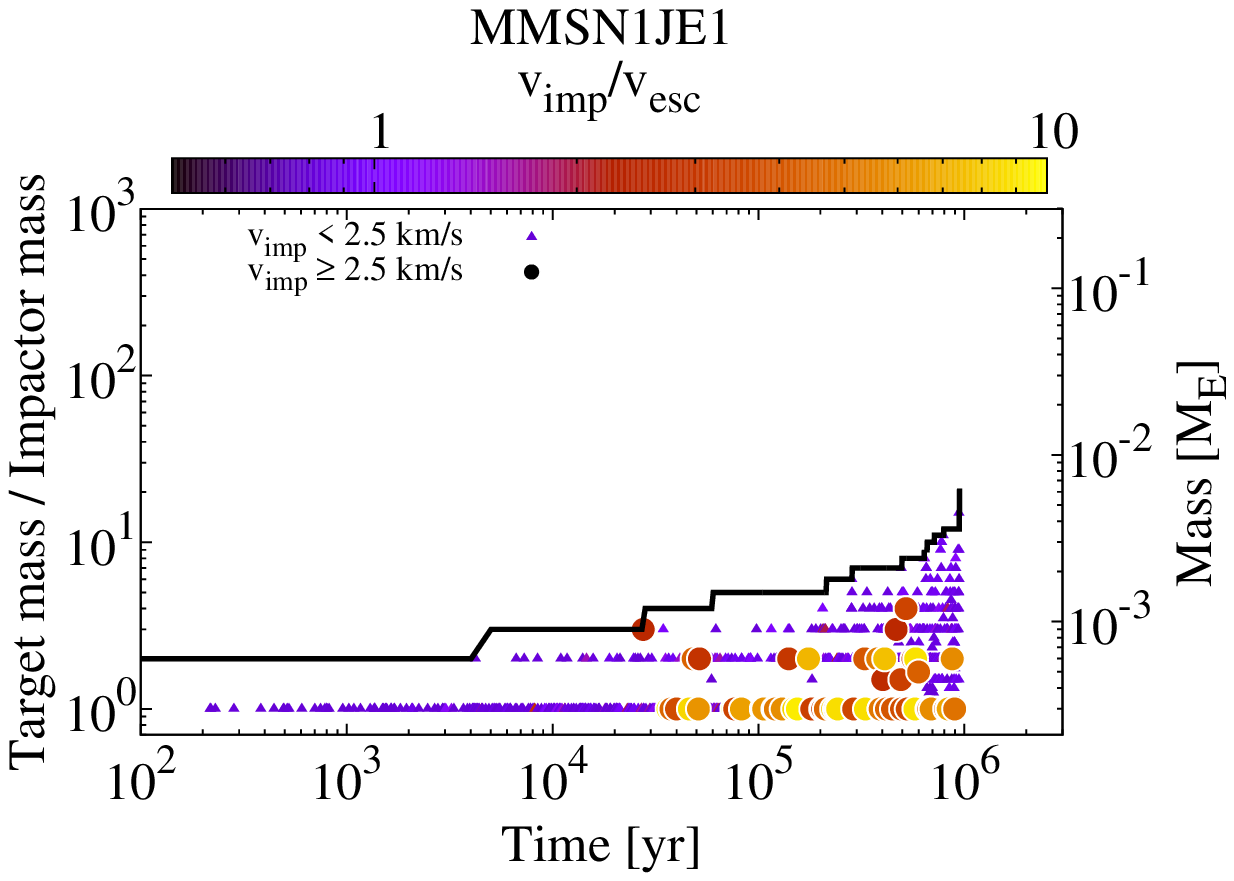}{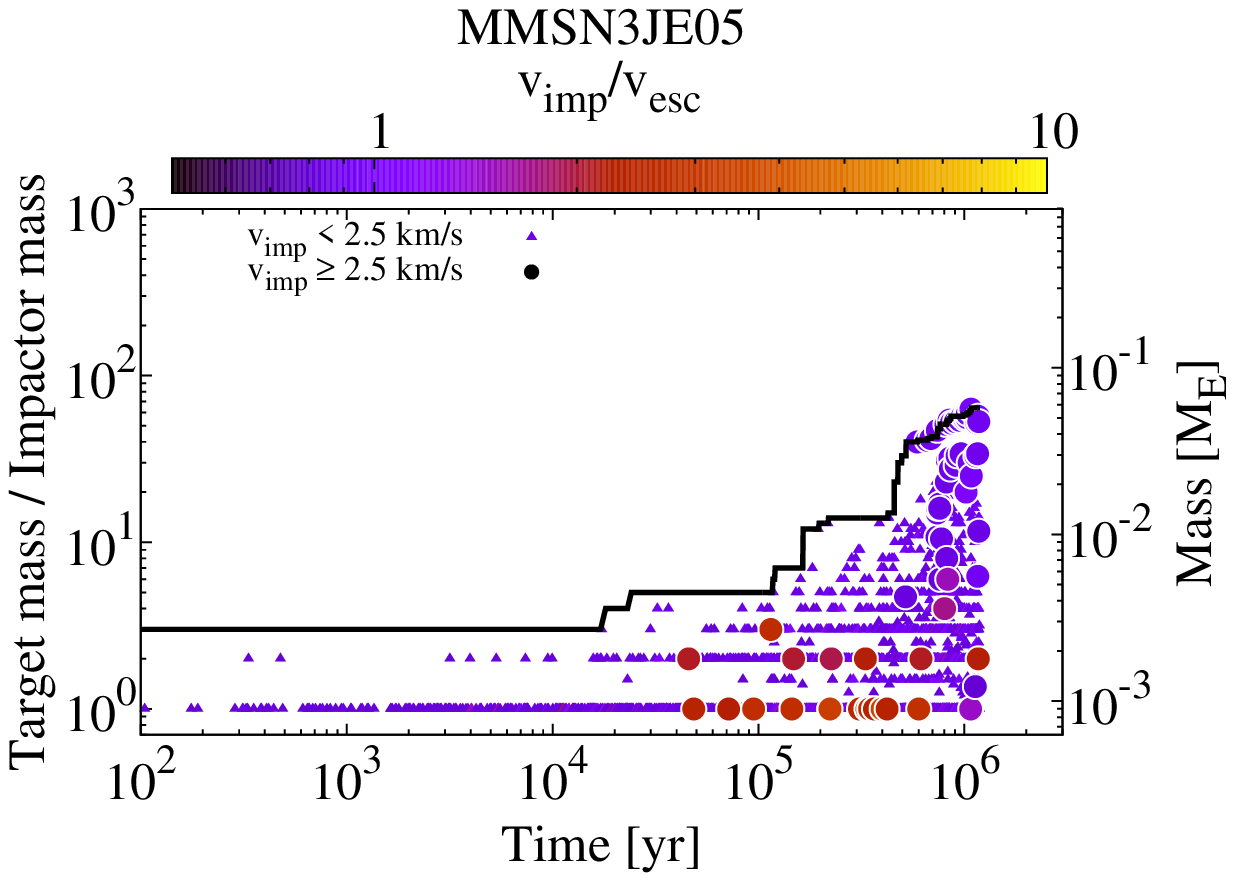}
\caption{The mass ratios of targets to impactors, the mass of the largest particle, and the ratio of $v_{\rm imp}$ to $v_{\rm esc}$
as a function of time as in Figure \ref{fig:ITmass-ratio-woJ}.
The results for the MMSN1JE0, MMSN1JE05, MMSN1JE1, and MMSN3JE05 cases are shown on the top left, top right, bottom left and bottom right panels, respectively.}
\label{fig:ITmass-ratio-wJ}
\end{figure*}

\begin{figure*}
\plottwo{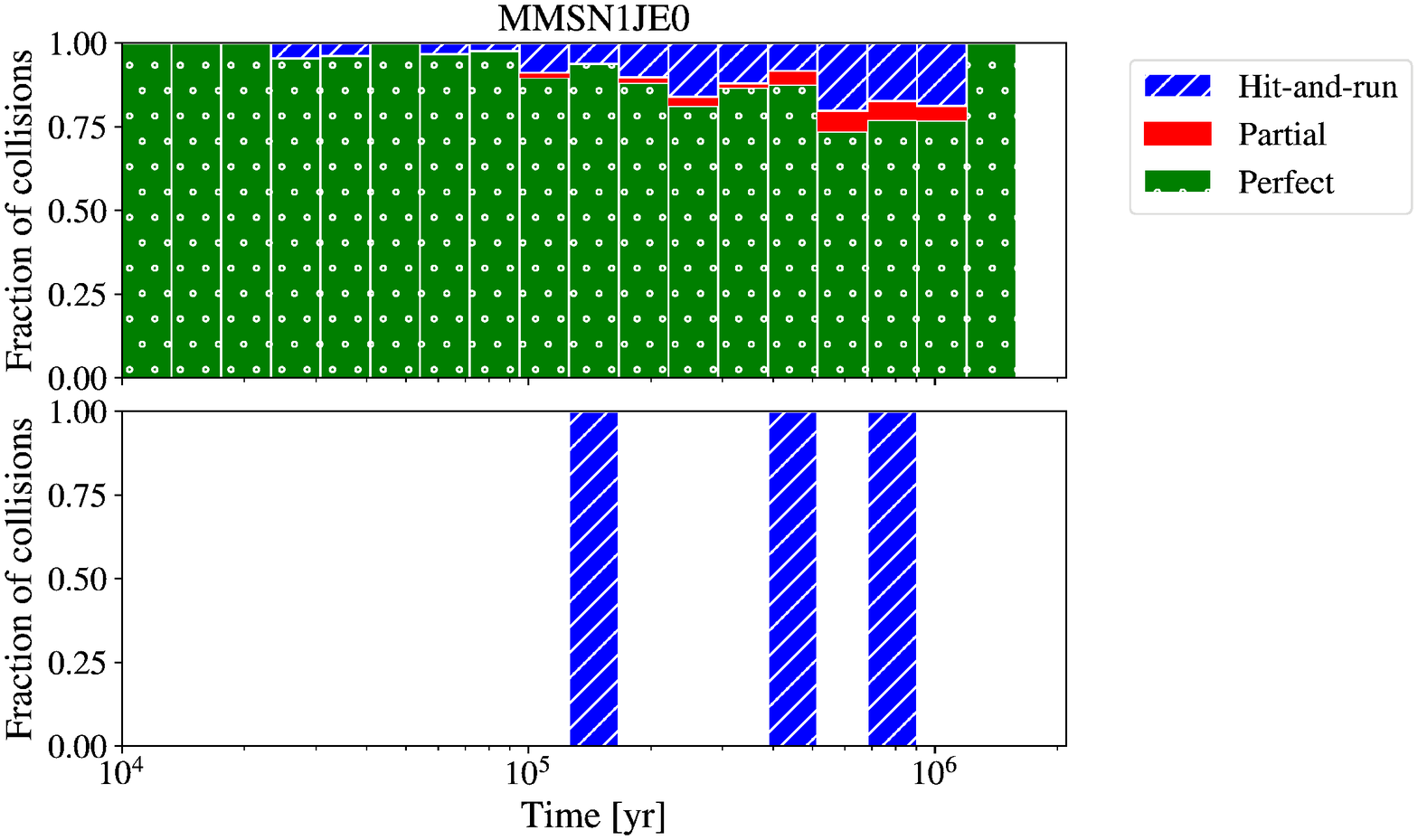}{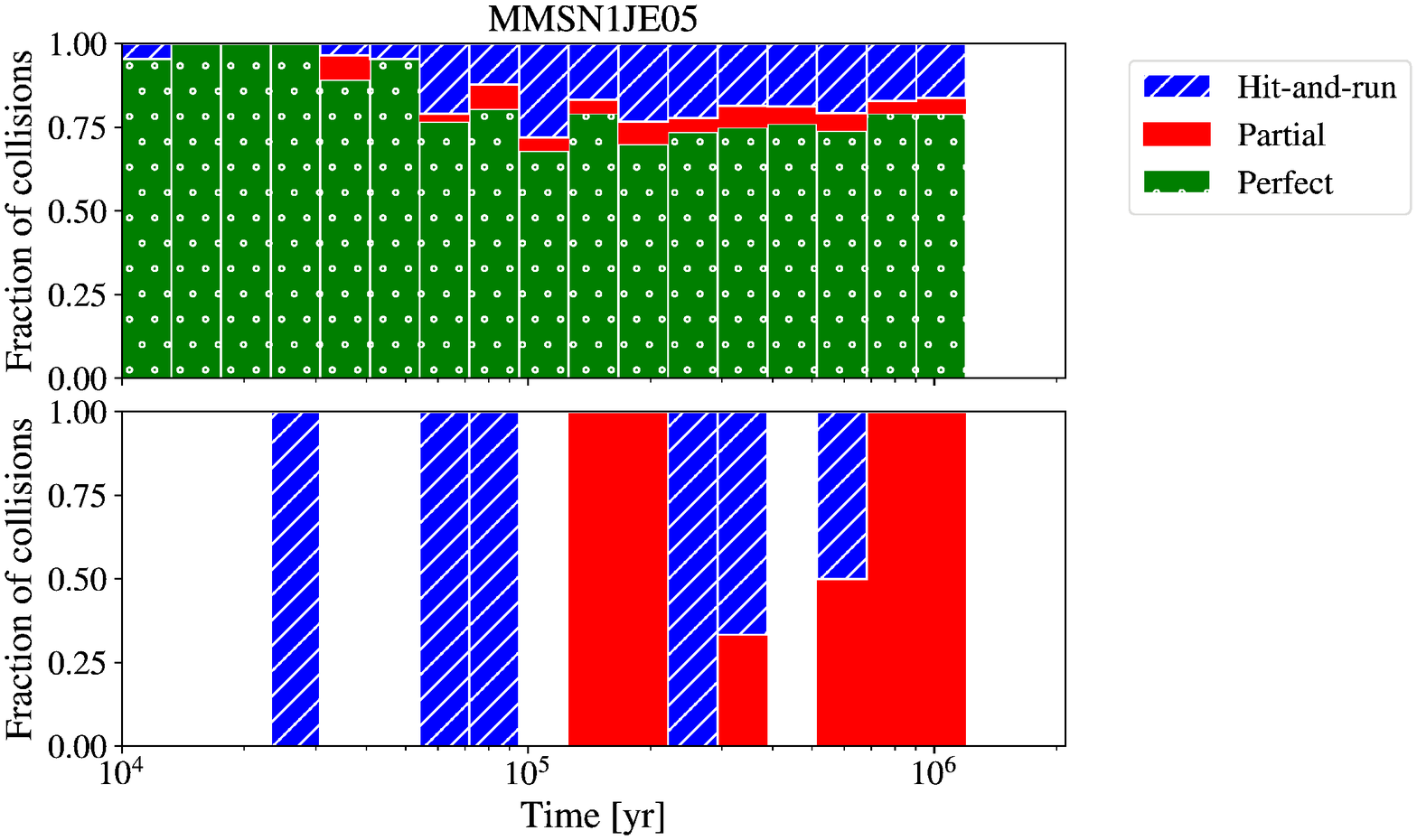}
\plottwo{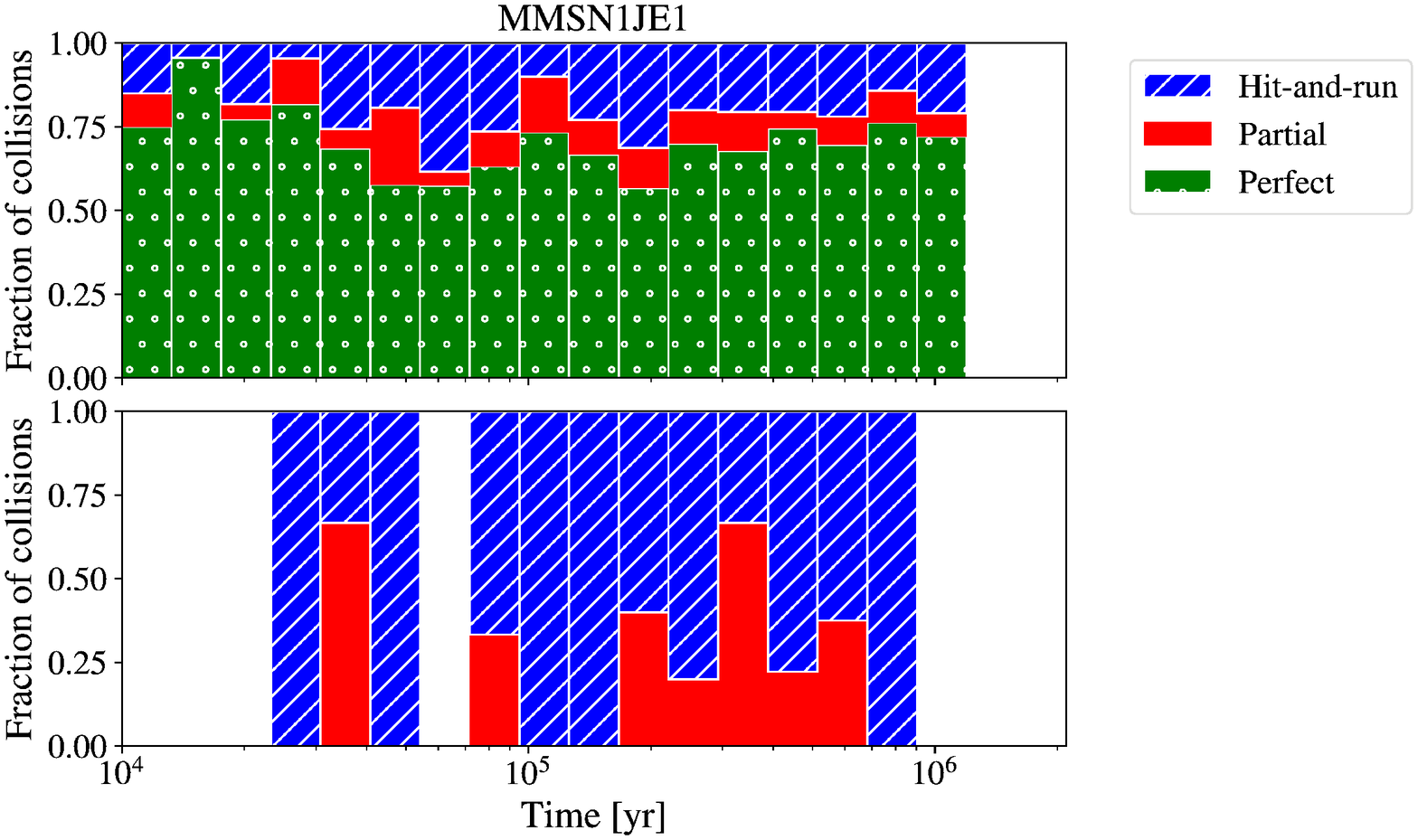}{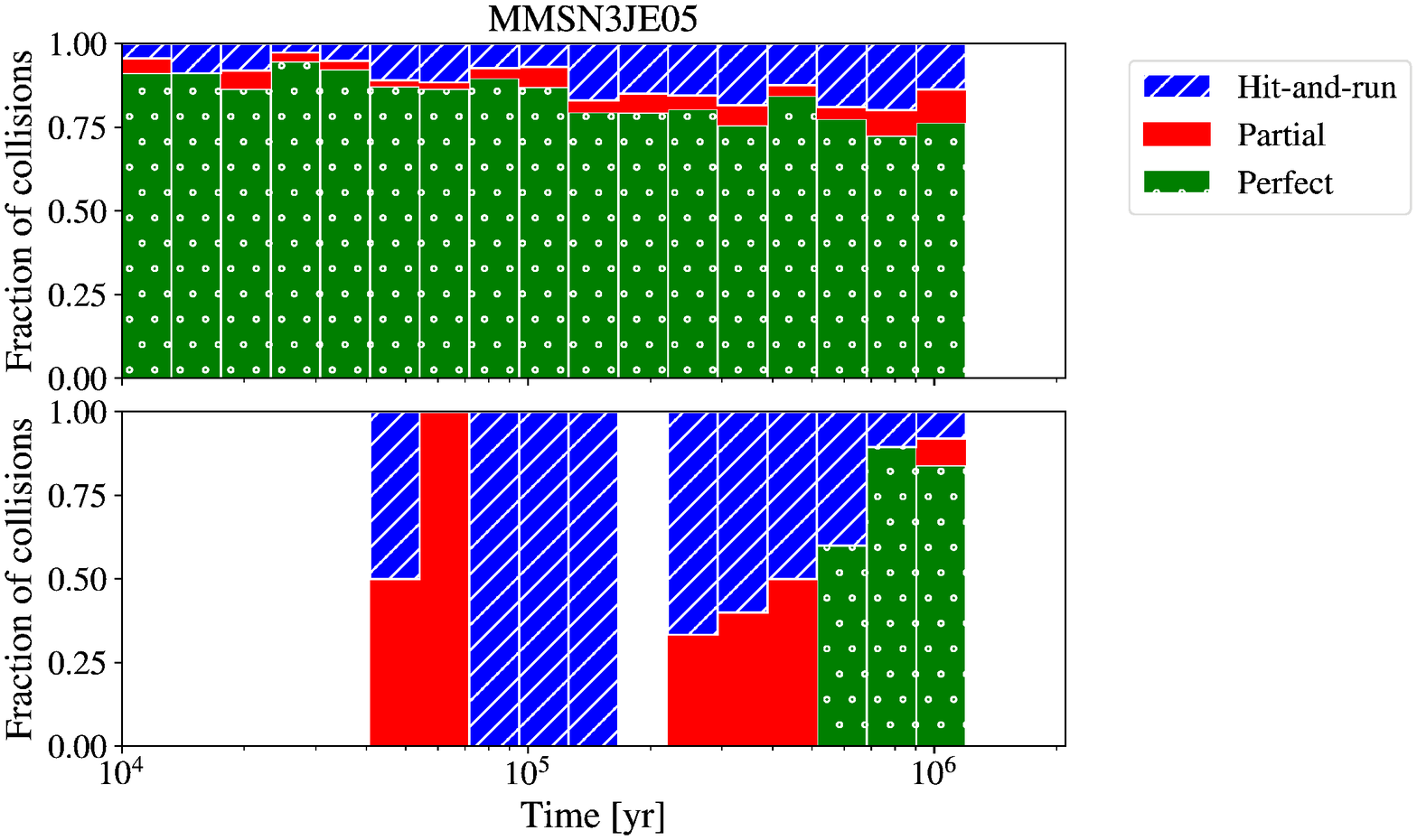}
\caption{
The fractional number of collisions as a function of time as in Figure \ref{fig:histgram-woJ}.
The results of MMSN1JE0, MMSN1JE05, MMSN1JE1, and MMSN3JE05 cases are shown on the top left, top right, bottom left and bottom right panels, respectively.
}
\label{fig:histgram-wJ}
\end{figure*}

We have demonstrated that if Jupiter is included in the simulations,
the high velocity collisions ($v_{\rm imp}>2.5$ km s$^{-1}$) becomes possible well before planetesimals grow up to protoplanets (see Figure \ref{fig:ITmass-ratio-wJ}).
This arises because the eccentricities of planetesimals are excited efficiently by the resonant and secular perturbations from Jupiter,
which leads to the high velocity collisions between planetesimals.
Thus, planetesimal-planetesimal collisions become the main source of energetic impacts in the early stage for the case with Jupiter (see Figure \ref{fig:histgram-wJ}).
As protoplanets grow, planetesimal-protoplanet collisions contribute to the population of energetic impacts as well.

\begin{figure*}
\plottwo{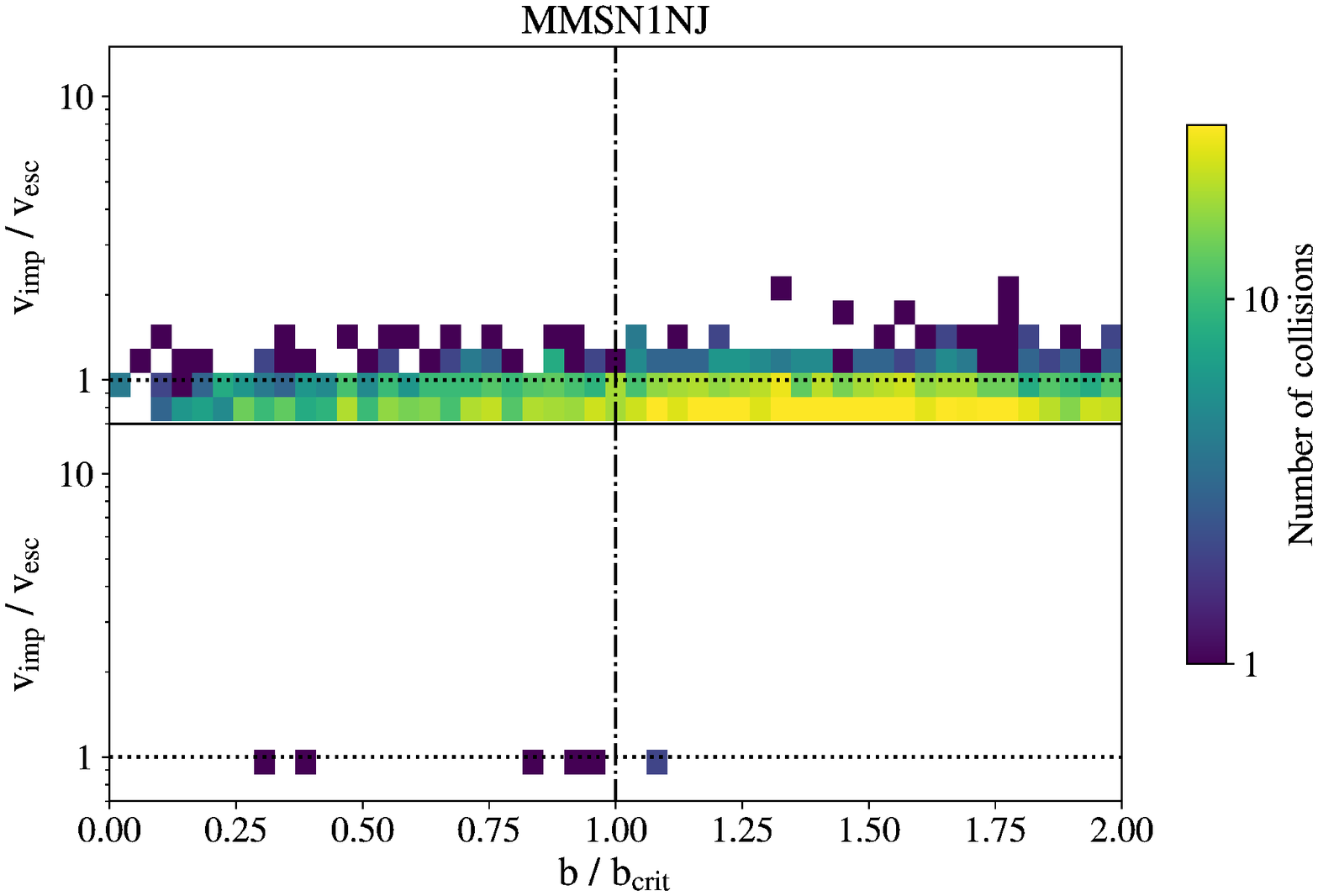}{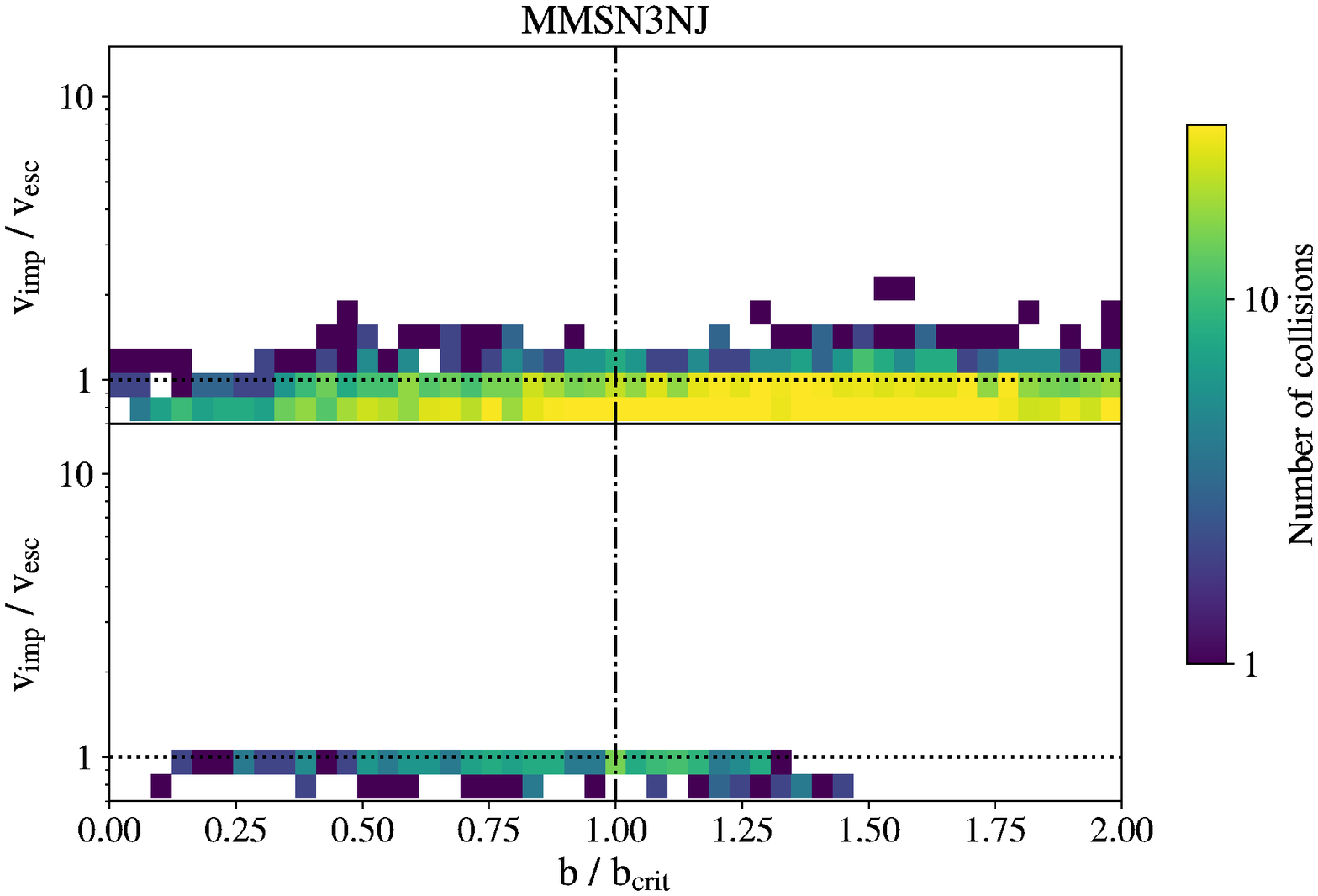}
\caption{
The distribution of collisions in the $v_{\rm imp}/v_{\rm esc}-b/b_{\rm crit}$ plane for the MMSN1NJ (the left panel) and MMSN3NJ (the right panel) cases.
The color bars denote the number of collisions.
As in Figure \ref{fig:histgram-woJ}, low velocity ($v_{\rm imp} \leq 2.5$ km s$^{-1}$) and high velocity ($v_{\rm imp} > 2.5$ km s$^{-1}$) collisions are shown 
on the top and bottom panels, respectively.}
\label{fig:heatmap-woJ}
\end{figure*}

\begin{figure*}
\plottwo{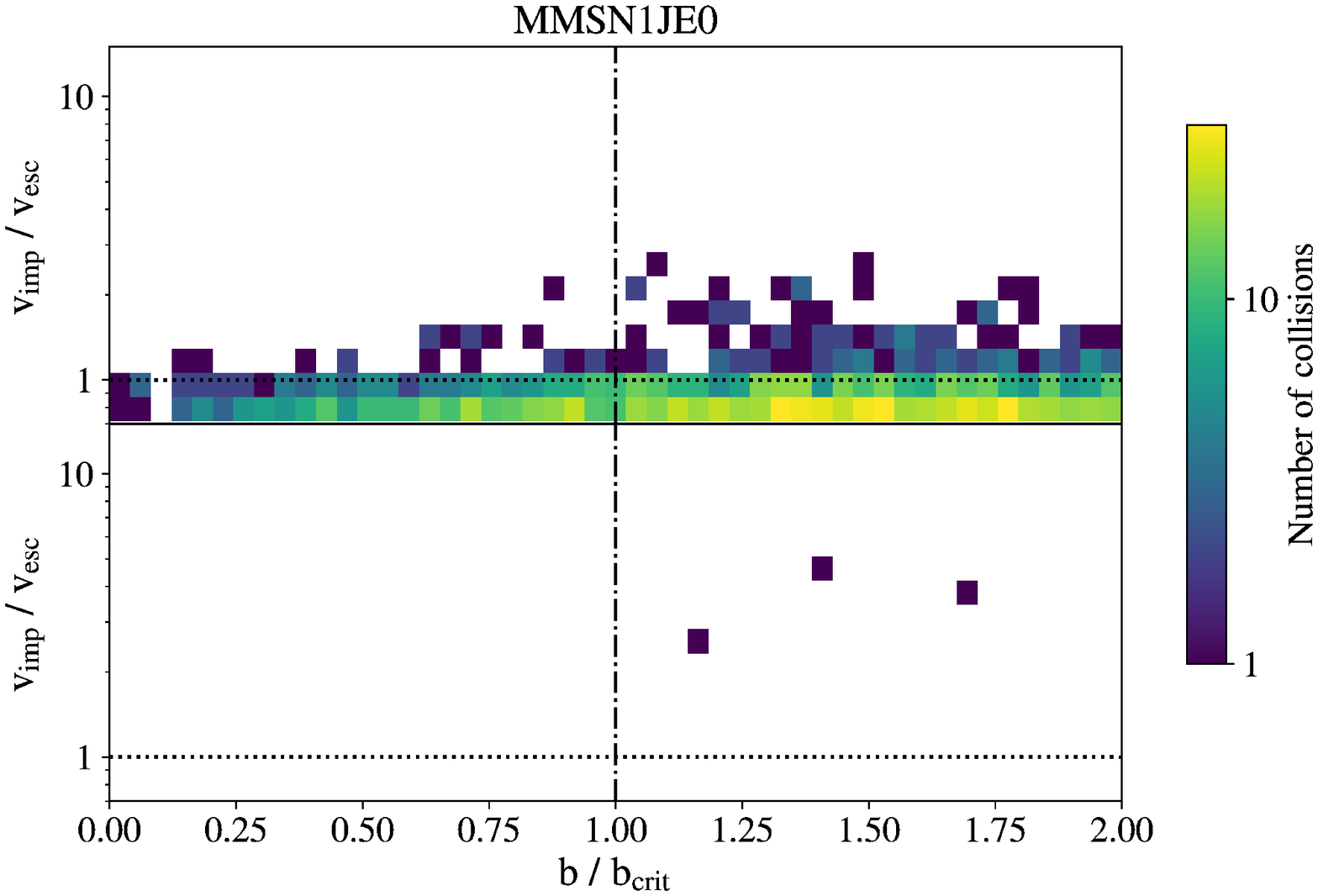}{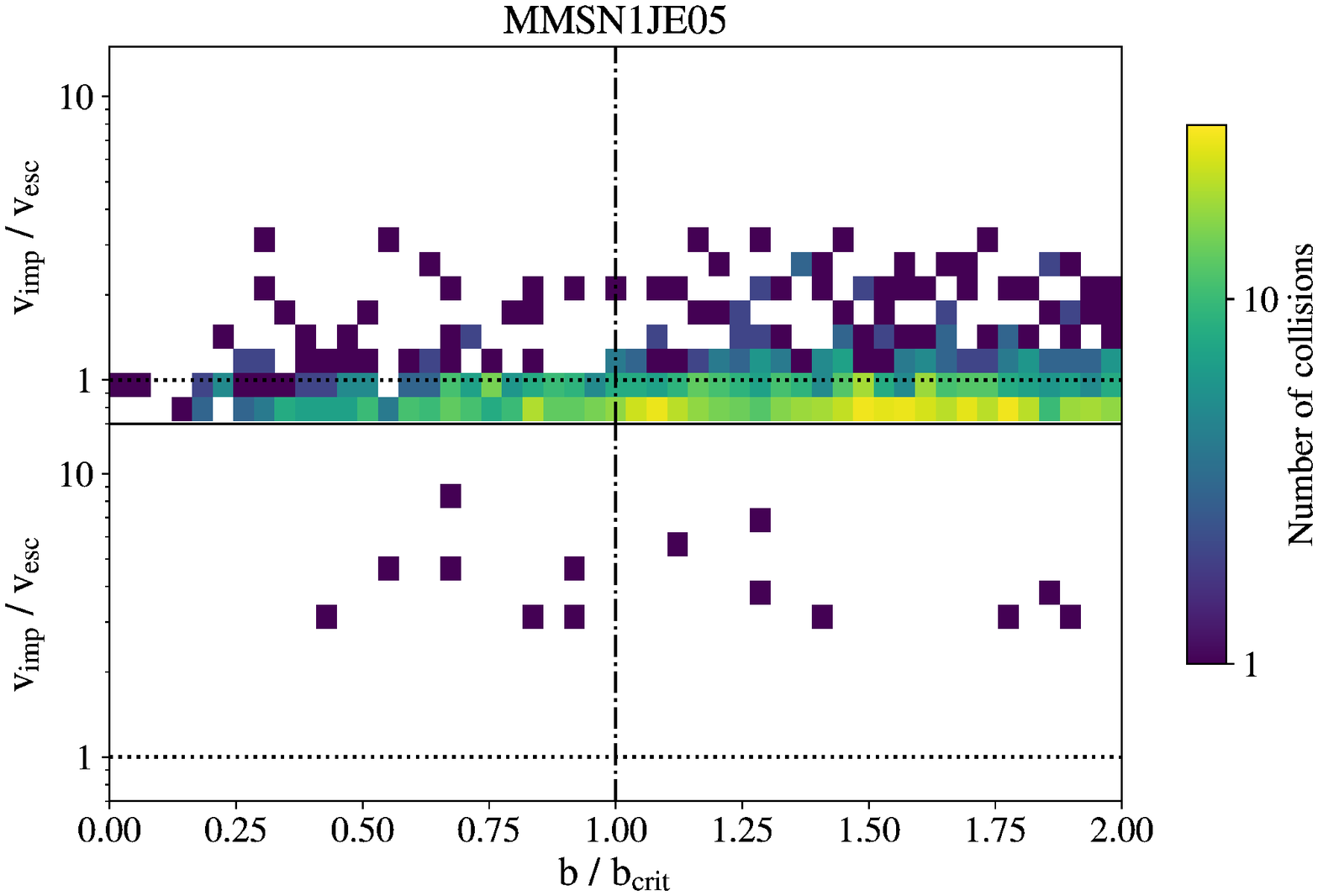}
\plottwo{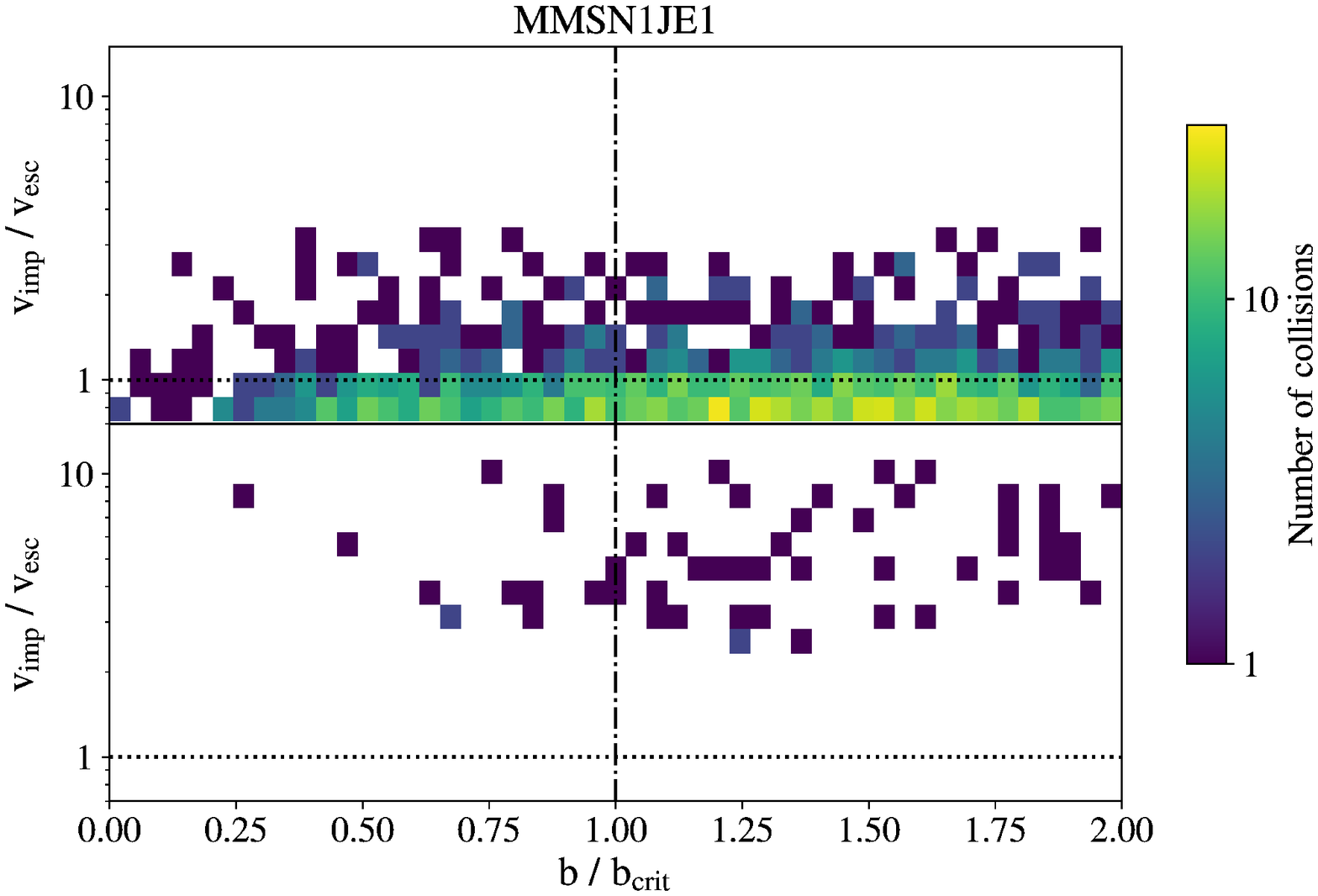}{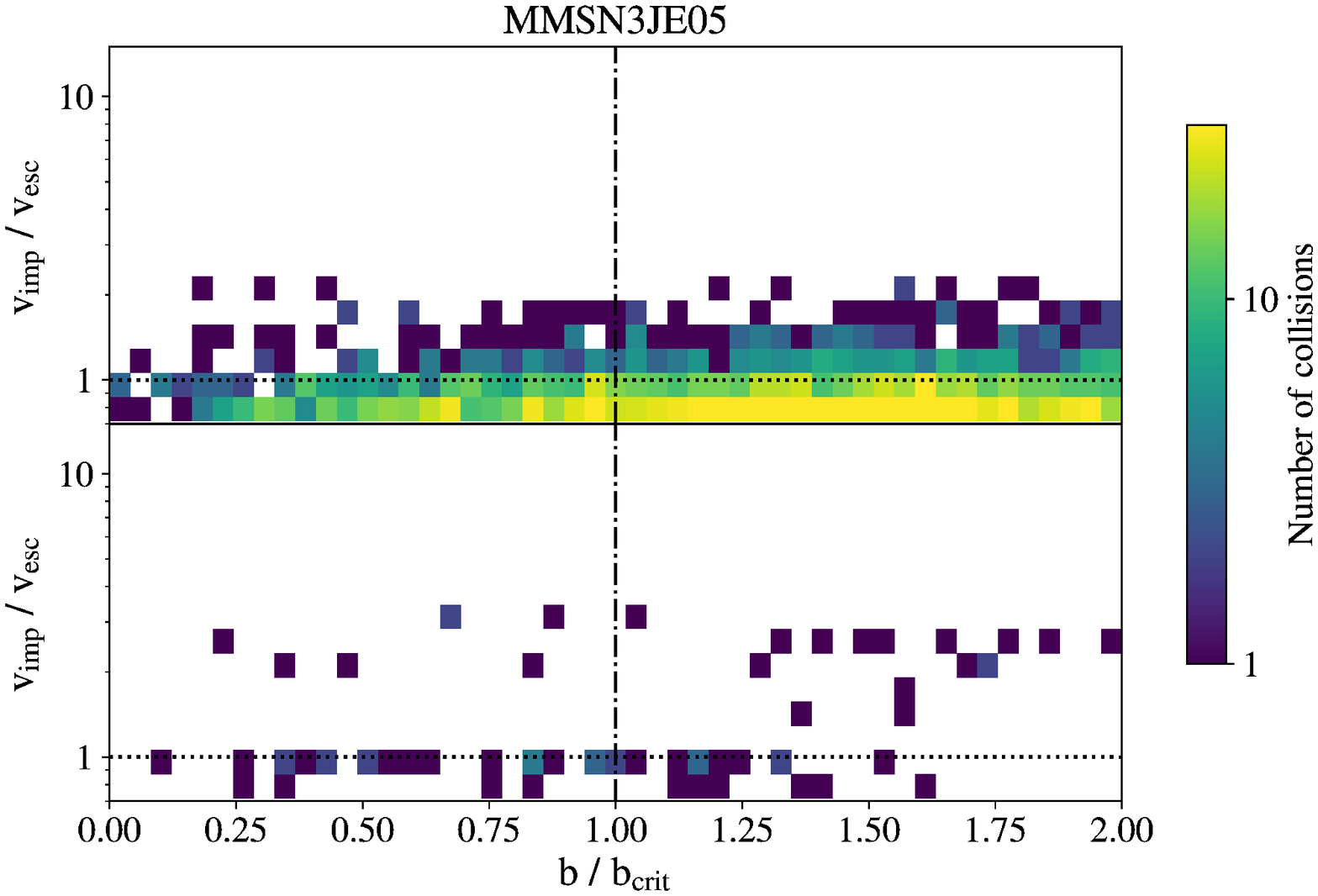}
\caption{
The distribution of collisions in the $v_{\rm imp}/v_{\rm esc}-b/b_{\rm crit}$ plane as in Figure \ref{fig:heatmap-woJ}.
The results for the MMSN1JE0, MMSN1JE05, MMSN1JE1, and MMSN3JE05 cases are shown on the top left, top right, bottom left and bottom right panels, respectively.
}
\label{fig:heatmap-wJ}
\end{figure*}

We have also classified planetesimal collisions, following the simple category (see Section \ref{sec:method_crit}).
We have found that most of high velocity collisions are grazing collisions in the cases with/without Jupiter (see Figures \ref{fig:heatmap-woJ} and \ref{fig:heatmap-wJ}).
Especially, planetesimal-planetesimal collisions are classified as hit-and-run, which can serve as chondrule-forming events.
We thus propose the importance of three-dimensional impact simulations for more accurately estimating the mass of impact melts as future work.

\begin{figure*}
\plottwo{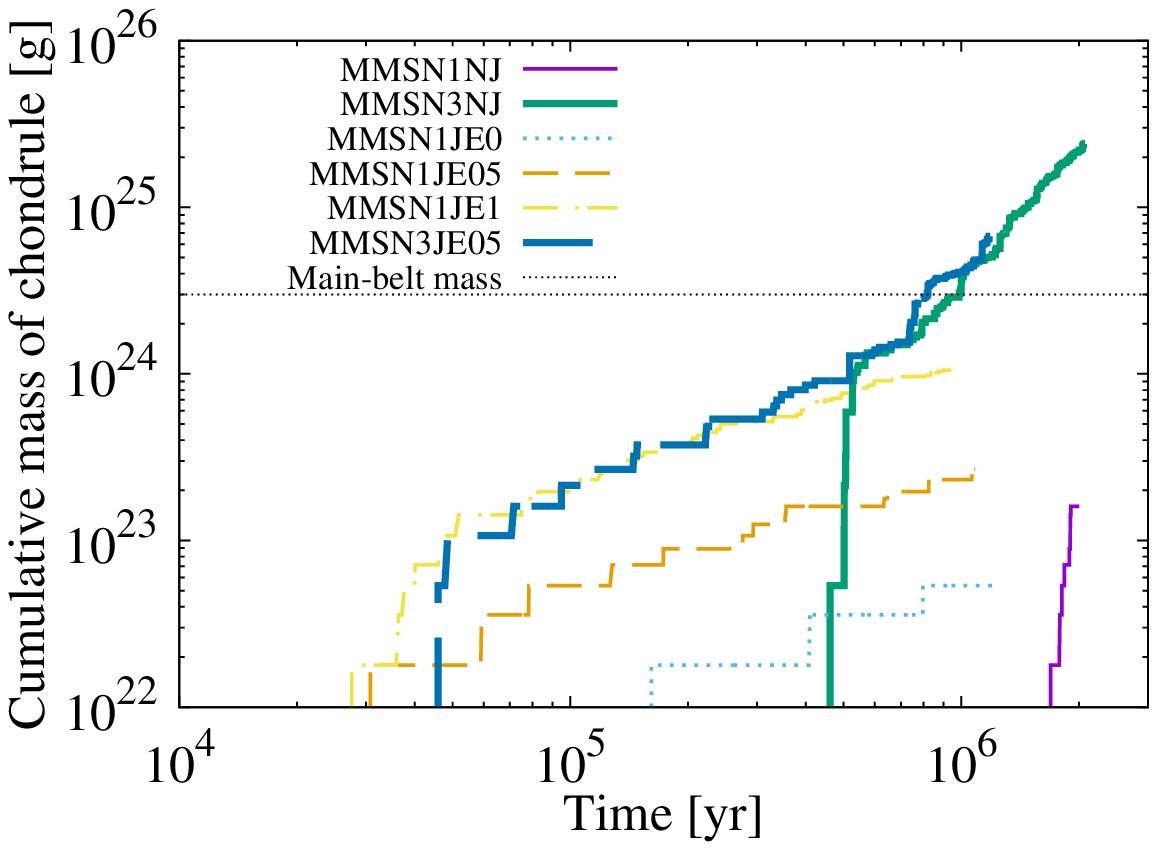}{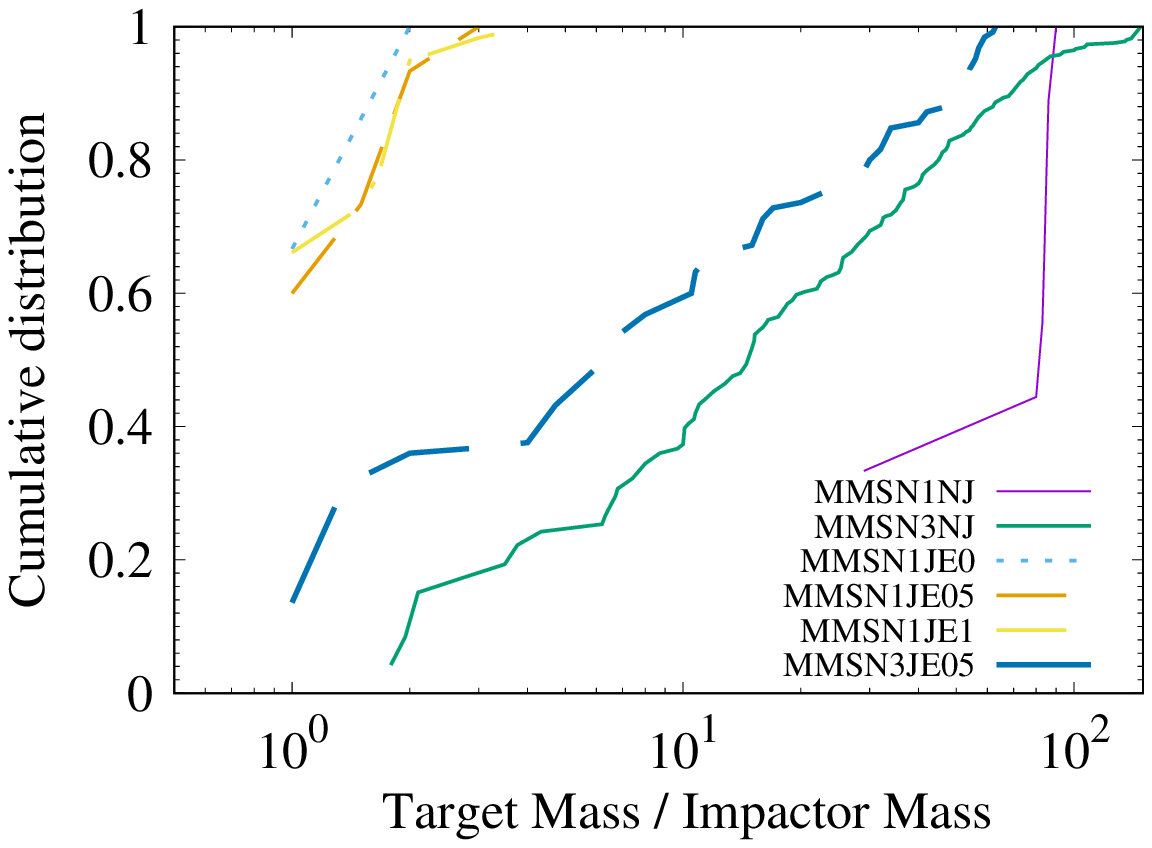}
\caption{Time evolution of the cumulative mass of chondrule, and its distribution as a function of the mass ratio between impactors and targets are shown in the left and right panels, respectively. 
Each line represents the results for each case.
The horizontal black dotted line on the left panel denotes the total mass of the current main asteroid belt, which is $\sim3\times 10^{24} \mathrm{\ g}$.}
\label{fig:ccm}
\end{figure*}

We have applied the results of our $N$-body simulations to chondrule formation via the impact jetting scenario.
We have shown that protoplanet-planetesimal collisions are most important for chondrule formation for the case without Jupiter.
We have also confirmed that chondrule formation occurs earlier in the massive disk (3 times of MMSN) than the standard disk (MMSN) (see Figure \ref{fig:ccm}).
When Jupiter is included in our simulations, planetesimal-planetesimal collisions become important for chondrule formation.
This is the outcome of Jupiter's perturbation, which excites the eccentricity of planetesimals.
Our results, therefore, indicate that the presence of Jupiter can accelerate the onset of chondrule formation via impact jetting.

In conclusion, the dynamics of planetesimals is the key to developing a better understanding of chondrule formation via planetesimal collisions
and hence the origin of the solar system.

\acknowledgments
We thank an anonymous referee for helpful comments and suggestions.
Numerical computations were in part conducted, using the GRAPE system at Center for Computational Astrophysics, National Astronomical Observatory of Japan.
The part of this research was carried out at the Jet Propulsion Laboratory, California Institute of Technology, 
under a contract with the National Aeronautics and Space Administration. 
S.O. thanks JPL/Caltech for hospitality during the visit in which a part of the manuscript was completed.
YH is supported by JPL/Caltech.

\bibliographystyle{apj}

\bibliography{apj-jour,bibliography}

\begin{thebibliography}{121}
\expandafter\ifx\csname natexlab\endcsname\relax\def\natexlab#1{#1}\fi

\bibitem[{{Adachi} {et~al.}(1976){Adachi}, {Hayashi}, \&
  {Nakazawa}}]{1976PThPh..56.1756A}
{Adachi}, I., {Hayashi}, C., \& {Nakazawa}, K. 1976, Progress of Theoretical
  Physics, 56, 1756

\bibitem[{{Amelin} {et~al.}(2010){Amelin}, {Kaltenbach}, {Iizuka}, {Stirling},
  {Ireland}, {Petaev}, \& {Jacobsen}}]{2010EPSL.300..343A}
{Amelin}, Y., {Kaltenbach}, A., {Iizuka}, T., {Stirling}, C.~H., {Ireland},
  T.~R., {Petaev}, M., \& {Jacobsen}, S.~B. 2010, Earth and Planetary Science
  Letters, 300, 343

\bibitem[{{Asphaug}(2010)}]{2010ChEG...70..199A}
{Asphaug}, E. 2010, Chemie der Erde / Geochemistry, 70, 199

\bibitem[{{Asphaug} {et~al.}(2011){Asphaug}, {Jutzi}, \&
  {Movshovitz}}]{2011E&PSL.308..369A}
{Asphaug}, E., {Jutzi}, M., \& {Movshovitz}, N. 2011, Earth and Planetary
  Science Letters, 308, 369

\bibitem[{{Bitsch} {et~al.}(2014){Bitsch}, {Morbidelli}, {Lega}, {Kretke}, \&
  {Crida}}]{2014AA...570A..75B}
{Bitsch}, B., {Morbidelli}, A., {Lega}, E., {Kretke}, K., \& {Crida}, A. 2014,
  \aap, 570, A75

\bibitem[{{Blackburn} {et~al.}(2017){Blackburn}, {Alexander}, {Carlson}, \&
  {Elkins-Tanton}}]{Blackburn:2017aa}
{Blackburn}, T., {Alexander}, C.~M.~O., {Carlson}, R., \& {Elkins-Tanton},
  L.~T. 2017, \gca, 200, 201

\bibitem[{Bollard {et~al.}(2015)Bollard, Connelly, \&
  Bizzarro}]{Bollard:2015aa}
Bollard, J., Connelly, J.~N., \& Bizzarro, M. 2015, Meteoritics \& Planetary
  Science, 50, 1197

\bibitem[{{Bollard} {et~al.}(2017){Bollard}, {Connelly}, {Whitehouse},
  {Pringle}, {Bonal}, {J{\o}rgensen}, {Nordlund}, {Moynier}, \&
  {Bizzarro}}]{2017Bollard}
{Bollard}, J., {Connelly}, J.~N., {Whitehouse}, M.~J., {Pringle}, E.~A.,
  {Bonal}, L., {J{\o}rgensen}, J.~K., {Nordlund}, {\AA}., {Moynier}, F., \&
  {Bizzarro}, M. 2017, Science Advances, 3, e1700407

\bibitem[{{Brouwers} {et~al.}(2018){Brouwers}, {Vazan}, \&
  {Ormel}}]{Brouwers:2018aa}
{Brouwers}, M.~G., {Vazan}, A., \& {Ormel}, C.~W. 2018, \aap, 611, A65

\bibitem[{{Chatterjee} {et~al.}(2008){Chatterjee}, {Ford}, {Matsumura}, \&
  {Rasio}}]{2008ApJ...686..580C}
{Chatterjee}, S., {Ford}, E.~B., {Matsumura}, S., \& {Rasio}, F.~A. 2008, \apj,
  686, 580

\bibitem[{{Connelly} {et~al.}(2012){Connelly}, {Bizzarro}, {Krot}, {Nordlund},
  {Wielandt}, \& {Ivanova}}]{2012Sci...338..651C}
{Connelly}, J.~N., {Bizzarro}, M., {Krot}, A.~N., {Nordlund}, {\AA}.,
  {Wielandt}, D., \& {Ivanova}, M.~A. 2012, Science, 338, 651

\bibitem[{{Connolly} \& {Jones}(2016)}]{Connolly:2016aa}
{Connolly}, H.~C. \& {Jones}, R.~H. 2016, Journal of Geophysical Research
  (Planets), 121, 1885

\bibitem[{{Crida} {et~al.}(2006){Crida}, {Morbidelli}, \&
  {Masset}}]{Crida+2006}
{Crida}, A., {Morbidelli}, A., \& {Masset}, F. 2006, \icarus, 181, 587

\bibitem[{{Davison} {et~al.}(2014){Davison}, {Ciesla}, {Collins}, \&
  {Elbeshausen}}]{2014MPS...49.2252D}
{Davison}, T.~M., {Ciesla}, F.~J., {Collins}, G.~S., \& {Elbeshausen}, D. 2014,
  Meteoritics and Planetary Science, 49, 2252

\bibitem[{{Dawson} \& {Johnson}(2018)}]{2018ARAA..56..175D}
{Dawson}, R.~I. \& {Johnson}, J.~A. 2018, \araa, 56, 175

\bibitem[{{Desch} \& {Cuzzi}(2000)}]{2000Icar..143...87D}
{Desch}, S.~J. \& {Cuzzi}, J.~N. 2000, \icarus, 143, 87

\bibitem[{{Desch} {et~al.}(2012){Desch}, {Morris}, {Connolly}, \&
  {Boss}}]{2012M&PS...47.1139D}
{Desch}, S.~J., {Morris}, M.~A., {Connolly}, H.~C., \& {Boss}, A.~P. 2012,
  Meteoritics and Planetary Science, 47, 1139

\bibitem[{Dullemond {et~al.}(2016)Dullemond, Harsono, Stammler, \&
  Johansen}]{Dullemond:2016aa}
Dullemond, C.~P., Harsono, D., Stammler, S.~M., \& Johansen, A. 2016, The
  Astrophysical Journal, 832, 91

\bibitem[{{Dullemond} {et~al.}(2014){Dullemond}, {Stammler}, \&
  {Johansen}}]{2014ApJ...794...91D}
{Dullemond}, C.~P., {Stammler}, S.~M., \& {Johansen}, A. 2014, \apj, 794, 91

\bibitem[{{Elkins-Tanton}(2012)}]{2012AREPS..40..113E}
{Elkins-Tanton}, L.~T. 2012, Annual Review of Earth and Planetary Sciences, 40,
  113

\bibitem[{{Fu} {et~al.}(2014){Fu}, {Weiss}, {Lima}, {Harrison}, {Bai}, {Desch},
  {Ebel}, {Suavet}, {Wang}, {Glenn}, {Le Sage}, {Kasama}, {Walsworth}, \&
  {Kuan}}]{2014Fu}
{Fu}, R.~R., {Weiss}, B.~P., {Lima}, E.~A., {Harrison}, R.~J., {Bai}, X.-N.,
  {Desch}, S.~J., {Ebel}, D.~S., {Suavet}, C., {Wang}, H., {Glenn}, D., {Le
  Sage}, D., {Kasama}, T., {Walsworth}, R.~L., \& {Kuan}, A.~T. 2014, Science,
  346, 1089

\bibitem[{{Gail} {et~al.}(2014){Gail}, {Trieloff}, {Breuer}, \&
  {Spohn}}]{Gail:2014aa}
{Gail}, H.-P., {Trieloff}, M., {Breuer}, D., \& {Spohn}, T. 2014, Protostars
  and Planets VI, 571

\bibitem[{{Genda} {et~al.}(2017){Genda}, {Fujita}, {Kobayashi}, {Tanaka},
  {Suetsugu}, \& {Abe}}]{2017Icar..294..234G}
{Genda}, H., {Fujita}, T., {Kobayashi}, H., {Tanaka}, H., {Suetsugu}, R., \&
  {Abe}, Y. 2017, \icarus, 294, 234

\bibitem[{{Genda} {et~al.}(2012){Genda}, {Kokubo}, \& {Ida}}]{Genda+2012}
{Genda}, H., {Kokubo}, E., \& {Ida}, S. 2012, \apj, 744, 137

\bibitem[{{Goldreich} \& {Tremaine}(1980)}]{1980ApJ...241..425G}
{Goldreich}, P. \& {Tremaine}, S. 1980, \apj, 241, 425

\bibitem[{{Goldstein} {et~al.}(2009){Goldstein}, {Scott}, \&
  {Chabot}}]{Goldstein:2009aa}
{Goldstein}, J.~I., {Scott}, E.~R.~D., \& {Chabot}, N.~L. 2009, Chemie der Erde
  / Geochemistry, 69, 293

\bibitem[{{Harrison} \& {Grimm}(2010)}]{Harrison:2010aa}
{Harrison}, K.~P. \& {Grimm}, R.~E. 2010, \gca, 74, 5410

\bibitem[{{Hasegawa} \& {Pudritz}(2011{\natexlab{a}})}]{2011MNRAS.413..286H}
{Hasegawa}, Y. \& {Pudritz}, R.~E. 2011{\natexlab{a}}, \mnras, 413, 286

\bibitem[{{Hasegawa} \& {Pudritz}(2011{\natexlab{b}})}]{2011MNRAS.417.1236H}
---. 2011{\natexlab{b}}, \mnras, 417, 1236

\bibitem[{{Hasegawa} \& {Pudritz}(2014)}]{2014ApJ...794...25H}
---. 2014, \apj, 794, 25

\bibitem[{{Hasegawa} {et~al.}(2016{\natexlab{a}}){Hasegawa}, {Turner},
  {Masiero}, {Wakita}, {Matsumoto}, \& {Oshino}}]{2016bHasegawa}
{Hasegawa}, Y., {Turner}, N.~J., {Masiero}, J., {Wakita}, S., {Matsumoto}, Y.,
  \& {Oshino}, S. 2016{\natexlab{a}}, \apjl, 820, L12

\bibitem[{{Hasegawa} {et~al.}(2016{\natexlab{b}}){Hasegawa}, {Wakita},
  {Matsumoto}, \& {Oshino}}]{2016aHasegawa}
{Hasegawa}, Y., {Wakita}, S., {Matsumoto}, Y., \& {Oshino}, S.
  2016{\natexlab{b}}, \apj, 816, 8

\bibitem[{{Hayashi}(1981)}]{1981PThPS..70...35H}
{Hayashi}, C. 1981, Progress of Theoretical Physics Supplement, 70, 35

\bibitem[{{Hern{\'a}ndez} {et~al.}(2007){Hern{\'a}ndez}, {Hartmann}, {Megeath},
  {Gutermuth}, {Muzerolle}, {Calvet}, {Vivas}, {Brice{\~n}o}, {Allen},
  {Stauffer}, {Young}, \& {Fazio}}]{2007ApJ...662.1067H}
{Hern{\'a}ndez}, J., {Hartmann}, L., {Megeath}, T., {Gutermuth}, R.,
  {Muzerolle}, J., {Calvet}, N., {Vivas}, A.~K., {Brice{\~n}o}, C., {Allen},
  L., {Stauffer}, J., {Young}, E., \& {Fazio}, G. 2007, \apj, 662, 1067

\bibitem[{{Ida} \& {Lin}(2004)}]{2004ApJ...604..388I}
{Ida}, S. \& {Lin}, D.~N.~C. 2004, \apj, 604, 388

\bibitem[{{Ida} \& {Makino}(1992)}]{1992Icar...96..107I}
{Ida}, S. \& {Makino}, J. 1992, \icarus, 96, 107

\bibitem[{{Iida} {et~al.}(2001){Iida}, {Nakamoto}, {Susa}, \&
  {Nakagawa}}]{2001Icar..153..430I}
{Iida}, A., {Nakamoto}, T., {Susa}, H., \& {Nakagawa}, Y. 2001, \icarus, 153,
  430

\bibitem[{{Johansen} {et~al.}(2014){Johansen}, {Blum}, {Tanaka}, {Ormel},
  {Bizzarro}, \& {Rickman}}]{2014prpl.conf..547J}
{Johansen}, A., {Blum}, J., {Tanaka}, H., {Ormel}, C., {Bizzarro}, M., \&
  {Rickman}, H. 2014, in Protostars and Planets VI, ed. H.~{Beuther}, R.~S.
  {Klessen}, C.~P. {Dullemond}, \& T.~{Henning}, 547

\bibitem[{Johansen {et~al.}(2015)Johansen, Low, Lacerda, \&
  Bizzarro}]{Johansen:2015aa}
Johansen, A., Low, M.-M.~M., Lacerda, P., \& Bizzarro, M. 2015, Science
  Advances, 1

\bibitem[{{Johnson} {et~al.}(2015){Johnson}, {Minton}, {Melosh}, \&
  {Zuber}}]{2015Natur.517..339J}
{Johnson}, B.~C., {Minton}, D.~A., {Melosh}, H.~J., \& {Zuber}, M.~T. 2015,
  \nat, 517, 339

\bibitem[{{Johnson} {et~al.}(2016){Johnson}, {Walsh}, {Minton}, {Krot}, \&
  {Levison}}]{Johnson:2016aa}
{Johnson}, B.~C., {Walsh}, K.~J., {Minton}, D.~A., {Krot}, A.~N., \& {Levison},
  H.~F. 2016, Science Advances, 2, e1601658

\bibitem[{{Kanagawa}(2019)}]{Kanagawa2019}
{Kanagawa}, K.~D. 2019, arXiv e-prints, arXiv:1906.06338

\bibitem[{{Kanagawa} {et~al.}(2018{\natexlab{a}}){Kanagawa}, {Muto}, {Okuzumi},
  {Tanigawa}, {Taki}, \& {Shibaike}}]{Kanagawa+2018}
{Kanagawa}, K.~D., {Muto}, T., {Okuzumi}, S., {Tanigawa}, T., {Taki}, T., \&
  {Shibaike}, Y. 2018{\natexlab{a}}, \apj, 868, 48

\bibitem[{{Kanagawa} {et~al.}(2015){Kanagawa}, {Muto}, {Tanaka}, {Tanigawa},
  {Takeuchi}, {Tsukagoshi}, \& {Momose}}]{Kanagawa+2015}
{Kanagawa}, K.~D., {Muto}, T., {Tanaka}, H., {Tanigawa}, T., {Takeuchi}, T.,
  {Tsukagoshi}, T., \& {Momose}, M. 2015, \apj, 806, L15

\bibitem[{{Kanagawa} {et~al.}(2016){Kanagawa}, {Muto}, {Tanaka}, {Tanigawa},
  {Takeuchi}, {Tsukagoshi}, \& {Momose}}]{Kanagawa+2016}
---. 2016, \pasj, 68, 43

\bibitem[{{Kanagawa} {et~al.}(2018{\natexlab{b}}){Kanagawa}, {Tanaka}, \&
  {Szuszkiewicz}}]{Kanagawa+2018a}
{Kanagawa}, K.~D., {Tanaka}, H., \& {Szuszkiewicz}, E. 2018{\natexlab{b}},
  \apj, 861, 140

\bibitem[{{Klahr} \& {Schreiber}(2016)}]{Klahr:2016aa}
{Klahr}, H. \& {Schreiber}, A. 2016, in IAU Symposium, Vol. 318, Asteroids: New
  Observations, New Models, ed. S.~R. {Chesley}, A.~{Morbidelli}, R.~{Jedicke},
  \& D.~{Farnocchia}, 1--8

\bibitem[{{Kley} \& {Nelson}(2012)}]{2012ARAA..50..211K}
{Kley}, W. \& {Nelson}, R.~P. 2012, \araa, 50, 211

\bibitem[{{Kobayashi} {et~al.}(2016){Kobayashi}, {Tanaka}, \&
  {Okuzumi}}]{Kobayashi+2016}
{Kobayashi}, H., {Tanaka}, H., \& {Okuzumi}, S. 2016, \apj, 817, 105

\bibitem[{{Kokubo} \& {Ida}(1996)}]{1996KI}
{Kokubo}, E. \& {Ida}, S. 1996, \icarus, 123, 180

\bibitem[{{Kokubo} \& {Ida}(1998)}]{1998Icar..131..171K}
---. 1998, \icarus, 131, 171

\bibitem[{{Kokubo} \& {Ida}(2000)}]{2000Icar..143...15K}
---. 2000, \icarus, 143, 15

\bibitem[{{Krot} {et~al.}(2005){Krot}, {Amelin}, {Cassen}, \&
  {Meibom}}]{Krot:2005aa}
{Krot}, A.~N., {Amelin}, Y., {Cassen}, P., \& {Meibom}, A. 2005, \nat, 436, 989

\bibitem[{{Kruijer} {et~al.}(2017){Kruijer}, {Burkhardt}, {Budde}, \&
  {Kleine}}]{2017PNAS..114.6712K}
{Kruijer}, T.~S., {Burkhardt}, C., {Budde}, G., \& {Kleine}, T. 2017,
  Proceedings of the National Academy of Science, 114, 6712

\bibitem[{{Kruijer} {et~al.}(2014){Kruijer}, {Touboul}, {Fischer-G{\"o}dde},
  {Bermingham}, {Walker}, \& {Kleine}}]{Kruijer:2014aa}
{Kruijer}, T.~S., {Touboul}, M., {Fischer-G{\"o}dde}, M., {Bermingham}, K.~R.,
  {Walker}, R.~J., \& {Kleine}, T. 2014, Science, 344, 1150

\bibitem[{{Lambrechts} \& {Johansen}(2012)}]{Lambrechts&Johansen2012}
{Lambrechts}, M. \& {Johansen}, A. 2012, \aap, 544, A32

\bibitem[{{Leinhardt} \& {Stewart}(2012)}]{Leinhardt&Stewart_ST2012}
{Leinhardt}, Z.~M. \& {Stewart}, S.~T. 2012, \apj, 745, 79

\bibitem[{{Levison} {et~al.}(2015){Levison}, {Kretke}, \&
  {Duncan}}]{2015Natur.524..322L}
{Levison}, H.~F., {Kretke}, K.~A., \& {Duncan}, M.~J. 2015, \nat, 524, 322

\bibitem[{{Lichtenberg} {et~al.}(2018){Lichtenberg}, {Golabek}, {Dullemond},
  {Sch{\"o}nb{\"a}chler}, {Gerya}, \& {Meyer}}]{2018Icar..302...27L}
{Lichtenberg}, T., {Golabek}, G.~J., {Dullemond}, C.~P.,
  {Sch{\"o}nb{\"a}chler}, M., {Gerya}, T.~V., \& {Meyer}, M.~R. 2018, \icarus,
  302, 27

\bibitem[{{Lichtenberg} {et~al.}(2019){Lichtenberg}, {Keller}, {Katz},
  {Golabek}, \& {Gerya}}]{Lichtenberg:2019aa}
{Lichtenberg}, T., {Keller}, T., {Katz}, R.~F., {Golabek}, G.~J., \& {Gerya},
  T.~V. 2019, Earth and Planetary Science Letters, 507, 154

\bibitem[{{Lin} {et~al.}(1996){Lin}, {Bodenheimer}, \&
  {Richardson}}]{1996Natur.380..606L}
{Lin}, D.~N.~C., {Bodenheimer}, P., \& {Richardson}, D.~C. 1996, \nat, 380, 606

\bibitem[{{Lin} \& {Papaloizou}(1986{\natexlab{a}})}]{Lin&Papaloizou1986a}
{Lin}, D.~N.~C. \& {Papaloizou}, J. 1986{\natexlab{a}}, \apj, 307, 395

\bibitem[{{Lin} \& {Papaloizou}(1986{\natexlab{b}})}]{Lin&Papaloizou1986b}
---. 1986{\natexlab{b}}, \apj, 309, 846

\bibitem[{{Lissauer} {et~al.}(2009){Lissauer}, {Hubickyj}, {D'Angelo}, \&
  {Bodenheimer}}]{2009Icar..199..338L}
{Lissauer}, J.~J., {Hubickyj}, O., {D'Angelo}, G., \& {Bodenheimer}, P. 2009,
  \icarus, 199, 338

\bibitem[{{Lissauer} \& {Stewart}(1993)}]{1993LS}
{Lissauer}, J.~J. \& {Stewart}, G.~R. 1993, in Protostars and Planets III, ed.
  E.~H. {Levy} \& J.~I. {Lunine}, 1061--1088

\bibitem[{{Lubow} \& {D'Angelo}(2006)}]{ld06}
{Lubow}, S.~H. \& {D'Angelo}, G. 2006, \apj, 641, 526

\bibitem[{{Makino}(1991)}]{1991ApJ...369..200M}
{Makino}, J. 1991, \apj, 369, 200

\bibitem[{{Makino} \& {Aarseth}(1992)}]{1992PASJ...44..141M}
{Makino}, J. \& {Aarseth}, S.~J. 1992, \pasj, 44, 141

\bibitem[{Makino {et~al.}(2007)Makino, Hiraki, \& Inaba}]{2007Makino}
Makino, J., Hiraki, K., \& Inaba, M. 2007, in Proceedings of the 2007 ACM/IEEE
  Conference on Supercomputing, SC '07 (New York, NY, USA: ACM), 18:1--18:11

\bibitem[{{Mamajek}(2009)}]{m09}
{Mamajek}, E.~E. 2009, in American Institute of Physics Conference Series, Vol.
  1158, American Institute of Physics Conference Series, ed. T.~{Usuda},
  M.~{Tamura}, \& M.~{Ishii}, 3--10

\bibitem[{{Matsumoto} {et~al.}(2017){Matsumoto}, {Oshino}, {Hasegawa}, \&
  {Wakita}}]{moh17}
{Matsumoto}, Y., {Oshino}, S., {Hasegawa}, Y., \& {Wakita}, S. 2017, \apj, 837,
  103

\bibitem[{{Morbidelli} {et~al.}(2009){Morbidelli}, {Bottke}, {Nesvorn{\'y}}, \&
  {Levison}}]{Morbidelli+2009}
{Morbidelli}, A., {Bottke}, W.~F., {Nesvorn{\'y}}, D., \& {Levison}, H.~F.
  2009, \icarus, 204, 558

\bibitem[{{Mordasini} {et~al.}(2016){Mordasini}, {van Boekel}, {Molli{\`e}re},
  {Henning}, \& {Benneke}}]{2016ApJ...832...41M}
{Mordasini}, C., {van Boekel}, R., {Molli{\`e}re}, P., {Henning}, T., \&
  {Benneke}, B. 2016, \apj, 832, 41

\bibitem[{{Morfill} {et~al.}(1993){Morfill}, {Spruit}, \&
  {Levy}}]{1993prpl.conf..939M}
{Morfill}, G., {Spruit}, H., \& {Levy}, E.~H. 1993, in Protostars and Planets
  III, ed. E.~H. {Levy} \& J.~I. {Lunine}, 939--978

\bibitem[{{Morishima} {et~al.}(2008){Morishima}, {Schmidt}, {Stadel}, \&
  {Moore}}]{2008Morishima}
{Morishima}, R., {Schmidt}, M.~W., {Stadel}, J., \& {Moore}, B. 2008, \apj,
  685, 1247

\bibitem[{{Morris} \& {Desch}(2010)}]{2010ApJ...722.1474M}
{Morris}, M.~A. \& {Desch}, S.~J. 2010, \apj, 722, 1474

\bibitem[{{Movshovitz} {et~al.}(2016){Movshovitz}, {Nimmo}, {Korycansky},
  {Asphaug}, \& {Owen}}]{Movshovitz:2016aa}
{Movshovitz}, N., {Nimmo}, F., {Korycansky}, D.~G., {Asphaug}, E., \& {Owen},
  J.~M. 2016, \icarus, 275, 85

\bibitem[{{Murray} \& {Dermott}(1999)}]{SSD}
{Murray}, C.~D. \& {Dermott}, S.~F. 1999, {Solar system dynamics}

\bibitem[{{Nagasawa} {et~al.}(2008){Nagasawa}, {Ida}, \&
  {Bessho}}]{2008ApJ...678..498N}
{Nagasawa}, M., {Ida}, S., \& {Bessho}, T. 2008, \apj, 678, 498

\bibitem[{{Nagasawa} {et~al.}(2000){Nagasawa}, {Tanaka}, \&
  {Ida}}]{2000Nagasawa}
{Nagasawa}, M., {Tanaka}, H., \& {Ida}, S. 2000, \aj, 119, 1480

\bibitem[{{Nelson} {et~al.}(2000){Nelson}, {Papaloizou}, {Masset}, \&
  {Kley}}]{2000MNRAS.318...18N}
{Nelson}, R.~P., {Papaloizou}, J.~C.~B., {Masset}, F., \& {Kley}, W. 2000,
  \mnras, 318, 18

\bibitem[{{O'Brien} {et~al.}(2006){O'Brien}, {Morbidelli}, \&
  {Levison}}]{2006Icar..184...39O}
{O'Brien}, D.~P., {Morbidelli}, A., \& {Levison}, H.~F. 2006, \icarus, 184, 39

\bibitem[{{Oka} {et~al.}(2011){Oka}, {Nakamoto}, \& {Ida}}]{2011Oka}
{Oka}, A., {Nakamoto}, T., \& {Ida}, S. 2011, \apj, 738, 141

\bibitem[{{Ormel} \& {Klahr}(2010)}]{Ormel&Klahr2010}
{Ormel}, C.~W. \& {Klahr}, H.~H. 2010, \aap, 520, A43

\bibitem[{{Paardekooper} {et~al.}(2010){Paardekooper}, {Baruteau}, {Crida}, \&
  {Kley}}]{2010MNRAS.401.1950P}
{Paardekooper}, S.-J., {Baruteau}, C., {Crida}, A., \& {Kley}, W. 2010, \mnras,
  401, 1950

\bibitem[{{Piso} {et~al.}(2015){Piso}, {{\"O}berg}, {Birnstiel}, \&
  {Murray-Clay}}]{2015ApJ...815..109P}
{Piso}, A.-M.~A., {{\"O}berg}, K.~I., {Birnstiel}, T., \& {Murray-Clay}, R.~A.
  2015, \apj, 815, 109

\bibitem[{{Pollack} {et~al.}(1996){Pollack}, {Hubickyj}, {Bodenheimer},
  {Lissauer}, {Podolak}, \& {Greenzweig}}]{1996Icar..124...62P}
{Pollack}, J.~B., {Hubickyj}, O., {Bodenheimer}, P., {Lissauer}, J.~J.,
  {Podolak}, M., \& {Greenzweig}, Y. 1996, \icarus, 124, 62

\bibitem[{{Rasio} \& {Ford}(1996)}]{1996Sci...274..954R}
{Rasio}, F.~A. \& {Ford}, E.~B. 1996, Science, 274, 954

\bibitem[{{Ricard} {et~al.}(2017){Ricard}, {Bercovici}, \&
  {Albar{\`e}de}}]{Ricard:2017aa}
{Ricard}, Y., {Bercovici}, D., \& {Albar{\`e}de}, F. 2017, \icarus, 285, 103

\bibitem[{{Rubin}(2010)}]{2010GeCoA..74.4807R}
{Rubin}, A.~E. 2010, \gca, 74, 4807

\bibitem[{{Sanders} \& {Scott}(2012)}]{Sanders:2012aa}
{Sanders}, I.~S. \& {Scott}, E.~R.~D. 2012, Meteoritics and Planetary Science,
  47, 2170

\bibitem[{{Schlichting} \& {Mukhopadhyay}(2018)}]{Schlichting:2018aa}
{Schlichting}, H.~E. \& {Mukhopadhyay}, S. 2018, \ssr, 214, 34

\bibitem[{{Scott}(2007)}]{s07}
{Scott}, E.~R.~D. 2007, Annual Review of Earth and Planetary Sciences, 35, 577

\bibitem[{{Scott} \& {Krot}(2005)}]{sk05}
{Scott}, E.~R.~D. \& {Krot}, A.~N. 2005, {Chondrites and their Components}, ed.
  A.~M. {Davis}, H.~D. {Holland}, \& K.~K. {Turekian} (Elsevier B), 143

\bibitem[{{Shu} {et~al.}(2001){Shu}, {Shang}, {Gounelle}, {Glassgold}, \&
  {Lee}}]{2001ApJ...548.1029S}
{Shu}, F.~H., {Shang}, H., {Gounelle}, M., {Glassgold}, A.~E., \& {Lee}, T.
  2001, \apj, 548, 1029

\bibitem[{{Shu} {et~al.}(1996){Shu}, {Shang}, \& {Lee}}]{1996Sci...271.1545S}
{Shu}, F.~H., {Shang}, H., \& {Lee}, T. 1996, Science, 271, 1545

\bibitem[{Simon {et~al.}(2017)Simon, Armitage, Youdin, \& Li}]{Simon:2017aa}
Simon, J.~B., Armitage, P.~J., Youdin, A.~N., \& Li, R. 2017, The Astrophysical
  Journal, 847, L12

\bibitem[{{Souli{\'e}} {et~al.}(2017){Souli{\'e}}, {Libourel}, \&
  {Tissandier}}]{Soulie:2017aa}
{Souli{\'e}}, C., {Libourel}, G., \& {Tissandier}, L. 2017, Meteoritics and
  Planetary Science, 52, 225

\bibitem[{{Standish} \& {Williams}(2010)}]{standish_ch8}
{Standish}, E. \& {Williams}, J. 2010, available at: {\tt
  ftp://ssd.jpl.nasa.gov/pub/eph/planets/ioms/ExplSupplChap8.pdf}

\bibitem[{{Standish} {et~al.}(1992){Standish}, {Newhall}, {Williams}, \&
  {Yeomans}}]{standish}
{Standish}, E.~M., {Newhall}, X.~X., {Williams}, J.~G., \& {Yeomans}, D.~K.
  1992, {Orbital Ephemerides of the Sun, Moon, and Planets}, ed. P.~K.
  Seidelmann (University Science Books)

\bibitem[{{Stewart} \& {Leinhardt}(2012)}]{2012ApJ...751...32S}
{Stewart}, S.~T. \& {Leinhardt}, Z.~M. 2012, \apj, 751, 32

\bibitem[{{Sugita} \& {Schultz}(1999)}]{Sugita:1999aa}
{Sugita}, S. \& {Schultz}, P.~H. 1999, \jgr, 104, 30825

\bibitem[{{Thommes} {et~al.}(2003){Thommes}, {Duncan}, \&
  {Levison}}]{2003Icar..161..431T}
{Thommes}, E.~W., {Duncan}, M.~J., \& {Levison}, H.~F. 2003, \icarus, 161, 431

\bibitem[{{Tsang} {et~al.}(2014){Tsang}, {Turner}, \&
  {Cumming}}]{2014ApJ...782..113T}
{Tsang}, D., {Turner}, N.~J., \& {Cumming}, A. 2014, \apj, 782, 113

\bibitem[{{Tsiganis} {et~al.}(2005){Tsiganis}, {Gomes}, {Morbidelli}, \&
  {Levison}}]{2005Natur.435..459T}
{Tsiganis}, K., {Gomes}, R., {Morbidelli}, A., \& {Levison}, H.~F. 2005, \nat,
  435, 459

\bibitem[{{Villeneuve} {et~al.}(2009){Villeneuve}, {Chaussidon}, \&
  {Libourel}}]{Villeneuve:2009aa}
{Villeneuve}, J., {Chaussidon}, M., \& {Libourel}, G. 2009, Science, 325, 985

\bibitem[{{Villeneuve} {et~al.}(2015){Villeneuve}, {Libourel}, \&
  {Souli{\'e}}}]{Villeneuve:2015aa}
{Villeneuve}, J., {Libourel}, G., \& {Souli{\'e}}, C. 2015, \gca, 160, 277

\bibitem[{{Visser} \& {Ormel}(2016)}]{Visser:2016aa}
{Visser}, R.~G. \& {Ormel}, C.~W. 2016, \aap, 586, A66

\bibitem[{Wakita {et~al.}(2018)Wakita, Hasegawa, \& Nozawa}]{Wakita:2018aa}
Wakita, S., Hasegawa, Y., \& Nozawa, T. 2018, Astrophysical Journal, 863

\bibitem[{{Wakita} {et~al.}(2017){Wakita}, {Matsumoto}, {Oshino}, \&
  {Hasegawa}}]{2017ApJ...834..125W}
{Wakita}, S., {Matsumoto}, Y., {Oshino}, S., \& {Hasegawa}, Y. 2017, \apj, 834,
  125

\bibitem[{{Walsh} {et~al.}(2011){Walsh}, {Morbidelli}, {Raymond}, {O'Brien}, \&
  {Mandell}}]{2011Natur.475..206W}
{Walsh}, K.~J., {Morbidelli}, A., {Raymond}, S.~N., {O'Brien}, D.~P., \&
  {Mandell}, A.~M. 2011, \nat, 475, 206

\bibitem[{{Weber} {et~al.}(2018){Weber}, {Ben{\'\i}tez-Llambay}, {Gressel},
  {Krapp}, \& {Pessah}}]{Weber+2018}
{Weber}, P., {Ben{\'\i}tez-Llambay}, P., {Gressel}, O., {Krapp}, L., \&
  {Pessah}, M.~E. 2018, \apj, 854, 153

\bibitem[{Weiss \& Elkins-Tanton(2013)}]{Weiss:2013aa}
Weiss, B.~P. \& Elkins-Tanton, L.~T. 2013, Annual Review of Earth and Planetary
  Sciences, 41, 529

\bibitem[{{Wetherill} \& {Stewart}(1989)}]{1989Icar...77..330W}
{Wetherill}, G.~W. \& {Stewart}, G.~R. 1989, \icarus, 77, 330

\bibitem[{{Williams} \& {Cieza}(2011)}]{2011ARA&A..49...67W}
{Williams}, J.~P. \& {Cieza}, L.~A. 2011, \araa, 49, 67

\bibitem[{Wilson \& Keil(2017)}]{Wilson:2017aa}
Wilson, L. \& Keil, K. 2017, Arguments for the non-existence of magma oceans in
  asteroids (Cambridge University Press), 159--179

\bibitem[{{Winn} \& {Fabrycky}(2015)}]{2015ARAA..53..409W}
{Winn}, J.~N. \& {Fabrycky}, D.~C. 2015, \araa, 53, 409

\bibitem[{{Wisdom}(1982)}]{1982Wisdom}
{Wisdom}, J. 1982, \aj, 87, 577

\bibitem[{{Wisdom}(1983)}]{1983Wisdom}
---. 1983, Meteoritics, 18, 422

\bibitem[{{Wood}(1963)}]{1963Icar....2..152W}
{Wood}, J.~A. 1963, \icarus, 2, 152

\bibitem[{{Young} {et~al.}(2019){Young}, {Shahar}, {Nimmo}, {Schlichting},
  {Schauble}, {Tang}, \& {Labidi}}]{Young:2019aa}
{Young}, E.~D., {Shahar}, A., {Nimmo}, F., {Schlichting}, H.~E., {Schauble},
  E.~A., {Tang}, H., \& {Labidi}, J. 2019, \icarus, 323, 1

\end{thebibliography}

\clearpage

\clearpage

\clearpage

\clearpage

\clearpage
\setlength\textfloatsep{5pt}

\clearpage

\clearpage

\clearpage

\clearpage

\clearpage

\clearpage

\clearpage

\clearpage

\clearpage

\clearpage

\end{document}